\documentclass[fleqn,usenatbib]{mnras}

\usepackage{newtxtext,newtxmath}

\usepackage[T1]{fontenc}

\DeclareRobustCommand{\VAN}[3]{#2}
\let\VANthebibliography\thebibliography
\def\thebibliography{\DeclareRobustCommand{\VAN}[3]{##3}\VANthebibliography}

\usepackage{graphicx}	
\usepackage{amsmath}	
\usepackage{siunitx}
\usepackage{comment}
\usepackage{soul}

\newcommand\Lya{Ly$\alpha$}

\newcommand\RLya{$R_{\rm Ly\alpha}$} 
\newcommand\RCM{$R_{21}$}

\newcommand\GHI{$\Gamma_{\rm HI}$}

\newcommand\HI{\hbox{H$\,\rm \scriptstyle I$}~}
\newcommand\HII{\hbox{H$\,\rm \scriptstyle II$}~} 

\newcommand\HeII{\hbox{He$\,\rm \scriptstyle II$}~}
\newcommand\HeIII{\hbox{He$\,\rm \scriptstyle III$}~} 
\newcommand\OI{\hbox{O$\,\rm \scriptstyle I$}~} 
\newcommand\MgII{\hbox{Mg$\,\rm \scriptstyle II$}~} 
\newcommand\CII{\hbox{[C$\,\rm \scriptstyle II$]}~} 
\newcommand\SiIV{\hbox{SI$\,\rm \scriptstyle IV$}~}

\DeclareSIUnit\erg{erg}
\DeclareSIUnit\parsec{pc}
\DeclareSIUnit\propMpc{pMpc}
\DeclareSIUnit\comMpc{cMpc}
\DeclareSIUnit\angstrom{\text {Å}}

\newcommand{\jsb}[1]{{\textcolor{red}{\bf JSB:#1}}}

\title[Quasar lifetimes and proximate 21-cm absorption]{Probing quasar lifetimes with proximate \boldmath{$21$}-centimetre absorption in the diffuse intergalactic medium at redshifts \boldmath{$z\geq 6$}}

\author[T. Šoltinský et al.]{Tomáš Šoltinský$^{1}$\thanks{E-mail: tomas.soltinsky@nottingham.ac.uk},
James S. Bolton$^{1}$,
Margherita Molaro$^{1}$,
Nina Hatch$^{1}$,
Martin G. Haehnelt$^{2}$,
\newauthor Laura C. Keating$^{3,4}$,
Girish Kulkarni$^{5}$
\& Ewald Puchwein$^{3}$
\\
$^{1}$School of Physics and Astronomy, University of Nottingham, University Park, Nottingham, NG7 2RD, UK\\
$^{2}$Kavli Institute for Cosmology and Institute of Astronomy, Madingley Road, Cambridge, CB3 0HA, UK\\
$^{3}$Leibniz-Institut f\"ur Astrophysik Potsdam, An der Sternwarte 16, 14482 Potsdam, Germany\\
$^{4}$Institute for Astronomy, University of Edinburgh, Blackford Hill, Edinburgh,
EH9 3HJ, UK\\
$^{5}$Tata Institute of Fundamental Research, Homi Bhabha Road, Mumbai 400005, India\\}

\date{Accepted XXX. Received YYY; in original form ZZZ}

\pubyear{2022}

\begin{document}
\label{firstpage}
\pagerange{\pageref{firstpage}--\pageref{lastpage}}
\maketitle

\begin{abstract}
Enhanced ionizing radiation in close proximity to redshift $z\gtrsim 6$ quasars creates short windows of intergalactic \Lya\ transmission blueward of the quasar \Lya\ emission lines. The majority of these \Lya\ near-zones are consistent with quasars that have optically/UV bright lifetimes of $t_{\rm Q}\sim 10^{5}-10^{7}\rm\,yr$.  However, lifetimes as short as $t_{\rm Q}\lesssim 10^{4}\rm\,yr$ appear to be required by the smallest \Lya\ near-zones.  These short lifetimes present an apparent challenge for the growth of $\sim 10^{9}\rm\,M_{\odot}$ black holes at $z\gtrsim 6$.  Accretion over longer timescales is only possible if black holes grow primarily in an obscured phase, or if the quasars are variable on timescales comparable to the equilibriation time for ionized hydrogen.  Distinguishing between very young quasars and older quasars that have experienced episodic accretion with \Lya\ absorption alone is challenging, however.  We therefore predict the signature of proximate 21-cm absorption around $z\gtrsim 6$ radio-loud quasars.  For modest pre-heating of intergalactic hydrogen by the X-ray background, where the spin temperature $T_{\rm S} \lesssim 10^{2}\rm\,K$ prior to any quasar heating, we find proximate 21-cm absorption should be observable in the spectra of radio-loud quasars.  The extent of the proximate 21-cm absorption is sensitive to the integrated lifetime of the quasar.  Evidence for proximate 21-cm absorption from the diffuse intergalactic medium within $2-3\rm\,pMpc$ of a (radio-loud) quasar would be consistent with a short quasar lifetime, $t_{\rm Q}\lesssim 10^{5}\rm\,yr$, and would provide a complementary constraint on models for high redshift black hole growth.

\end{abstract}



\section{Introduction}

The intergalactic medium (IGM) becomes opaque to \Lya~photons approaching the end stages of reionization at $z\gtrsim 5.5$, when the average neutral hydrogen fraction $\langle x_{\rm HI}\rangle \gtrsim 10^{-4}$ \citep[for a review see][]{Becker_Bolton_Lidz_2015}.   However, in close proximity to highly luminous quasars at $z\gtrsim 5.5$, local enhancements in the ionizing radiation field leave short windows of \Lya\ transmission blueward of the quasar \Lya\ emission line.  These regions -- referred to as \Lya\ near-zones or proximity zones -- are typically $1$--$10$ proper Mpc (pMpc) in extent \citep{Fan_2006,Carilli_2010,Willott_2010,Venemans_2015,Reed_2015,Eilers_2017,Eilers_2021,Mazzucchelli_2017,Ishimoto_2020}.  Several near-zones at $z\simeq 7$  also exhibit evidence for \Lya\ damping wings that extend redward of the quasar systemic redshift \citep{Mortlock_2011,Banados_2018_z754QSO,Wang_2020,Yang_2020_z75}, which is expected if the surrounding IGM is substantially neutral \citep{MiraldaEscudeRees_1998}.  Early work suggested that \Lya\ near-zones may be tracing quasar \HII regions embedded in an otherwise largely neutral IGM \citep[e.g.][]{ShapiroGiroux1987,CenHaiman2000,MadauRees2000,WyitheLoeb2004}.  Subsequent radiative transfer modelling \citep{Bolton_Haehnelt_2007,Maselli2007,Lidz_2007,Wyithe2008_NZ} demonstrated a more complex picture, where the \Lya\ near-zones at $z\simeq 6$ may also be explained if the quasars are surrounded by a highly ionized IGM -- analogous to the classical proximity effect at lower redshift \citep[e.g.][]{Murdoch1986,Bajtlik1988}.  

In the last decade the number of $z\gtrsim 6$ quasar spectra with well measured \Lya\ near-zone sizes has grown considerably. Over $280$ quasars at $z>6$ have now been discovered \citep[see e.g.][]{Bosman_2022}.  Submillimetre observations have provided improved measurements of quasar systemic redshifts, yielding better estimates of the \Lya\ near-zone sizes \citep{Eilers_2021}.  After correcting for differences in the intrinsic luminosity of the quasars, the scatter in the $\sim 80$ published \Lya\ near-zone sizes can be largely explained by a combination of cosmic variance \citep{Keating_2015}, differences in the optically/UV bright lifetime of the quasars \citep{Morey_2021}, and perhaps the occasional proximate high column density absorption system \citep{Chen_Gnedin_2021_DistEvoPZ}.  The observed \Lya\ near-zone size distribution is reasonably well reproduced if a highly ionized IGM surrounds the quasars at $z\simeq 6$ \citep{Wyithe2008_NZ,Morey_2021}.  However, the \Lya\ damping wings in the spectra of several $z>7$ quasars are suggestive of a substantially more neutral IGM by $z\simeq 7$, such that $\langle x_{\rm HI}\rangle >0.1$,  \citep[][but see also \cite{BosmanBecker_2015}]{Bolton2011,Greig_2017,Greig_2022,Davies_2018}. 

Several recent studies have focused on constraining optically/UV bright quasar lifetimes, $t_{\rm Q}$, from the \Lya\ near-zone data at $z\simeq 6$.  \citet{Morey_2021} find an average optically/UV bright lifetime of $t_{\rm Q}\sim 10^{6}\rm\,yr$ is consistent with the transmission profiles of most \Lya\ near-zones at $z\simeq 6$.   \cite{Eilers_2017, Eilers_2021} have furthermore presented several very small \Lya\ near-zones with luminosity corrected sizes of $\lesssim 1\rm\,pMpc$, consistent with optically/UV bright lifetimes of $t_{\rm Q} \lesssim 10^{4}$--$10^{5}\rm\,yr$.  These small \Lya\ near-zones represent $\lesssim 10$ per cent of all quasar \Lya\ near-zones at $z\simeq  6$. However, if the black holes powering these quasars accrete most of their mass when the quasars are optically/UV bright, such a short average lifetime is in significant tension with the build up of $\sim 10^{9}\rm\,M_{\odot}$ supermassive black holes by $z=6$; the e-folding time for Eddington limited accretion is at least an order of magnitude larger.  Possible solutions are radiatively inefficient, mildly super-Eddington accretion \citep{Madau_2014,Davies_2019,Kroupa_2020}, black holes that grow primarily in an obscured, optically/UV faint phase \citep{Hopkins_2005,Ricci_2017obsc} or episodic accretion that produces  ``flickering'' quasar light curves \citep{Schawinski_2015,Davies_2020}. 
  
Observationally distinguishing between very young quasars and older quasars that have experienced episodic or obscured accretion with \Lya\ near-zones is challenging, however.  Another possibility is detecting the 21-cm signal from neutral hydrogen around the quasars.  In principle, if the foregrounds can be accurately removed, the sizes of quasar \HII regions may be measured directly with 21-cm tomography; the neutral, X-ray heated hydrogen outside of the quasar \HII region should appear in emission against the radio background \citep[e.g.][]{Wyithe_2004,Kohler_2005,RhookHaehnelt_2006,GeilWyithe_2008,Majumdar_2012,Datta_2012,Kakiichi_2017,Ma_2020,Davies_2021}.  Assuming the recombination timescale $t_{\rm rec}\gg t_{\rm Q}$, 21-cm tomography measurements  would enable a direct determination of the quasar age, because the \HII region size $R_{\rm HII}\propto t_{\rm Q}^{1/3}$ (see e.g. Eq.~(\ref{eq:RHII}) later).  A related approach that has received less attention is to instead consider the forest of redshifted 21-cm absorption expected from the neutral IGM in the spectra of radio-loud background sources at $z\gtrsim 6$ \citep[for recent examples of potential background sources, see e.g.][]{Belladitta_2020,Ighina_2021,Banados_2021,Liu_2021}.   Unlike tomography, observing the IGM in 21-cm absorption allows small-scale IGM structure to be resolved and it is (in principle) a simpler observation that does not rely on the removal of challenging foregrounds \citep[see e.g.][]{Carilli_2002,Furlanetto_2002,Furlanetto_2006a,Meiksin_2011,Xu_2011,Ciardi_2013,Semelin_2016,VillanuevaDomingo2021}.  

\cite{Soltinsky_2021} recently discussed the detectability of the 21-cm forest in the context of the late ($z \simeq 5.3$)  reionization models \citep[e.g.][]{Kulkarni_2019,Keating_2020,Nasir_2020,Qin_2021,Choudhury_2021} that appear to be favoured by the large variations found in the \Lya\ forest effective optical depth at $z>5$ \citep{Becker_2015,Eilers_2018,Yang_2020_z63,Bosman_2018,Bosman2022}.  \cite{Soltinsky_2021} noted that, for modest X-ray pre-heating, such that the IGM spin temperature $T_{\rm S}\lesssim 10^{2}\rm\,K$, strong 21-cm forest absorption with optical depths $\tau_{21}\geq 10^{-2}$ will persist until $z=6$ in late reionization models.   A null detection of the 21-cm forest at $z=6$ would also place useful limits on the soft X-ray background.  Toward higher redshifts, $z>7$, strong 21-cm forest absorbers will become significantly more abundant, particularly if the spin and kinetic temperatures are not tightly coupled (see e.g. fig. 7 in \cite{Soltinsky_2021}).  

In this context, \citet{Banados_2021} have recently reported the discovery of a radio-loud quasar PSO J172$+$18 at $z=6.82$, with an absolute AB magnitude $M_{1450}=-25.81$ and an optical/near-infrared spectrum that exhibits a \Lya\ near-zone size $R_{\rm Ly\alpha}=3.96\pm 0.48\rm\,pMpc$.  This raises the intriguing possibility of also obtaining a radio spectrum from this or similar objects with low frequency radio interferometry arrays \citep[see also e.g.][]{Gloudemans2022}.   For spin temperatures of $T_{\rm S}\sim 10^{2}\rm\,K$ in the pre-reionization IGM, in late reionization scenarios there will be proximate 21-cm absorption from neutral islands in the diffuse IGM that will approximately trace the extent of the quasar \HII region.  If this proximate 21-cm absorption is detected, either for an individual radio-loud quasar or within a population of objects, it would provide another possible route to constraining the lifetime of high redshift quasars.  In particular, when combined with \Lya\ near-zone sizes, such a measurement could help distinguish between quasars that are very young (as is suggested if taking the \citet{Eilers_2017,Eilers_2021} \Lya\ near-zone data at face value), or that are much older and have only recently transitioned to an optically/UV bright phase.  

Our goal is to explore this possibility by modelling the properties of proximate 21-cm absorbers in the diffuse IGM around (radio-loud) quasars.  We do this by building on the simulation framework presented in \citet{Soltinsky_2021}, who used the Sherwood-Relics simulations \citep[see][]{Puchwein_2022} of inhomogeneous, late reionization to predict the properties of the 21-cm forest.  In this work, we now additionally couple Sherwood-Relics with a line of sight radiative transfer code that simulates the photo-ionization and photo-heating around bright quasars \citep[for similar approaches see e.g.][]{Bolton_Haehnelt_2007,Lidz_2007,Davies_2020,Chen_Gnedin_2021_DistEvoPZ,Satyavolu_2022}.   

We begin by describing our fiducial quasar spectral energy distribution and the effect of the quasar UV and soft X-ray radiation on proximate \Lya\ and 21-cm absorption using a simplified, homogeneous IGM model in Section \ref{sec:model}.  We then introduce a more realistic model by using the Sherwood-Relics simulations in Section~\ref{sec:inhomog_model}, and validate our model by comparing the predicted \Lya\ near-zone sizes in our simulations to observational data.  Our predictions for the extent of the proximate 21-cm absorption around $z\geq 6$ quasars for a constant ``light bulb'' quasar emission model are presented in Section~\ref{sec:21cmNZ}.  In Section~\ref{sec:flicker} we then extend this model to include ``flickering'' quasar light curves that may be appropriate for episodic black hole accretion, and discuss the implications for constraining quasar lifetimes and black hole growth.  Finally, we summarise and conclude in Section~\ref{sec:conclusions}.  Supplementary information may be found in the Appendices at the end of the paper.


\section{Quasar radiative transfer model}\label{sec:model}
\subsection{The quasar spectral energy distribution}\label{sec:quasar_model}

\begin{figure*}
    \begin{minipage}{2\columnwidth}
 	  \centering
 	  \includegraphics[width=\linewidth]{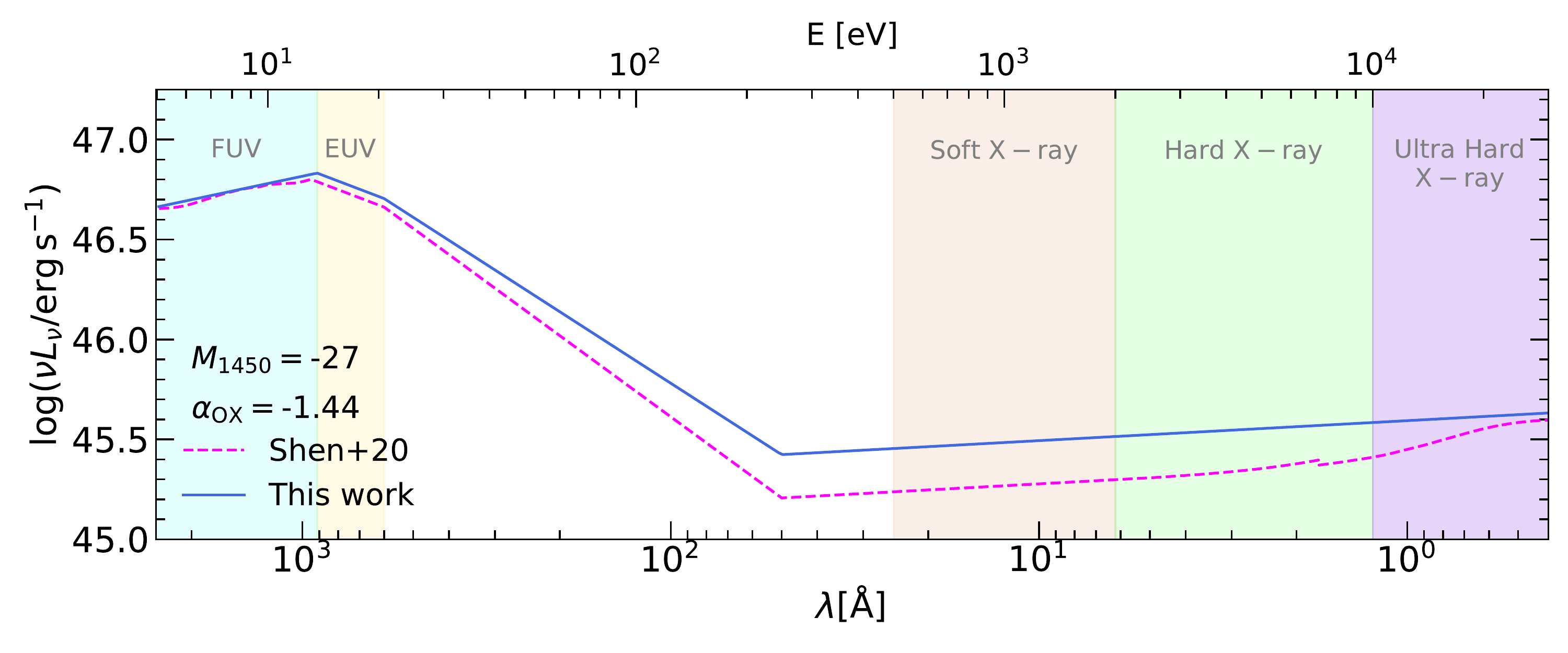}
	\end{minipage}
	\vspace{-0.3cm}
    \caption{The fiducial power-law quasar SED used in this work (solid blue curve) compared to the SED template from \citet[][]{Shen_2020} (dashed fuchsia curve). Both SEDs are normalised at $1450\,\si{\angstrom}$ to correspond to an absolute AB magnitude $M_{1450}=-27$. The SED is modelled as a broken power law, $f_{\nu}\propto \nu^{\alpha}$, with spectral index $\alpha_{\rm FUV}=-0.61$ between $\lambda=912\,\si{\angstrom}-2500\,\si{\angstrom}$ (far UV), $\alpha_{\rm EUV}=-1.70$ between $\lambda=600\,\si{\angstrom}-912\,\si{\angstrom}$ (extreme UV) and $\alpha_{\rm X}=-0.9$ at $\lambda\leq50\,\si{\angstrom}$ (X-ray). The X-ray part of the spectrum is normalized with an optical-to-X-ray spectral index of $\alpha_{\rm OX}=-1.44$. The SED between $\lambda=50\,\si{\angstrom}-600\,\si{\angstrom}$ connects the UV and X-ray sections of the spectrum.  The shaded regions indicate common wavelength bands.   Our fiducial model corresponds to an ionizing photon emission rate of $\dot{N}=1.64\times 10^{57}\rm\,s^{-1}$.}
    \label{fig:SED_range}
\end{figure*}

\begin{figure*}
    \begin{minipage}{1.9\columnwidth}
 	  \centering
 	  \includegraphics[width=\linewidth]{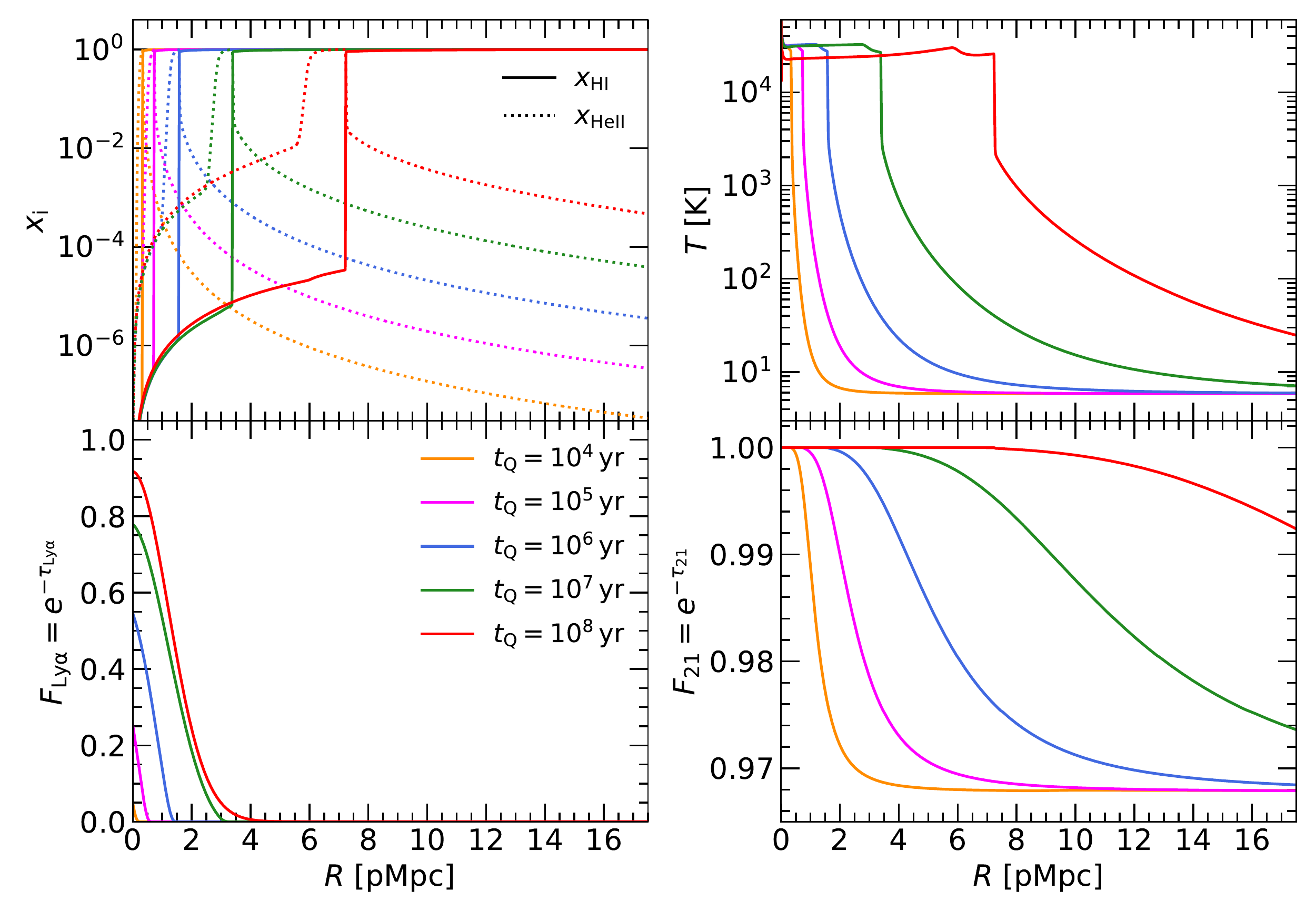}
	\end{minipage}
	\vspace{-0.3cm}
    \caption{Radiative transfer simulation of UV and X-ray photons emitted by a quasar into a uniform density field with $\Delta=\rho/\langle \rho \rangle=1$ at $z=7$.   The hydrogen and helium gas is assumed to be initially cold and neutral, and the quasar has an absolute AB magnitude $M_{\rm 1450}=-27$ (corresponding to an ionizing photon emissivity of $\dot N=1.64\times 10^{57}\rm\,s^{-1}$ for our fiducial SED in Fig.~\ref{fig:SED_range}).  Curves with different colours show different values for the optically/UV bright lifetime of the quasar, $t_{\rm Q}$, as indicated in the lower left panel.  \emph{Upper left:} the \HI fraction (solid curves, $x_{\rm HI}=n_{\rm HI}/n_{\rm H}$) and \HeII fraction (dotted curves, $x_{\rm HeII}=n_{\rm HeII}/n_{\rm He}$).  \emph{Upper right:} the gas kinetic temperature $T$.  We assume strong coupling of the spin temperature in the vicinity of the quasar, such that the spin temperature $T_{\rm S}=T$.   \emph{Lower left:} the \Lya\ transmission, $F_{\rm Ly\alpha}=e^{-\tau_{\rm Ly\alpha}}$.  \emph{Lower right:}  the 21-cm transmission, $F_{21}=e^{-\tau_{21}}$. }
    \label{fig:uniform_profiles}
\end{figure*}

The effect of UV and X-ray ionizing photons emitted by quasars on the high redshift IGM is simulated using the 1D multi-frequency radiative transfer (RT) calculation first described by \citet[][]{Bolton_Haehnelt_2007}, and subsequently updated in \citet{Knevitt2014} to include X-rays and secondary ionizations by fast photo-electrons \citep{FurlanettoStoever2010}.  In brief, as an input this model takes the gas overdensity $\Delta$, peculiar velocity $v_{\rm pec}$, neutral hydrogen fraction $x_{\rm HI}$, gas temperature $T$, and background photo-ionization rate \GHI, from sight lines drawn through a hydrodynamical simulation (see Section~\ref{sec:simulations} for further details).  We assume a spectral energy distribution (SED) for the quasar, and follow the RT of ionizing photons through hydrogen and helium gas along a large number of individual sight lines, all of which start at the position of a halo.  Our RT simulations track ionizing photons emitted by the quasar at energies between $13.6\rm\,eV$ and $30\rm\,keV$,  using 80 logarithmically spaced photon energy bins.

We model the quasar SED as a broken power law, $f_{\nu}\propto \nu^{\alpha}$, as shown in Fig. \ref{fig:SED_range} (blue solid curve).  Our choice of SED is similar to the template from \citet[][]{Shen_2020} (dashed fuchsia curve).   To construct the UV part of the SED, we follow \citet[][]{Lusso_2015} and assume a spectral index  $\alpha_{\rm FUV}=-0.61$  at $912\,\si{\angstrom}\leq\lambda\leq2500\,\si{\angstrom}$ and $\alpha_{\rm EUV}=-1.70$ at $600\,\si{\angstrom}\leq\lambda\leq912\,\si{\angstrom}$.   We choose the spectral index at X-ray energies ($\lambda\leq50\,\si{\angstrom}$) to be $\alpha_{\rm X}=-0.9$, to approximately match the shape of the \citet[][]{Shen_2020} SED. 
The X-ray part of the SED is normalised using the observed correlation between the specific luminosities $L_{\nu}(2500\,\si{\angstrom})$ and $L_{\nu}(2\rm\,keV)$, typically parameterised by the optical-to-X-ray spectral index   \citep{Steffen_2006,Lusso_2010}
\begin{equation}\alpha_{\rm OX}=\frac{\mathrm{log}(L_{\nu}(2\rm\,keV))-\mathrm{log}(L_{\nu}(2500\,\si{\angstrom}))}{\mathrm{log}(\nu(2\rm\,keV))-\mathrm{log}(\nu(2500\,\si{\angstrom}))}. \end{equation}  
\noindent
We assume a fiducial value of $\alpha_{\rm OX}=-1.44$ in this work, but vary this by $\Delta \alpha_{\rm OX}=0.3$ to account for a range of $L_{\nu}(2500\,\si{\angstrom})$ values.  Our fiducial $\alpha_{\rm OX}$ is similar to the best fit value of $\alpha_{\rm OX}=-1.45 \pm 0.11$ recently inferred by \cite{Connor_2021} for a radio-loud quasar at $z=5.831$.    Finally, the spectral shape at $\lambda=50\,\si{\angstrom}-600\,\si{\angstrom}$ is obtained by connecting the UV and X-ray parts of the SED.  

For ease of comparison with previous literature \citep[][]{Eilers_2017,Davies_2020}, we adopt a normalisation for the quasar SED corresponding to an absolute AB magnitude at $1450\,\si{\angstrom}$ of $M_{1450}=-27$ and a specific luminosity $L_{\nu}(2500\,\si{\angstrom})=3.8\times10^{31}\rm\,ergs^{-1}Hz^{-1}$. 
The ionizing photon (i.e. $E>13.6\rm\,eV$) emission rate of the quasar is given by
\begin{equation}\dot{N}=\int_{\nu(13.6\rm\,eV)}^{\nu(30\rm\,keV)}\frac{L_{\nu}}{h_{\rm p}\nu}\mathrm{d}\nu, \end{equation}
\noindent
where $h_{\rm p}$ is the Planck constant. For $\alpha_{\rm OX}=-1.44$, this results in $\dot{N}=1.64\times10^{57}\rm\,s^{-1}$.  For most of this study we will furthermore assume a constant luminosity ``light bulb'' model for the quasar light curve \citep[e.g.][]{Bolton_Haehnelt_2007}.    However, in Section \ref{sec:flicker} we will also consider a model where the quasar luminosity varies with time \citep[cf.][]{Davies_2020}.


\begin{figure}
    \begin{minipage}{\columnwidth}
 	  \centering
 	  \includegraphics[width=\linewidth]{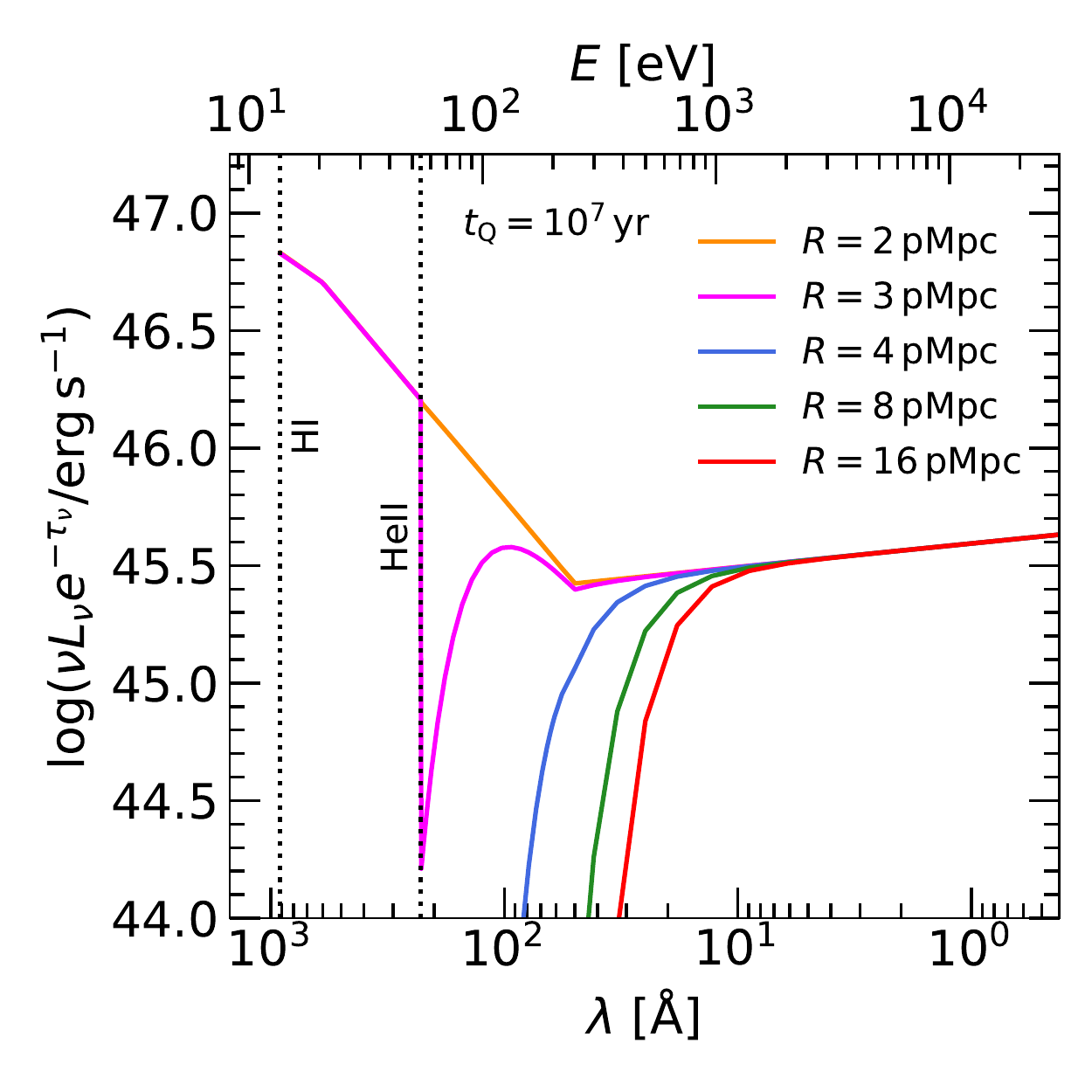}
	\end{minipage}
	\vspace{-0.4cm}
    \caption{Ionizing spectrum (i.e. photon energies $E>13.6\rm\,eV$) at different distances from a $M_{1450}=-27$ quasar after following the radiative transfer of the intrinsic quasar SED displayed in Fig.~\ref{fig:SED_range} through a uniform, neutral IGM with density $\Delta=1$ at $z=7$.  The spectrum corresponds to the model shown by the green curves in Fig.~\ref{fig:uniform_profiles} for an optically/UV bright lifetime of $t_{\rm Q}=10^{7}\rm\,yr$.   The ionization thresholds for \HI and \HeII are displayed as dotted vertical lines.  Note that only X-ray photons propagate unimpeded beyond the \HII ionization front, which is located at $R=3.5\rm\,pMpc$ in Fig.~\ref{fig:uniform_profiles}.}
    \label{fig:uniform_spectra}
\end{figure}

\subsection{Ly\texorpdfstring{$\alpha$}{alpha} and 21-cm  absorption in a homogeneous medium}\label{sec:homog_model}

We examine the \Lya\ and 21-cm absorption in the vicinity of bright quasars by constructing mock absorption spectra from the sight lines extracted from our RT simulations.   We calculate the \Lya~optical depth, $\tau_{\rm Ly\alpha}$, along each quasar sight line following \citet[][]{Bolton_Haehnelt_2007} (see their eq. (15)), where we use the  \citet[][]{TepperGarcia_2006} approximation for the Voigt line profile.  To compute the 21-cm forest optical depth, $\tau_{21}$, we follow the approach described in \citet[][]{Soltinsky_2021} and assume a Gaussian line profile (see their eq. (9)).  We shall assume strong \Lya~coupling when calculating the 21-cm optical depths, such that the hydrogen spin temperature, $T_{\rm S}$, is equal to the gas kinetic temperature, $T$.  At the redshifts ($z\leq 8$) and typical gas kinetic temperatures ($T\lesssim 10^2\rm\,K)$ considered in this work,  the hydrogen spin temperature, $T_{\rm S}$, should be strongly coupled to the gas kinetic temperature, $T$, for reasonable assumptions regarding the \Lya\ background,  even in the absence of a nearby quasar \citep[see fig. 3 of][]{Soltinsky_2021}. Although we do not model the \Lya~photons emitted by the quasar, these would promote even stronger coupling of $T_{\rm S}$ and $T$ in the proximate gas by locally enhancing the \Lya\ background.  For reference, in the absence of redshift space distortions, the optical depth to 21-cm photons at redshift $z$ is then 
\begin{equation} \tau_{21}(z) =  0.19 x_{\rm HI}\left(\frac{\Delta}{10}\right)\left(\frac{T_{\rm S}}{10\rm \,K}\right)^{-1}\left(\frac{1+z}{8}\right)^{3/2},  \end{equation}
\noindent
where $\Delta=\rho/\langle \rho \rangle$ is the ratio of the gas density to the mean background value, and the factor of $0.19$ is cosmology dependent \citep{Madau_1997}. Strong absorption will therefore arise from dense, cold and significantly neutral hydrogen gas.

First, to develop intuition, we shall consider the propagation of ionizing radiation from a quasar into a homogeneous medium.  We assume $\Delta=\rho/\langle \rho \rangle = 1$, ignore peculiar velocities, and assume the gas is initially cold and neutral.  Fig.~\ref{fig:uniform_profiles} shows the results from an RT simulation for a quasar at $z=7$ with $M_{\rm AB}=-27$, assuming our fiducial SED.  The outputs for different optically/UV bright lifetimes, $t_{\rm Q}$, for the quasar are shown by the coloured curves and are labelled in the lower left panel.

The top left panel in Fig.~\ref{fig:uniform_profiles} shows the neutral hydrogen ($x_{\rm HI}$, solid curves) and singly-ionized helium ($x_{\rm HeII}$, dotted curves) fractions around the quasar. One can see the \HII and \HeIII ionization fronts expanding with time. The hydrogen within the quasar \HII region is highly ionized ($x_{\rm HI}<10^{-4}$), and the gas is optically thin to \Lya\ photons. This is demonstrated in the bottom left panel of Fig.~\ref{fig:uniform_profiles} where we show the \Lya\ transmission, $F_{\rm Ly\alpha}=e^{-\tau_{\rm Ly\alpha}}$.   Note, however, that the \Lya\ transmission does not saturate at the position of the \HII ionization front.  This is particularly apparent for larger optically/UV bright lifetimes, $t_{\rm Q}>10^{7}\rm\,yr$.  This is in part due to the IGM \Lya\ damping wing from the neutral IGM that is evident in the \Lya\ transmission profile \citep{MiraldaEscudeRees_1998,MesingerFurlanetto2008,Bolton2011}, but also because the residual neutral hydrogen density close to the \HII ionization front has already risen above the threshold required for saturated \Lya\ absorption \citep[see e.g.][for further details]{Bolton_Haehnelt_2007,Lidz_2007,Maselli2007,Keating_2015,Eilers_2017,Davies_2020,Chen_Gnedin_2021_DistEvoPZ}.

The gas temperature around the quasar, displayed in the top right panel of Fig.~\ref{fig:uniform_profiles}, is $T\sim 2$--$3\times 10^{4}\rm\,K$ behind the \HII and \HeIII ionization fronts \citep[e.g.][]{D_Aloisio_2019}.  However, there is also heating of the neutral gas \emph{ahead} of the \HII ionization front.  For example, for $t_{\rm Q}=10^{7}\rm\,yr$ (green curve), the average gas temperature ahead of the \HII ionization front position at $R=3.5\rm\,pMpc$ is $\langle T\rangle \sim 100 \rm\,K$.  This heating is due to soft X-ray photons with long mean free paths, $\lambda_{\rm X}$, that can penetrate into the neutral IGM.  For an \HI photo-ionization cross section $\sigma_{\rm HI}=6.34\times 10^{-18}\mathrm{\,cm^{2}}\,(E/13.6\rm\,eV)^{-2.8}$ we obtain
\begin{equation} \lambda_{\rm X} = \frac{1}{n_{\rm HI}\sigma_{\rm HI}} \simeq 1.0\mathrm{\,pMpc}\, x_{\rm HI}^{-1} \Delta^{-1}\left(\frac{E}{0.2 \rm\, keV}\right)^{2.8}\left(\frac{1+z}{8}\right)^{-3}. \end{equation}
The role of X-rays is further evident from Fig.~\ref{fig:uniform_spectra}, which shows the IGM attenuated quasar luminosity, $L_{\nu}e^{-\tau_{\nu}}$, at different distances, $R$, from the quasar assuming an optically/UV bright lifetime of $t_{\rm Q}=10^{7}\rm\,yr$ (the green curves in Fig. \ref{fig:uniform_profiles}).   Beyond the \HII ionization front (i.e. $R \geq3.5\rm\,pMpc$) only X-ray photons  penetrate into the neutral IGM surrounding the quasar \HII region.   This long range X-ray heating acts to suppress the $21$-cm absorption from neutral gas by increasing the \HI spin temperature \cite[see e.g.][]{Xu_2011,Mack_2012,Soltinsky_2021} and thus lowering the 21-cm optical depth.  Note also that at $R=2\rm\,pMpc$ (orange curve in Fig.~\ref{fig:uniform_spectra}) the IGM is optically thin and the quasar spectrum matches the intrinsic SED in Fig.~\ref{fig:SED_range}, while the spectrum at $R=3\rm\,pMpc$ (fuchsia curve) lies between the \HII and \HeIII ionization front and therefore exhibits a strong absorption edge at the \HeII ionization potential, $E=54.4\rm\,eV$. 

The lower right panel of Fig.~\ref{fig:uniform_profiles} shows the resulting $21$-cm transmission, $F_{21}=e^{-\tau_{21}}$, around the quasar.  Here $\tau_{21}\ll 1$ behind the \HII ionization front because the gas is hot and ionized, but where the gas (and spin) temperature decrease to $T=T_{\rm S} <100\rm\, K$, some 21-cm absorption (i.e. $F_{21}<1$) is apparent.  For longer optically/UV bright lifetimes the quasar \HII region expands and X-ray heating extends further into the neutral IGM.  The 21-cm absorption close to the quasar then becomes partially or completely suppressed even if the gas ahead of the \HII ionization front remains largely neutral.   

In summary, we expect the \Lya\  transmission arising from the highly ionized hydrogen around quasars to be influenced by UV photons, but for neutral hydrogen, the $21$-cm forest absorption will be very sensitive to long range heating by the X-ray photons emitted by the quasar. We now turn to consider more detailed simulations of \Lya\ and $21$-cm absorption around quasars using realistic density, peculiar velocity and ionization fields extracted from the Sherwood-Relics simulations.


\section{Near-zones in inhomogeneous reionization simulations}\label{sec:inhomog_model}

\subsection{Hydrodynamical simulations}\label{sec:simulations}

We use a subset of simulations drawn from the Sherwood-Relics project \citep{Puchwein_2022} to generate realistic \Lya~and 21-cm forest spectra around bright quasars. The Sherwood-Relics models are high resolution cosmological hydrodynamical simulations performed with a modified version of \textsc{P-Gadget-3} \citep[][]{Springel_2005}. These are combined with 3D RT simulations of (stellar photon driven) inhomogeneous reionization performed with the moment based, M1-closure code \textsc{ATON} \citep{Aubert_2008}.  Unlike many other radiation-hydrodynamical simulations of patchy reionization \citep[e.g.][]{Gnedin2014,Finlator2018,Ocvirk2021,Lewis2022,Garaldi2022}, Sherwood-Relics uses a novel, hybrid approach for self-consistently coupling the pressure response of the gas on small scales to the inhomogeneous heating from reionization \citep[see also][]{Onorbe_2019}.  The ATON RT simulations are performed first, and the resulting three dimensional maps for the photo-ionization and photo-heating rates are then applied on-the-fly to the hydrodynamical simulations \citep[see][for further details]{Puchwein_2022,Gaikwad_2020,Soltinsky_2021,Molaro_2022}. The main advantage that Sherwood-Relics offers for this work is it provides a model for the spatial variations expected in the \HI fraction and photo-ionization rates around the dark matter haloes hosting bright quasars at $z\geq 6$ \citep[see also][]{Lidz_2007,Satyavolu_2022}. 

All the simulations follow $2\times2048^3$ dark matter and baryon particles in a $(40h^{-1}$cMpc$)^3$ volume, and have a flat $\Lambda$CDM cosmology with $\Omega_{\Lambda}=0.692$, $\Omega_{\rm m}=0.308$, $\Omega_{\rm b}=0.0482$, $\sigma_8=0.829$, $n_{\rm s} =0.961$, $h=0.678$, consistent with \citet[][]{planck2014}.  The assumed primordial helium fraction by mass is $Y=0.24$ \citep[][]{Hsyu_2020}. Gas particles with density $\Delta>10^{3}$ and kinetic temperature $T<10^5 \rm\, K$ are converted into collisionless star particles \citep[][]{Viel_2004}.  Our chosen mass resolution, corresponding to a dark matter particle mass of $7.9\times 10^{5}\rm\,M_{\odot}$, is sufficient for resolving the \Lya\ forest and 21-cm absorption from the diffuse IGM \citep{Gaikwad_2020,Soltinsky_2021}, although note it will not resolve dark  matter haloes with masses $\lesssim 2.5\times 10^{7}\rm\,M_{\odot}$.   

In this work we analyse Sherwood-Relics runs that use the three reionization histories first described by \cite{Molaro_2022} (see their fig. 2), in which reionization completes at $z_{\rm R}=5.3$, $z_{\rm R}=6.0$ and $z_{\rm R}=6.6$ (labelled RT-late, RT-mid and RT-early, respectively).  Here we define $z_{\rm R}$ as the redshift where the volume averaged neutral fraction first falls below $\langle x_{\rm HI}\rangle \sim 10^{-3}$.  The volume averaged \HI\ fractions in the simulations at $z=6,\,7$ and $8$ are listed in Table~\ref{tab:xHI}.  All three models are consistent with existing constraints on $\langle x_{\rm HI}\rangle$ at $z>6$ and the CMB electron scattering optical depth, but the RT-late model in particular is chosen to match the $z_{\rm R}$ required by the large scale fluctuations observed in the \Lya\ forest effective optical depth at $z\gtrsim 5$ \citep{Becker_2015,Kulkarni_2019,Keating_2020,Bosman2022,Zhu2022}.  We use RT-late for our fiducial reionization model in this work.

\begin{table}
    \centering
    \caption{The volume averaged \HI fraction in the IGM, $\langle x_{\rm HI} \rangle$, at redshift $z=6,\,7$ and $8$ for the three Sherwood-Relics simulations used in this work: RT-late, RT-mid and RT-early \citep[see][for further details]{Molaro_2022}.}
    \label{tab:xHI}
    \begin{tabular}{c c c c}
    \hline
    Model & $\langle x_{\rm HI} \rangle,\, z=6$ & $\langle x_{\rm HI} \rangle,\, z=7$  & $\langle x_{\rm HI} \rangle,\, z=8$  \\     
    \hline
    RT-late & $1.42\times10^{-1}$ & $4.75\times10^{-1}$ & $7.07\times10^{-1}$  \\
    RT-mid & $2.39\times10^{-3}$ & $4.44\times10^{-1}$ & $7.12\times10^{-1}$ \\
    RT-early & $7.70\times10^{-6}$ & $1.56\times10^{-1}$ &  $5.49\times10^{-1}$ \\
    \hline
    \end{tabular}
\end{table}

In order to construct realistic quasar sight-lines from Sherwood-Relics simulations, we first use a friends-of-friends halo finder to identify dark matter haloes in the simulations.  We select haloes with mass $>10^{10}\rm\,M_{\odot}$ and extract sight lines in three orthogonal directions around them.  The mass of the dark matter haloes that host supermassive black holes is uncertain, although clustering analyses at lower redshift suggest $\sim 10^{12}\rm\,M_{\odot}$ \cite[e.g.][]{Shen2007},  which is significantly larger than our minimum halo mass.  However, as discussed by \cite{Keating_2015} and \cite{Satyavolu_2022}, the choice of halo mass has a very limited impact on the sizes of quasar \Lya\ near-zones.  This is because the halo bias at $\gtrsim 2\rm\,pMpc$ from a halo at $z\gtrsim 6$ is very small \citep[see also][]{Calverley_2011,Chen2022}.  We have confirmed this is also true for the 21-cm absorption from the diffuse IGM we consider in this work.  Next, we splice these halo sight lines (consisting of the gas overdensity $\Delta$, gas peculiar velocity $v_{\rm pec}$, gas temperature $T$, neutral hydrogen fraction $x_{\rm HI}$, and UV background photo-ionization rate $\Gamma_{\rm HI}$) with skewers drawn randomly through the simulation volume to give a total sight line length of $100h^{-1}\rm\,cMpc$.  Each of the randomly drawn skewers is taken from simulation outputs sampled every $\Delta z=0.1$ to account for the redshift evolution along the quasar line of sight. Individual skewers are connected at pixels where $\Delta$, $T$, $x_{\rm HI}$ and $v_{\rm pec}$ agree within $<10$ per cent.  For every model parameter variation, we then construct $2000$ unique sight lines for performing the 1D quasar RT calculations.  

 Finally although our hydrodynamical simulations follow heating from adiabatic compression, shocks and photo-ionization by an inhomogeneous UV radiation field, they do not model neutral gas heated and ionized by the high redshift X-ray background.   We therefore follow \cite{Soltinsky_2021} (see section 2.2 and appendix B in that work) and include the pre-heating of the neutral IGM by assuming a uniform X-ray background emissivity
 \begin{align} \epsilon_{\rm X, \nu}(z) =~& 3.5\times 10^{21}f_{\rm X} \rm\,erg\,s^{-1}\,Hz^{-1}\,cMpc^{-3} \nonumber \\  &\times \left(\frac{\nu}{\nu_{\rm 0.2\rm\,keV}}\right)^{-\alpha_{\rm Xb}}\left(\frac{\rho_{\rm SFR}(z)}{10^{-2} \rm\,M_{\odot}\rm\,yr^{-1}\,cMpc^{-3}}\right), \label{eq:epsilonX} \end{align}
for photons with $E>0.2\rm\,keV$, where $f_{\rm X}$ is the uncertain X-ray efficiency \citep[e.g.][]{Furlanetto_2006b}, $\alpha_{\rm Xb}=1.5$ and $\rho_{\rm SFR}(z)$ is the comoving star formation rate density from \cite{Puchwein_2019}. We consider $0.01 \leq f_{\rm X} \leq 0.1$ in this work, which is equivalent to $10^{36.2} \rm\,erg\,s^{-1}\,cMpc^{-3}\leq \epsilon_{\rm X,0.5-2\rm\,keV} \leq 10^{37.2} \rm\,erg\,s^{-1}\,cMpc^{-3}$ at $z=7$.  This is consistent with the $1\sigma$ lower limit of $\epsilon_{\rm X,0.5-2\rm\,keV}\gtrsim 10^{34.5}\rm\,erg\,s^{-1}\,cMpc^{-3}$ at $6.5<z<8.7$ inferred from Murchison Widefield Array data \citep{Greig_2020_MWA}.  

\begin{figure*}
    \begin{minipage}{2.\columnwidth}
 	  \centering
 	  \includegraphics[width=\linewidth]{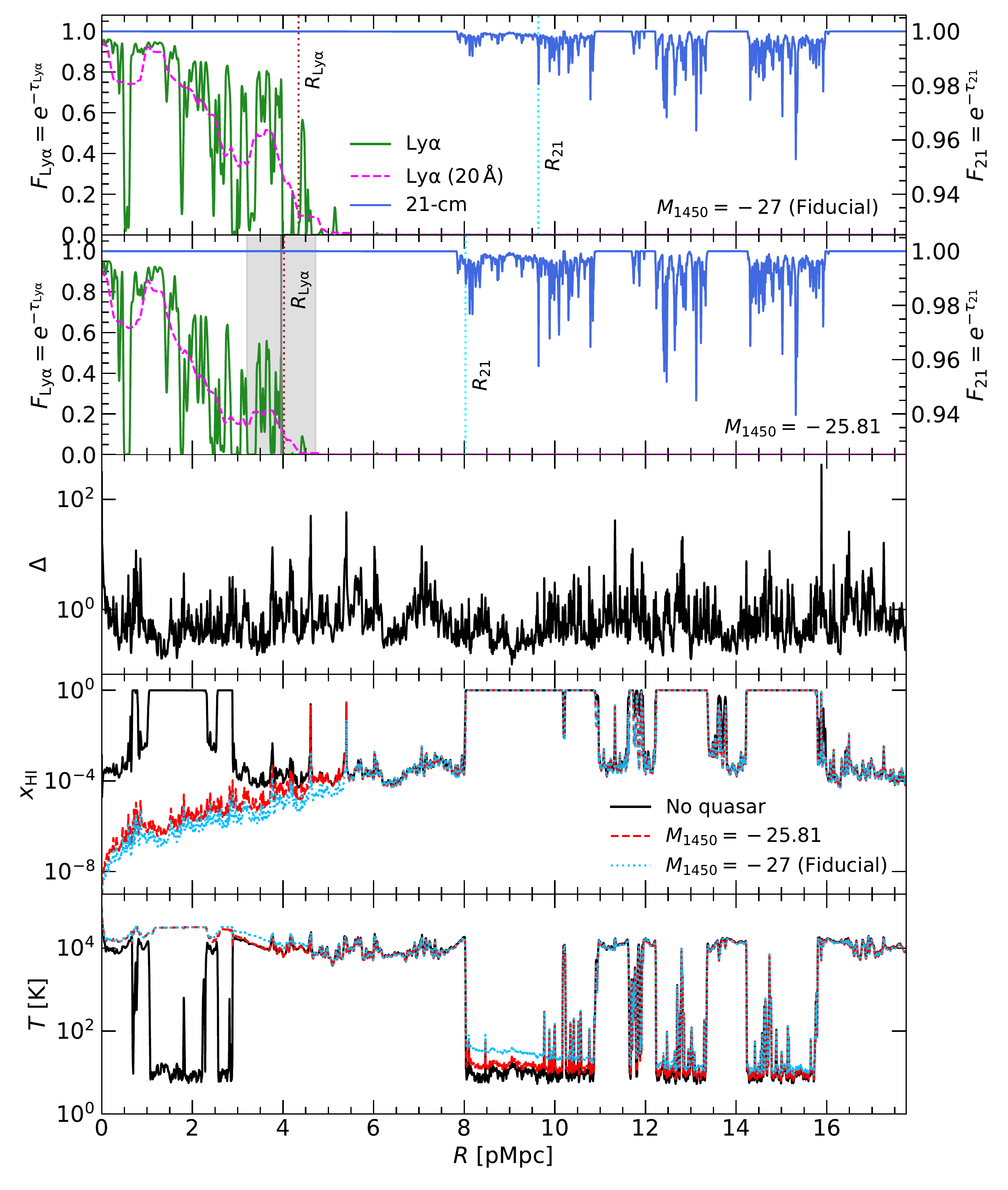}
	\end{minipage}
	\vspace{-0.2cm}
    \caption{An example of simulated \Lya\ and 21-cm absorption in the vicinity of a bright quasar at $z=7$, obtained from the RT-late Sherwood-Relics simulation with $\langle x_{\rm HI}\rangle=0.48$ combined with a 1D RT calculation for the quasar radiation.  The quasar has an optically/UV bright lifetime of $t_{\rm Q}=10^{7}\rm\,yr$ and an X-ray background efficiency of $f_{\rm X}=0.01$ is assumed.  \emph{Top panel:} The \Lya\ (green curves) and  21-cm (blue curves) transmission for our fiducial quasar SED with $M_{\rm 1450}=-27$.  Note the scale for $F_{21}$ is shown on the right vertical axis.  The dashed fuchsia curve shows the \Lya\ transmission after smoothing by a boxcar of width $20\,\si{\angstrom}$, with the \Lya\ near-zone size, $R_{\rm Ly\alpha}$, shown by the vertical brown dotted line.  The 21-cm forest spectrum is smoothed by a boxcar of width $5\rm\,kHz$ and the cyan vertical line, labelled with $R_{21}$, shows the distance from the quasar where the 21-cm absorption first reaches $F_{21}=0.99$. \emph{Second panel:} As for the top panel, but for a fainter quasar absolute magnitude of $M_{1450}=-25.81$, matching the $z=6.82$ radio-loud quasar PSO J172+18 \citep[][]{Banados_2021}.  The grey band shows the observed $R_{\rm Ly\alpha}$ for PSO J172+18.  \emph{Middle panel:}  Gas overdensity, $\Delta=\rho/\langle \rho \rangle$, along the sight line.  \emph{Fourth panel:} Neutral hydrogen fraction, $x_{\rm HI}$, for the case of no quasar (black curve), the fiducial quasar model (cyan dotted curve) and for the fainter quasar that mimics PSO J172$+$18 (red dashed curve).  \emph{Bottom panel:} Gas temperature, where the line styles match those in the panel above.}
    \label{fig:profiles_fid_B21}
\end{figure*}

\begin{figure*}
    \begin{minipage}{2.\columnwidth}
 	  \centering
 	  \includegraphics[width=\linewidth]{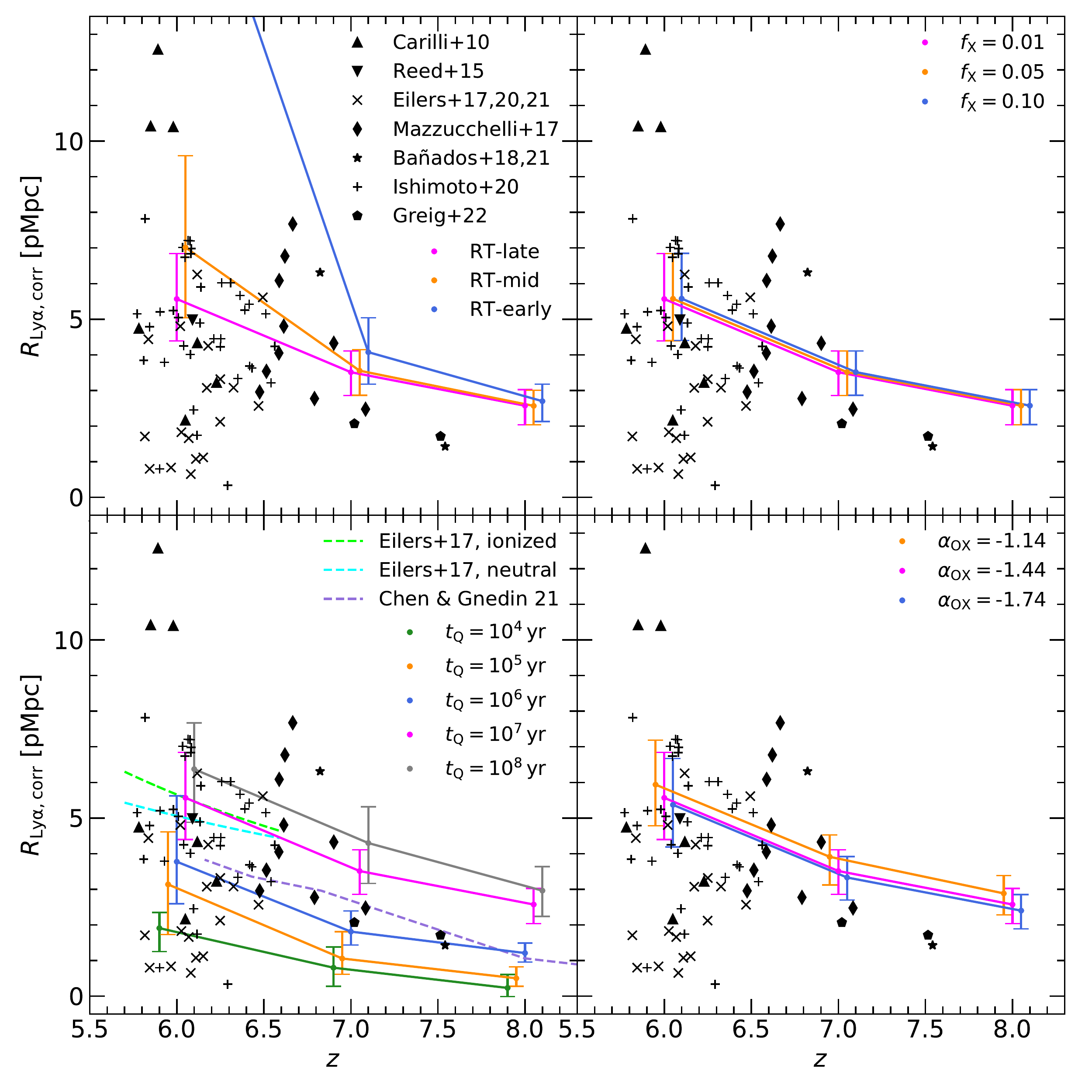}
	\end{minipage}
	\vspace{-0.35cm}
    \caption{The redshift evolution of observed and simulated \Lya\ near-zone sizes.  The filled circles at $z=6,\,7$ and $8$ connected by solid lines show the median \RLya\ and $68$ per cent scatter from $2000$ simulated quasar sight lines.   Clockwise from the top left, each panel shows the effect of varying one parameter around our fiducial model value:  the reionization history of the Sherwood-Relics model (and hence the initial volume averaged \HI fraction in the IGM, see Table~\ref{tab:xHI}), the efficiency parameter for the X-ray background, $f_{\rm X}$, the optical-to-X-ray spectral index of the quasar, $\alpha_{\rm OX}$, and the optically/UV bright lifetime of the quasar, $t_{\rm Q}$, assuming a ``light bulb'' model for the quasar light curve.  Note that the data point at $z=6$ for the RT-early model (blue, top left panel) is outside the range shown here.  Results from the 1D RT simulations performed by \citet{Eilers_2017} for an optically/UV bright lifetime of $t_{\rm Q}=10^{7.5}\rm\,yr$ are also shown for an initially highly ionized IGM (dashed green line) or fully neutral IGM (dashed cyan line) in the bottom left panel. In this panel we also show results from the 1D RT simulations from \citet[][]{Chen_Gnedin_2021_DistEvoPZ} for $t_{\rm Q}=10^{6}\rm\,yr$ and an inhomogeneously reionized IGM (dashed purple curve).  The observed \RLya\  \citep[black data points,][]{Carilli_2010,Reed_2015,Eilers_2017,Eilers_2020,Eilers_2021,Mazzucchelli_2017,Banados_2018_z754QSO,Banados_2021,Ishimoto_2020,Greig_2022} have been rescaled to correspond to an absolute magnitude of $M_{1450}=-27$ (see Eq.~\ref{eq:RNZ_cor}).}
    \label{fig:RLya}
\end{figure*}

\subsection{Example Ly\texorpdfstring{$\alpha$}{alpha} and 21-cm absorption spectrum} \label{sec:example}

A simulated quasar spectrum at $z=7$ constructed from the RT-late simulation is displayed in Fig.~\ref{fig:profiles_fid_B21}.  The upper two panels show the \Lya~(solid green curves) and 21-cm absorption (solid blue curves) for our fiducial SED with $M_{1450}=-27$, and for a fainter quasar with $M_{1450}=-25.81$, corresponding to an ionizing photon emissivity of $\dot{N}=5.48\times10^{56}\rm\,s^{-1}$.   Both models assume an optically/UV bright quasar lifetime of $t_{\rm Q}=10^{7}\rm\,yr$ and an X-ray background efficiency $f_{\rm X}=0.01$.  The fainter absolute magnitude is chosen to match the radio-loud quasar PSO J172$+$18 at $z=6.82$, recently presented by \cite{Banados_2021}.  The lower three panels display the gas overdensity, $\Delta=\rho/\langle \rho \rangle$, neutral hydrogen fraction, $x_{\rm HI}$, and gas temperature, $T$, for the case of no quasar (black curves), the fiducial quasar model (cyan dotted curves) and for the fainter quasar that mimics PSO J172$+$18 (red dashed curves).  Note the pre-existing neutral and ionized regions associated with patchy reionization, and the heating of neutral gas ahead of the large ionized region at $R>8\rm\,pMpc$ by the X-ray emission from the quasar.  The 21-cm absorption is only present where the gas is neutral, and it is stronger for the $M_{1450}=-25.81$ quasar due to the lower gas (and \HI spin) temperature at $R>8\rm\,pMpc$.  There is also a proximate Lyman limit system at $R\sim 5.4\rm\,pMpc$ that terminates the quasar \HII ionization front, beyond which the neutral hydrogen fractions are very similar for all the three cases  \citep[see also][]{Chen_Gnedin_2021_DistEvoPZ}.

We obtain the size of the simulated \Lya\ near-zones, $R_{\rm Ly\alpha}$, following the definition introduced by \cite{Fan_2006}.  This is the point where the normalised transmission first drops below $F_{\rm Ly\alpha}=0.1$ after smoothing the \Lya\ spectrum with a boxcar of width $20\,\si{\angstrom}$.  The smoothed spectrum is shown by the fuchsia dashed curves in the upper panels of Fig.~\ref{fig:profiles_fid_B21}.  For our fiducial quasar SED we obtain $R_{\rm Ly\alpha}=4.34\rm\,pMpc$ (shown by the vertical brown dotted line in Fig.~\ref{fig:profiles_fid_B21}), and for the fainter quasar with $M_{1450}=-25.81$ we find $R_{\rm Ly\alpha}=4.03\rm\,pMpc$.\footnote{Note that the dependence of $R_{\rm Ly\alpha}$ on $\dot{N}$ for this example is much weaker than the expected scaling of between $R_{\rm Ly\alpha}\propto \dot{N}^{1/3}$ and $R_{\rm Ly\alpha}\propto \dot{N}^{1/2}$ \citep{Bolton_Haehnelt_2007,Eilers_2017}.  This is due to the effect of the proximate Lyman limit system.}   In this case we have deliberately chosen a simulated quasar sight line that matches the observed \Lya\ near-zone size of \RLya~$=3.96\pm0.48\rm\,pMpc$ for PSO J172+18 \citep{Banados_2021}, shown by the grey band in the second panel of Fig.~\ref{fig:profiles_fid_B21}.  As noted by \cite{Banados_2021}, after correcting for the quasar luminosity, the \Lya\ near-zone size for PSO J172+18 is in the top quintile of $R_{\rm Ly\alpha}$ for quasars at $z\gtrsim 6$.  Our modelling suggests a possible explanation is that PSO J172+18 is surrounded by an IGM that is (locally) highly ionized due to UV emission from galaxies, despite the \emph{average} \HI fraction in the IGM being much larger.  For example, for the model displayed in Fig.~\ref{fig:profiles_fid_B21}, the average IGM neutral fraction is $\langle x_{\rm HI}\rangle = 0.48$, but there is a pre-existing highly ionized region with $x_{\rm HI}\sim 10^{-4}$ close to the quasar halo at $R\sim 3$--$8\rm\,pMpc$.

In Fig.~\ref{fig:profiles_fid_B21} we have also marked the distance from the quasar, $R_{21}$, where the proximate 21-cm absorption first reaches a threshold of $F_{21,\rm th}=e^{-\tau_{21}}=0.99$ after smoothing the spectrum with a boxcar of width $5\rm\,kHz$ (vertical cyan dotted lines).  This occurs at $R_{21}=9.63\rm\,pMpc$ ($R_{21}=8.03\rm\,pMpc$) for the $M_{\rm AB}=-27$ ($M_{\rm AB}=-25.81$) quasar.  In what follows, we will use this as our working definition of what we term the ``21-cm near-zone'' size, although we discuss this choice further in Appendix~\ref{app:R21}. Note that -- in analogy to the \Lya\ near-zone -- because of X-ray heating beyond the ionization front and the patchy ionization state of the IGM, $R_{21}$ does not always correspond to the position of the quasar \HII ionization front.  We find that when averaging over $2000$ sight-lines $R_{21}$ does, however, scale with the quasar ionizing photon emission rate as $R_{21}\propto \dot{N}^{1/3}$.  This is the same scaling expected for the size of the quasar \HII region (see Appendix~\ref{app:R21_Nion} for details).   

Lastly, given our definition for $R_{21}$, we may also estimate the minimum radio source flux density, $S_{\rm min}$, required to detect an absorption feature with $F_{21,\rm th}=0.99$ for a signal-to-noise ratio, $\rm S/N$.  Using eq. (13) in  \citet{Soltinsky_2021} and adopting values representative for SKA1-low \citep{Braun_2019}, we find
\begin{align}\label{eq:Smin}
S_{\rm min} =~& 17.2\rm\,mJy \left(\frac{0.01}{1-\textit{F}_{21,\rm th}}\right)\left(\frac{\rm S/N}{5}\right)\left(\frac{5\rm\,kHz}{\Delta \nu}\right)^{1/2}\left(\frac{1000\rm\,hr}{\textit{t}_{\rm int}}\right)^{1/2} \nonumber \\ 
&\times \left(\frac{600\rm\,m^{2}\,K^{-1}}{A_{\rm eff}/T_{\rm sys}}\right), 
\end{align}
\noindent
where $T_{\rm sys}$ is the system temperature, $\Delta \nu$ is the bandwidth, $A_{\rm eff}$ is the effective area of the telescope and $t_{\rm int}$ is the integration time.  For a sensitivity appropriate for SKA1-low (SKA2), $A_{\rm eff}/T_{\rm sys}\simeq600\rm\,m^{2}\,K^{-1}$ ($5500\rm\,m^{2}\,K^{-1}$) \citep[][]{Braun_2019}, an integration time of $t_{\rm int}=1000\rm\,hr$ ($100\rm\,hr$) and $\rm S/N=5$, we obtain $S_{\rm min}=17.2\rm\,mJy$ ($5.9\rm\,mJy$).  For comparison, PSO J172+18 has a $3\sigma$ upper limit on the flux density at $147.5\rm\,MHz$ of $S_{147.5\rm\,MHz}<8.5\rm\,mJy$ \citep{Banados_2021}.  The brightest known radio-loud blazar at $z>6$, PSO J0309+27 at $z=6.1$ with $M_{\rm 1450}=-25.1$, instead has a flux density $S_{147\rm\, MHz}= 64.2\pm6.2\rm\,mJy$ \citep[][]{Belladitta_2020}.  Both objects are therefore potential targets for detecting proximate 21-cm absorption from the diffuse IGM, although note the shape of their SEDs will be rather different.  

\subsection{Comparison to observed Ly\texorpdfstring{$\alpha$}{alpha} near-zone sizes}\label{sec:comparison_to_obs}

Next, as a consistency check of our model, we compare the \Lya\ near-zone sizes predicted in our simulations to the observed distribution in Fig.~\ref{fig:RLya}.  We have compiled a sample of \Lya~near-zone sizes measured from the spectra of 76 $z>5.77$ quasars \citep[][]{Carilli_2010,Reed_2015,Eilers_2017,Eilers_2020,Eilers_2021,Mazzucchelli_2017,Banados_2018_z754QSO,Banados_2021,Ishimoto_2020,Greig_2022}.  We use the (model dependent) $R_{\rm Ly\alpha}$--$M_{1450}$ scaling relation derived by \citet{Eilers_2017} to  approximately correct for differences in the quasar absolute magnitudes.   For an observed absolute magnitude of $M_{\rm 1450,obs}$, this gives a corrected \Lya\ near-zone size of
\begin{equation}\label{eq:RNZ_cor}
R_{\rm Ly\alpha,corr}=R_{\rm Ly\alpha,obs} \times 10^{0.4\left(27+M_{\rm 1450,obs}\right)/2.35} \propto \dot{N}^{0.43}.
\end{equation}
\noindent
In this work we rescale the observed sizes, $R_{\rm Ly\alpha,obs}$, to obtain a corrected size, $R_{\rm Ly\alpha,corr}$, at our fiducial absolute magnitude $M_{1450}=-27$.

In each panel of Fig.~\ref{fig:RLya} we vary one parameter around our fiducial model values  and compare the simulated \Lya\ near-zone sizes at $z=6,\,7$ and $8$ to the observed $R_{\rm Ly\alpha,corr}$.   Clockwise from the upper left, the parameters varied are: the reionization history of the Sherwood-Relics model (and hence the initial volume averaged \HI fraction in the IGM, see Table~\ref{tab:xHI}), the efficiency parameter for the X-ray background, $f_{\rm X}$, the optical-to-X-ray spectral index of the quasar, $\alpha_{\rm OX}$, and the optically/UV bright lifetime of the quasar, $t_{\rm Q}$, assuming a ``light bulb" model for the quasar light curve.  At each redshift, we show the median \RLya\ and the $68$ per cent distribution from $2000$ simulated sight lines.   For comparison, in the lower left panel we also show the results from the 1D RT simulations performed by \citet{Eilers_2017} for an optically/UV bright lifetime of $t_{\rm Q}=10^{7.5}\rm\,yr$, assuming either a highly ionized IGM (dashed green line) or fully neutral IGM (dashed cyan line).  The results for our fiducial parameters (i.e. RT-late, $f_{\rm X}=0.01$, $\alpha_{\rm OX}=-1.44$ and $t_{\rm Q}=10^{7}\rm\,yr$) are consistent with the \cite{Eilers_2017} models within the $68$ per cent scatter.   Similarly, the dashed purple curve shows the 1D RT simulations from \cite{Chen_Gnedin_2021_DistEvoPZ} for $t_{\rm Q}=10^{6}\rm\,yr$, which -- allowing for the somewhat larger $\langle x_{\rm HI} \rangle$ we have assumed in the RT-late reionization model --  are again similar to this work if using the same optically/UV bright quasar lifetime.

In general, the simulated \RLya\ decreases with increasing redshift \citep[e.g.][]{Fan_2006,Wyithe_2008,Carilli_2010} and, as shown in the upper left panel of Fig.~\ref{fig:RLya}, models with a larger initial IGM \HI fraction produce slightly smaller \Lya\ near-zone sizes.  Note, however, that any inferences regarding $\langle x_{\rm HI} \rangle$ from \RLya\ will be correlated with the assumed optically/UV bright lifetime \citep[e.g.][]{Bolton2011,Keating_2015}.  Furthermore, at $z=6$  the $R_{\rm Ly\alpha}$ for RT-early (blue data points), which has a volume averaged \HI fraction of $\langle x_{\rm HI} \rangle = 7.7\times 10^{-6}$ at this redshift, is outside the range displayed.  This is because many sight lines in this model are highly ionized and do not have ($20\,\si{\angstrom}$ smoothed) \Lya\ transmission that falls below $F_{\rm Ly\alpha}=0.1$.  For RT-early at $z=6$, we instead obtain a $68$ per cent lower limit of $R_{\rm Ly\alpha}>18.33\rm\,pMpc$, suggesting that the UV background at $z\simeq 6$ is significantly overproduced by the RT-early model.   In contrast, varying the X-ray heating of the IGM, either by changing $f_{\rm X}$ or $\alpha_{\rm OX}$ (upper and lower right panels, respectively), has very little  effect on the \Lya\ near-zone sizes.  As already discussed in Section~\ref{sec:homog_model}, this is because the ionization and heating by X-rays is important only for the cold, neutral IGM, and not the ionized gas observed in \Lya\ transmission.  

Finally, in the lower left panel of Fig.~\ref{fig:RLya}, we observe that some of the scatter in the observational data may be reproduced by varying the optically/UV bright lifetime of the quasar.  Indeed, \cite{Morey_2021} have recently demonstrated that the majority of $R_{\rm Ly\alpha,\rm corr}$ measurements at $z\simeq 6$ are reproduced assuming a median optically/UV bright lifetime of $t_{\rm Q}=10^{5.7}\rm\,yr$ with a $95$ per cent confidence interval $t_{\rm Q}=10^{5.3}$--$10^{6.5}\rm\,yr$.\footnote{See also \citet{Khrykin_2019,Khrykin_2021} and \citet{Worseck_2021} for closely related results obtained with the \HeII proximity effect at $z\simeq 3$--$4$.}  We have independently checked this with our own modelling and found broadly similar results (see Appendix~\ref{app:tq_obs}), although there is a hint that slightly larger quasar lifetimes may be favoured within our late reionization model \citep[see also][]{Satyavolu_2022}. On the other hand, the largest \Lya\ near-zones with $R_{\rm Ly\alpha,corr}\geq 10\rm\,pMpc$ reported by \cite{Carilli_2010} are not reproduced by the RT-late simulation even for $t_{\rm Q}=10^{8}\rm\,yr$, suggesting the IGM along these sight lines may be more ionized than assumed in the RT-late model.  It is also possible our small box size of $40 h^{-1}\rm\,cMpc$ fails to correctly capture large ionized regions near the quasar host haloes at the tail-end of reionization \citep[cf.][]{Iliev_2014,Kaur_2020}, and may therefore miss sight lines with the largest \RLya. Of particular interest here, however, are the quasars with $R_{\rm Ly\alpha,corr}\lesssim 2\rm\,pMpc$ \citep{Eilers_2020,Eilers_2021}, which correspond to $\lesssim 10$ per cent of the observational data at $z\simeq 6$.   As noted by \citet{Eilers_2021}, a very short optically/UV bright quasar lifetime of $t_{\rm Q}\lesssim 10^4$--$10^5\rm\,yr$ is required to reproduce these \Lya\ near-zone sizes.   The implied average optically/UV bright lifetime of $t_{\rm Q}\sim 10^{6}\rm\,yr$, consistent with \citet{Morey_2021}, therefore presents an apparent challenge for black hole growth at $z\geq 6$.  We discuss this further in Section~\ref{sec:flicker_model}.

In summary, the \Lya~forest near-zone sizes predicted by our simulations assuming a late end to reionization at $z\simeq 5.3$ are consistent with both independent modelling and the observational data if we allow for a distribution of optically/UV bright quasar lifetimes  \citep[e.g.][]{Morey_2021}.  We now use this model to explore the expected proximate 21-cm forest absorption around (radio-loud) quasars at $z\geq 6$.


\section{Predicted extent of proximate 21-cm absorption}\label{sec:21cmNZ}

\begin{figure*}
    \begin{minipage}{2.\columnwidth}
 	  \centering
 	  \includegraphics[width=\linewidth]{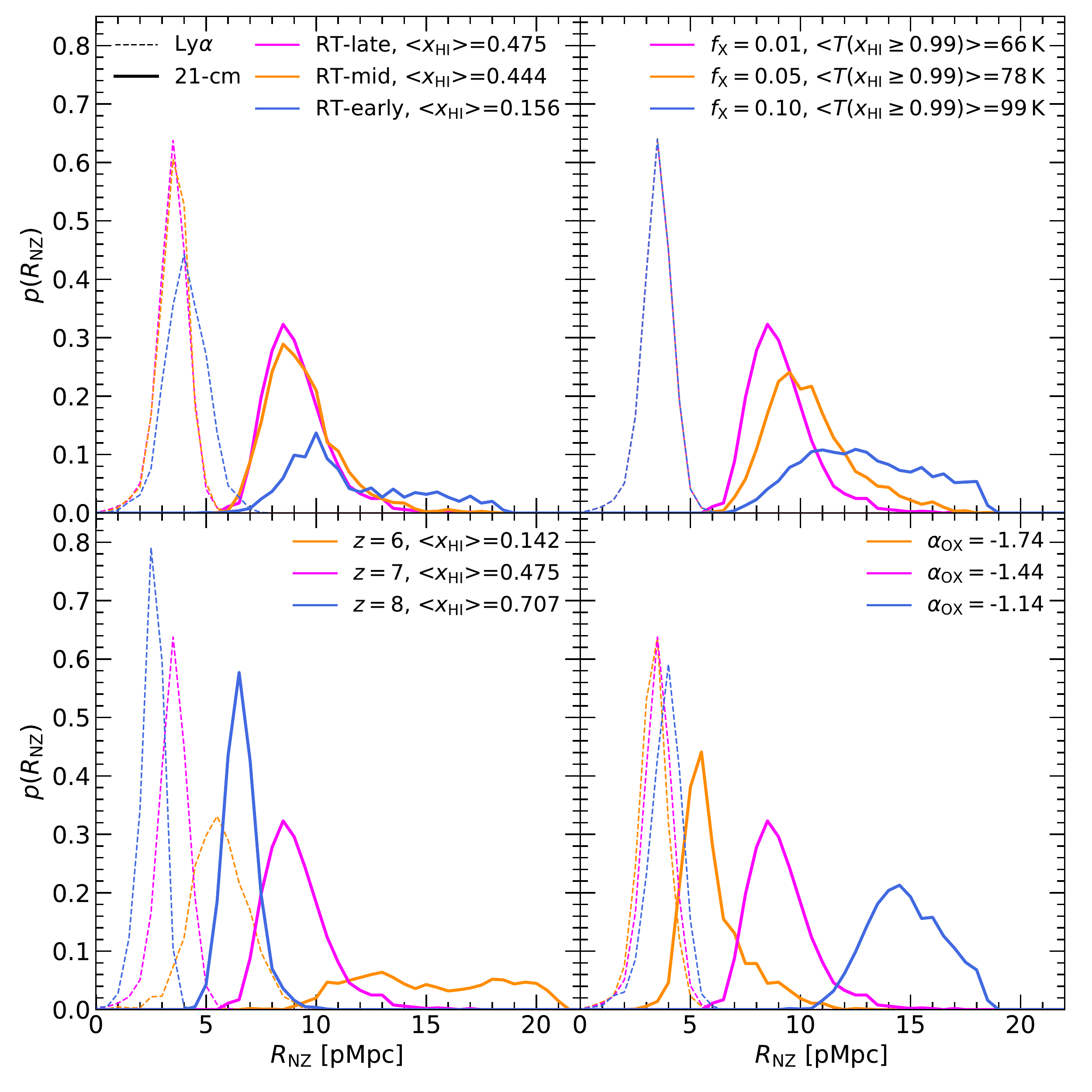}
	\end{minipage}
	\vspace{-0.3cm}
    \caption{Probability distributions for \Lya~(dashed thin curves) and 21-cm (solid thick curves) near-zone sizes obtained from $2000$ simulated quasar sight lines (see Section~\ref{sec:example} for the definition of $R_{\rm Ly\alpha}$ and $R_{21}$).    The distributions show the effect of varying parameters around our fiducial model.  Clockwise from the top left, these parameters are: the reionization history, the X-ray background  efficiency $f_{\rm X}$, the quasar optical-to-X-ray spectral index $\alpha_{\rm OX}$, and the redshift of the quasar. We also list the mean neutral hydrogen fraction (left panels) and the mean temperature in pixels with $x_{\rm HI}\geq0.99$ (top right panel) prior to any quasar heating. The fiducial values at $z=7$ are RT-late with $\langle x_{\rm HI} \rangle=0.48$, $f_{\rm X}=0.01$ and $\alpha_{\rm OX}=-1.44$.  All models furthermore assume an absolute magnitude of $M_{1450}=-27$ and an optically/UV bright lifetime of $t_{\rm Q}=10^{7}\rm\,yr$.  Note that while $R_{\rm Ly\alpha}$ is insensitive to $f_{\rm X}$ or $\alpha_{\rm OX}$, $R_{21}$ has a strong dependence on the X-ray heating around the quasar.  Both $R_{\rm Ly\alpha}$ and $R_{21}$ are sensitive to the IGM neutral fraction.}
    \label{fig:R21_RLya_hist_param}
\end{figure*}

\begin{figure*}
    \begin{minipage}{2\columnwidth}
 	  \centering
 	  \includegraphics[width=\linewidth]{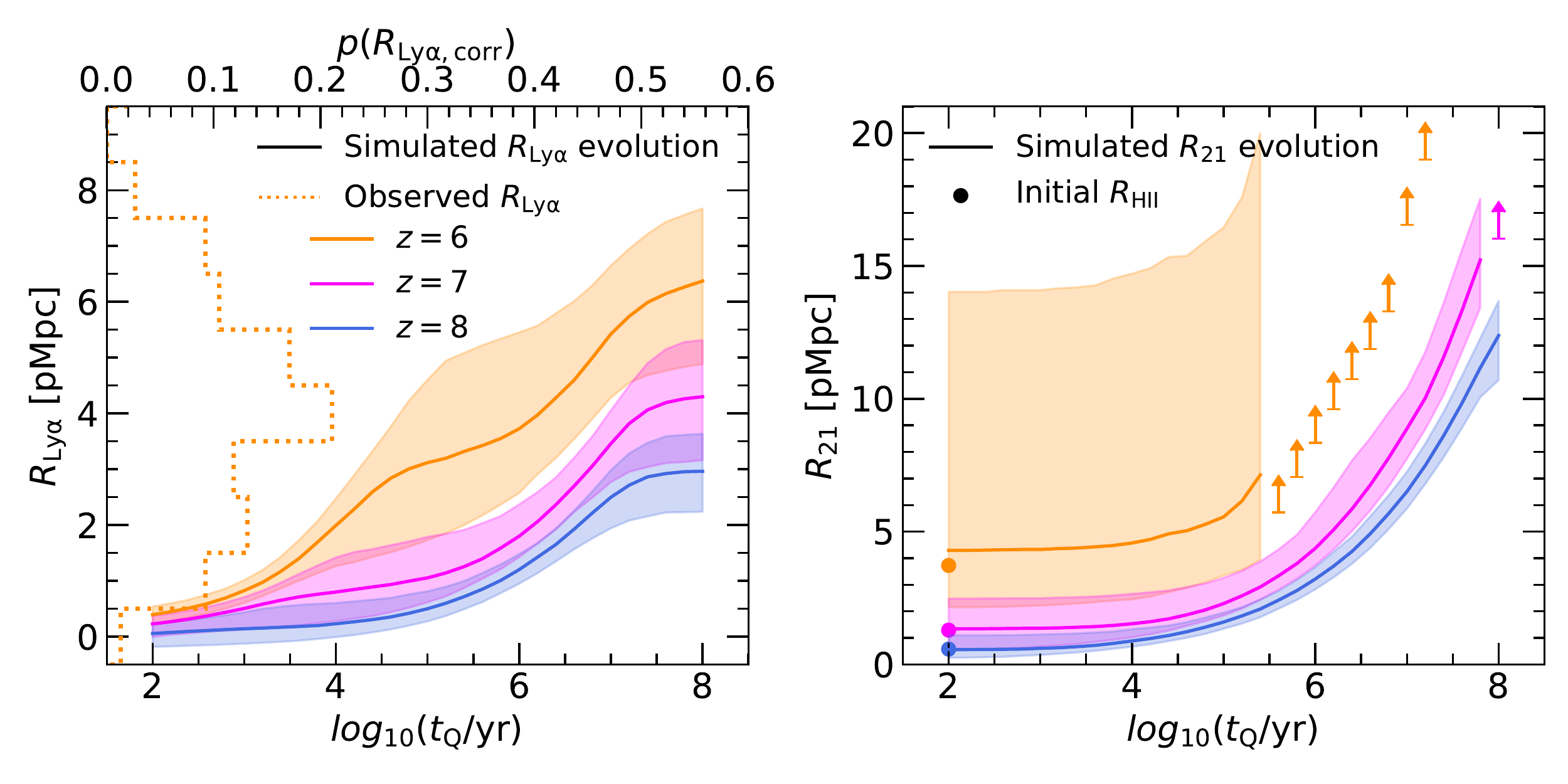}
	\end{minipage}
	\vspace{-0.3cm}
    \caption{The dependence of the \Lya~(left panel) and 21-cm (right panel) near-zone sizes on the optically/UV bright quasar lifetime, $t_{\rm Q}$, at $z=6$ (orange), $z=7$ (fuchsia) and $z=8$ (blue).  Note the different scales on the vertical axes of the panels.  The curves show the median value obtained from 2000 mock spectra, while the shaded regions mark the $68$ per cent range around the median. Upward pointing arrows give the lower 68 per cent bound on $R_{21}$ in the cases where some of the sight-lines have no pixels with $F_{21}<0.99$. The dotted orange histogram in the left panel shows the observed distribution of $R_{\rm Ly\alpha,corr}$, with a mean quasar redshift of $z=6.26$. The filled circles at $t_{\rm Q}=10^{2}\rm\,yr$ in the right panel show the median size, $R_{\rm HII}$, of the pre-existing \HII region surrounding the quasar host halo.  All models are drawn from the RT-late simulation and assume $M_{1450}=-27$, $f_{\rm X}=0.01$ and $\alpha_{\rm OX}=-1.44$.}
    \label{fig:tevo_light bulb}
\end{figure*}

\subsection{The effect of X-ray heating and IGM neutral fraction}\label{sec:21cmNZ_Xray_reion}

The effect of X-ray heating and the IGM neutral fraction on the distribution of ``21-cm near zone'' sizes, $R_{21}$, predicted by our simulations is displayed in Fig.~\ref{fig:R21_RLya_hist_param} (solid curves).  In all cases we assume $M_{1450}=-27$ and a light bulb quasar model with an optically/UV bright lifetime of $t_{\rm Q}=10^{7}\rm\,yr$.  For comparison, the \RLya~distributions from the same models are given by the dashed curves.  The top left panel shows the effect of varying the reionization model, and hence the initial volume averaged neutral fraction in the IGM, $\langle x_{\rm HI}\rangle$.  At $z=7$, the $\langle x_{\rm HI}\rangle$ values for RT-late (fuchsia curves) and RT-mid (orange curves) are very similar, and we find little difference between these models for $R_{21}$ or $R_{\rm Ly\alpha}$.   For the more highly ionized RT-early simulation, the near-zone sizes are slightly larger, although note almost half of the $2000$ quasar spectra do not have any pixels with $F_{21}<0.99$ at $z=7$.   In the bottom left panel, we instead show results from the RT-late simulation at three different redshifts, $z=6,\,7$ and $8$.  The \Lya\ and 21-cm near-zone sizes are larger toward lower redshift, again due to the smaller \HI fraction in the IGM, but also now because of the decrease in the proper gas density (i.e. $n_{\rm H}\propto (1+z)^{3}$).  However, once again, at $z=6$ (fuchsia curves) around half the quasar sight-lines do not exhibit 21-cm absorption with $F_{21}<0.99$.  This suggests that observing 21-cm absorption from the diffuse IGM in close proximity to radio-loud quasars will be more likely if reionization is late ($z_{\rm R}\simeq 5.3$) as suggested by \citet{Kulkarni_2019}, and if suitably bright radio-loud quasars can be identified at $z\gtrsim 7$.

The effect of X-ray heating on the near-zone sizes is displayed in the right panels of Fig.~\ref{fig:R21_RLya_hist_param}.   The top right panel shows the heating by the X-ray background, while the bottom right panel shows the effect of quasar X-ray heating when varying the optical-to-X-ray spectral index, $\alpha_{\rm OX}$.   As noted earlier, $R_{\rm Ly\alpha}$ is insensitive to $f_{\rm X}$ and $\alpha_{\rm OX}$, but $R_{21}$ is sensitive to both; the average 21-cm near-zone size increases as the spin temperature of the neutral gas is raised by X-ray photo-heating.   For example, for $f_{\rm X}=0.01$ the average temperature of hydrogen with $x_{\rm HI}>0.99$ (i.e. neutral gas ahead of the \HII ionization front) is $T=66\rm\,K$, but this increases to $T=99\rm\,K$ for $f_{\rm X}=0.1$. Here, the average temperature of neutral gas is consistent with the recent constraint of $15.6\mathrm{\,K} < T_{\rm S} < 656.7\rm\,K$ (95 per cent confidence) from  \citet{HERA_2022_constraints_new} in all three cases. For $T\gg 10^{2}\rm\,K$ (or equivalently, $f_{\rm X}\gg 0.1$), we expect very little 21-cm absorption will be detectable at all \citep[e.g.][]{Soltinsky_2021}.   A similar situation holds for $\alpha_{\rm OX}$, with a harder quasar X-ray spectrum producing larger $R_{21}$.\footnote{One could also vary the spectral index at $\lambda\leq 50\,\si{\angstrom} $ away from our fiducial value of $\alpha_{\rm X}=-0.9$. However, the effect of changing $\alpha_{\rm X}$ and $\alpha_{\rm OX}$ on gas temperature is degenerate.  For a reasonable range of values, $\alpha_{\rm X}=-0.9\pm0.5$ \citep[][]{Vito_2019,Wang_2021} we find the effect of changing $\alpha_{\rm X}$ on the gas temperature is smaller than the effect of varying $\alpha_{\rm OX}$, where we consider $\alpha_{\rm OX}=-1.44\pm 0.3$}.  Deep X-ray observations may be used to constrain $\alpha_{\rm OX}$ for at least some $z\gtrsim 6$ radio-loud quasars \citep{Connor_2021}.  Prior knowledge of the quasar X-ray spectrum could therefore help break some of the degeneracy between $R_{21}$ and the X-ray heating parameters $f_{\rm X}$ and $\alpha_{\rm OX}$.   As already discussed, however, the location of the expanding quasar \HII region and the spin temperature beyond the \HII ionization front determine the optical depth of neutral gas, where $\tau_{21}\propto x_{\rm HI}/T_{\rm S} \sim T_{\rm S}^{-1}$.   This means $R_{21}$ is also sensitive to the  optically/UV bright lifetime of the quasar, $t_{\rm Q}$.

\subsection{The effect of the optically/UV bright lifetime}\label{sec:21cmNZ_tQ}

In Fig.~\ref{fig:tevo_light bulb}, for our fiducial model  we examine how \RLya\ and \RCM\ evolve with the optically/UV bright lifetime of the quasar at redshift $z=6$ (orange curves), $z=7$ (fuchsia curves) and $z=8$ (blue curves).  The shaded regions bound $68$ per cent of the data around the median for $2000$ simulated sight-lines.   The behaviour of \RLya\ at $z=6$, displayed in the left panel, is qualitatively similar to the results of other recent work \citep[e.g.][]{Eilers_2018,Eilers_2021,Davies_2020,Satyavolu_2022}.  There are three distinct phases in the evolution of \RLya\ at $z=6$.  For a highly ionized IGM, when the optically/UV bright lifetime of the quasar is shorter than the equilibriation timescale, $t_{\rm Q}<t_{\rm eq}$, we expect \RLya\ to increase with $t_{\rm Q}$.  The equilibriation timescale is approximately
\begin{equation} t_{\rm eq} = \frac{x_{\rm HI,\rm\,eq}}{n_{\rm e} \alpha_{A}(T)} \simeq \frac{10^{5.0}\rm\, yr}{\Delta} \left(\frac{x_{\rm HI,\rm\,eq}}{10^{-4}}\right)\left(\frac{T}{10^{4}\rm\,K}\right)^{0.72}\left(\frac{1+z}{7}\right)^{-3}, \label{eq:teq} \end{equation}
where $x_{\rm HI,\,eq}$ is the \HI fraction in ionization equilibrium, we have used a case-A recombination coefficient $\alpha_{\rm A}=4.06\times 10^{-13}\mathrm{\,cm^{3}\,s^{-1}}(T/10^{4}\rm\,K)^{-0.72}$ and assumed $n_{\rm e}=1.158 n_{\rm H}$ for a fully ionized hydrogen and helium IGM.  For $t_{\rm Q}>t_{\rm eq}$, the growth of the \Lya\ near-zone size slows and becomes largely insensitive to $t_{\rm Q}$ \citep[see e.g.][]{Bolton_Haehnelt_2007}.  In this regime the near-zone size is set by the \Lya\ absorption from the residual \HI in the IGM, rather than the growth of the \HII region around the quasar.  Finally, at $t_{\rm Q}\gtrsim 10^{6.5}\rm\,yr$, the \Lya\ near-zone starts to grow again.  As noted by \citet{Eilers_2018}, the late growth of \RLya\ is due to the propagation of the \HeIII ionization front into the IGM.  The associated \HeII photo-heating raises the IGM temperature and hence further lowers the \HI fraction in the IGM \citep[see also][]{Bolton_2012}.  We also point out that the median \RLya\ we obtain at $z=6$ for $10^{5}\mathrm{\,yr}<t_{\rm Q}<10^{6.5}\rm\,yr$ are slightly smaller than those reported in fig. 2 of \citet{Davies_2020}.  This is because we use our RT-late simulation with $\langle x_{\rm HI}\rangle=0.14$ at $z=6$, instead of assuming a highly ionized IGM as \cite{Davies_2020} do.  In the RT-late model, neutral islands will persist in underdense regions at $z=6$ and hence slow the growth of the near-zones.  Further discussion of this point can also be found in \citet{Satyavolu_2022}.

For reference, we also show the distribution of observed $R_{\rm Ly\alpha,corr}$ in the left panel of Fig.~\ref{fig:tevo_light bulb}, which has a mean quasar redshift of $z=6.26$.  Once again, note that reproducing the \Lya\ near-zones with $R_{\rm Ly\alpha,corr}<2\rm\,pMpc$ at $z\simeq 6$ requires $t_{\rm Q}\lesssim 10^{4}$--$10^{5}\rm\,yr$.   As expected, at $z=7$ and $z=8$, the \Lya\ near-zones are smaller.  Here the initial \HI fractions in the IGM for RT-late are $\langle x_{\rm HI}\rangle=0.48$ and  $\langle x_{\rm HI}\rangle=0.71$, respectively.  The large IGM \HI fractions also produce a strong \Lya\ damping wing that suppresses \Lya\ near-zone sizes.  For reference, the $z=7.54$ quasar ULAS J1342+0928 has $R_{\rm Ly\alpha,corr}=1.43\rm\,pMpc$ \citep{Banados_2018_z754QSO}, whereas the $z=7.08$ quasar ULAS J1120$+$0641 has $R_{\rm Ly\alpha, corr}=2.48\pm0.2\rm\,pMpc$ \citep{Mortlock_2011,Mazzucchelli_2017}.  We find our simulations are consistent with these sizes for optically/UV bright lifetimes in the range $10^{4}\mathrm{\,yr}\leq t_{\rm Q}\leq 10^{6.8}\rm\,yr$.

In the right panel of Fig.~\ref{fig:tevo_light bulb} we show the dependence of the 21-cm near-zone size on the optically/UV bright lifetime, $t_{\rm Q}$.  Note in particular the filled circles in Fig.~\ref{fig:tevo_light bulb} at $t_{\rm Q}=10^{2}\rm\,yr$, which show the median size, $R_{\rm HII}$, of the pre-existing \HII regions created by the galaxies surrounding the quasar host haloes.\footnote{We define $R_{\rm HII}$ as the distance from the quasar host halo where $x_{\rm HI}=0.9$ is first exceeded, and have verified that choosing larger values of $x_{\rm HI}$ up to $0.999$ does not change $R_{\rm HII}$ significantly.}   The initial value of $R_{21}$ is very similar to $R_{\rm HII}$, suggesting the typical size of these pre-existing \HII regions will set the 21-cm near-zone sizes for short optically/UV bright lifetimes.  We find $R_{21}\sim R_{\rm HII}$ for $t_{\rm Q}\lesssim 10^{4}\rm\,yr$.  However, for $t_{\rm Q}\gtrsim 10^{4}\rm\,yr$ (i.e. exceeding the local photo-ionization timescale at $R_{\rm HII}$, where $t_{\rm ion}=\Gamma_{\rm HI}^{-1}\sim 10^{4}$--$10^{5}\rm\,yr$), the quasar starts to expand the pre-existing \HII region and X-rays begin to photo-heat the neutral gas  ahead of the quasar \HII ionization front to $T>10^{2}\rm\,K$.  The 21-cm near-zone then grows.  Note also that at $z=6$, there is a large $68$ per cent scatter around the median $R_{21}$, and for $t_{\rm Q}>10^{5.5}\rm\,yr$, many of the simulated sight-lines at $z=6$ have no pixels with $F_{21}<0.99$.  In this case we instead show lower limits for $R_{21}$ that bound $68$ per cent of the simulated sight-lines.  At $z=7$ and $z=8$, the median $R_{21}$ is smaller with significantly less scatter, which (as for the case for the \Lya\ near-zones) is primarily because the average \HI fraction in the IGM is larger at these redshifts. 

In summary, our results suggest two intriguing possibilities. First, if there is a population of very young quasars at $z\geq 6$, as observed \Lya\ near-zones with $R_{\rm Ly\alpha}<2\rm\,pMpc$ imply \citep[e.g.][]{Eilers_2017}, then if $f_{\rm X}\lesssim 0.01$, a measurement of $R_{21}$ around these objects should constrain the size of the \HII region created by the galaxies clustered around the quasar host halo.  Such a measurement would be complimentary to similar proposed measurements of $R_{\rm HII}$ from 21-cm tomography \citep[e.g.][]{Furlanetto_2004,Wyithe_2004,GeilWyithe_2008,Datta_2012,Kakiichi_2017,Ma_2020,Davies_2021}, and would provide a strong constraint on the reionization sources.  Second, once the quasar begins to heat the IGM ahead of the \HII ionization front to $T\gtrsim 10^{2}\rm\,K$, the 21-cm absorption is suppressed and $R_{21}$ increases monotonically.  In the absence of significant ionization, the cooling timescale for this gas is the adiabatic cooling timescale, where 
\begin{equation} t_{\rm ad}= \frac{1}{2H(z)}\simeq 10^{8.8}\mathrm{\,yr}\, \left(\frac{1+z}{8}\right)^{-3/2},\end{equation}  
\noindent
and $H(z)\simeq H_{0}\Omega_{\rm m}^{1/2}(1+z)^{3/2}$ is the Hubble parameter.  Hence, in general $R_{21}$ should always increase and it will be sensitive to the integrated lifetime of the quasar, because we typically expect $t_{\rm Q} \lesssim t_{\rm ad}$ \citep[e.g][]{Haehnelt_1998,YuTremaine_2002, Martini_2004}.  We now turn to explore the consequence of this for variable quasar emission, with particular emphasis on the possible implications for black hole growth at $z\gtrsim 6$ \citep[cf.][]{Eilers_2018,Eilers_2021}.


\section{Probing integrated quasar lifetimes with proximate 21-cm absorption}\label{sec:flicker}

\begin{figure*}
    \begin{minipage}{2.\columnwidth}
 	  \centering
 	  \includegraphics[width=\linewidth]{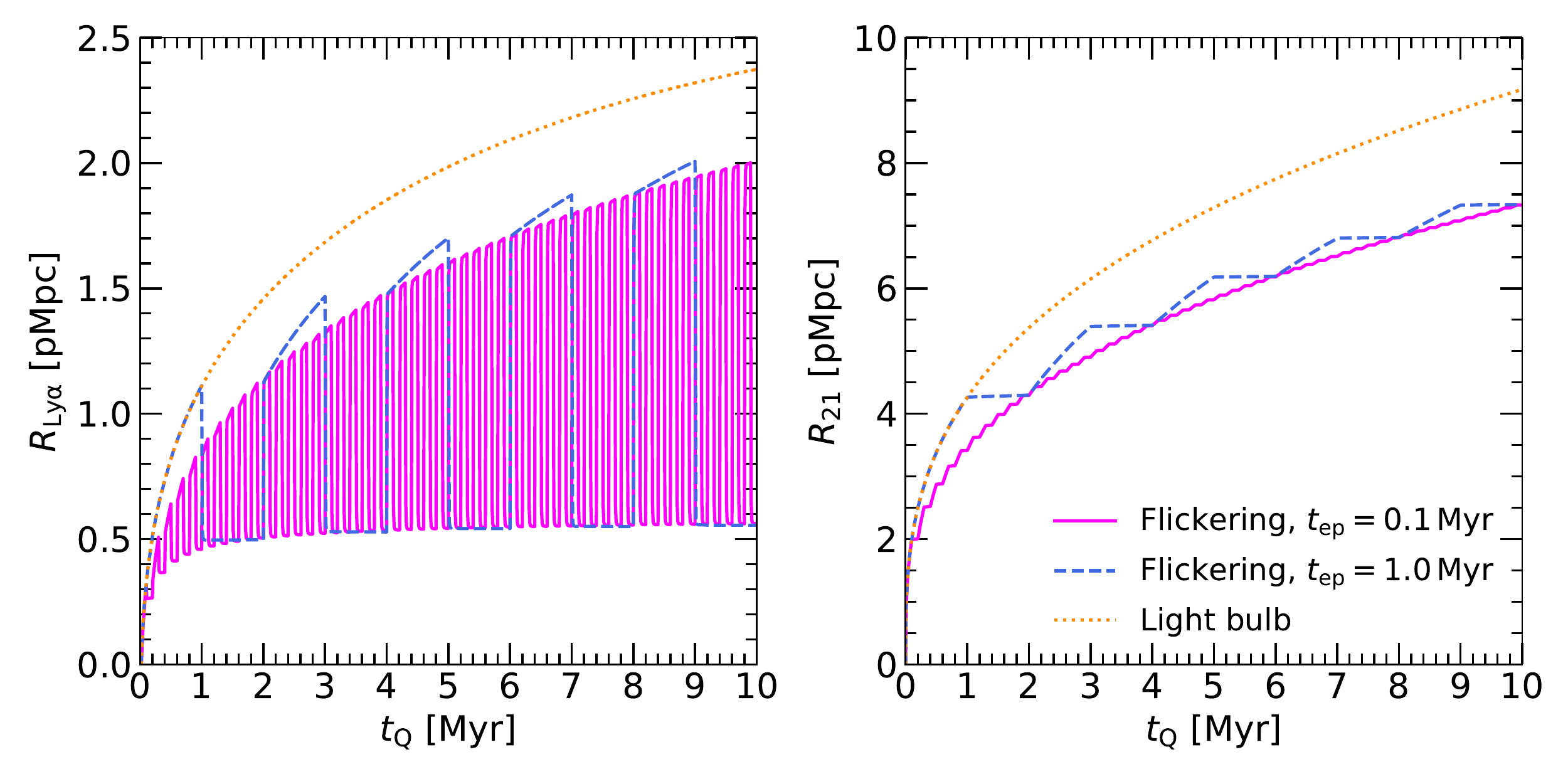}
	\end{minipage}
	\vspace{-0.3cm}
    \caption{The dependence of \RLya~(left panel) and \RCM\ (right panel) on the integrated quasar lifetime, $t_{\rm Q}$, for a quasar at $z=7$ that varies between a bright phase with $M_{1450}=-27$ and faint phase with $M_{1450}=-23$.   We assume an optically/UV bright duty cycle of $f_{\rm duty}=0.5$ and consider episodic lifetimes of $t_{\rm ep}=10^{5}\rm\,yr$ (fuchsia solid curves) and $t_{\rm ep}=10^{6}\rm\,yr$ (blue dashed curves).  The IGM surrounding the quasar is initially cold and neutral.  The near-zone size for a light bulb quasar emission model (dotted orange curves) is shown for comparison.  Note in particular that while $R_{\rm Ly\alpha}$ decreases on the equilibriation timescale during the faint phase, $t_{\rm eq}$, $R_{21}$ remains almost constant due to the much longer adiabatic cooling timescale for the neutral gas, where the 21-cm optical depth $\tau_{21}\propto x_{\rm HI}/T_{\rm S}$.}
    \label{fig:flicker_profiles}
\end{figure*}

\subsection{A simple model for flickering quasar emission}\label{sec:flicker_model}

\citet{Morey_2021} have recently pointed out that the typical optically/UV bright lifetime of $t_{\rm Q}\sim 10^{6}\rm\,yr$ implied by the observed $R_{\rm Ly\alpha}$ is a challenge for the growth of $\sim 10^{9}\rm \,M_{\odot}$ black holes observed at $z\gtrsim 6$ \citep{Mortlock_2011,Banados_2018_z754QSO,Yang_2020_z75,Wang_2020,Farina_2022}.  Further discussion of this point in the context of \Lya\ near-zones can be found in \citet{Eilers_2018} and \citet{Eilers_2021}, but we briefly repeat the argument here. For a quasar with bolometric luminosity $L$, the \citet{Salpeter_1964} (or e-folding) timescale if the black hole is accreting at the Eddington limit is 
\begin{equation} t_{\rm S}=\frac{\epsilon}{1-\eta} \frac{c \sigma_{\rm T}}{4\pi G \mu m_{\rm p}} = 4.33\times10^{7}\mathrm{\,yr}\, \left(\frac{L}{L_{\rm E}}\right)^{-1}\left(\frac{\epsilon}{0.1}\right)\left(\frac{1-\eta}{0.9}\right)^{-1},  \end{equation}
\noindent
where $L_{\rm E}$ is the Eddington luminosity, $\sigma_{\rm T}$ is the Thomson cross-section, $\mu=1.158$ is the mean molecular weight for fully ionized hydrogen and helium with $Y=0.24$, $\eta$ is the accretion efficiency, and $\epsilon$ is the radiative efficiency \citep[e.g.][]{ShakuraSunyaev_1973} where we assume $\epsilon=\eta$.  For a black hole seed of mass $M_{\rm seed}$ and a constant accretion rate, the black hole mass, $M_{\rm BH}$, after $t_{\rm Q}=[10^{6},\,10^{7},\,10^{8}]\rm\,yr$ is then
\begin{equation} M_{\rm BH} = M_{\rm seed} \exp\left(\frac{t_{\rm Q}}{t_{\rm S}}\right) = [1.0,\,1.3,\,10.1] M_{\rm seed}. \label{eq:MBH} \end{equation}
If $t_{\rm Q}\sim 10^{6}\rm\,yr$ there is insufficient time for the black hole to grow; Eq.~(\ref{eq:MBH}) requires $M_{\rm BH}\sim M_{\rm seed}\sim 10^{9}\rm\, M_{\odot}$ , yet the largest theoretically plausible seed mass is $M_{\rm seed}\sim 10^{5}$--$10^{6}\rm\,M_{\odot}$ \citep[e.g. from the direct collapse of atomically cooled halo gas,][]{LoebRasio_1994,Dijkstra_2008,Regan_2017,Inayoshi_2020}.  

As discussed by \citet{Eilers_2021}, there are two possible solutions to this apparent dilemma; the $z\gtrsim 6$ quasars are indeed very young and have grown rapidly from massive seeds by radiatively inefficient ($\epsilon \sim 0.01$), mildly super-Eddington accretion \citep[e.g.][]{Madau_2014,Volonteri_2015,Davies_2019} or the quasars are much older than the $R_{\rm Ly\alpha}$ measurements imply, such that $t_{\rm Q}\gtrsim 10^{7}\rm\,yr$.  This is possible if the black holes have grown primarily in an optically/UV obscured phase and the quasars have only recently started to ionize their vicinity, perhaps due to the evacuation of obscuring material by feedback processes \citep{Hopkins_2005}.  Alternatively, quasar luminosity may vary between optically/UV bright and faint phases over an episodic lifetime of $t_{\rm ep}\sim 10^{4}$--$10^{6}\rm\,yr$, likely as a result of variable accretion onto the black hole \citep{Schawinski_2015,King_2015,AnglesAlcazar_2017,Shen_2021}.  In this scenario, when the quasars are faint the ionized hydrogen in their vicinity recombines on the equilibriation timescale (see Eq.~\ref{eq:teq}).  This produces an initially small \Lya\ near-zone size that regrows over a timescale $t_{\rm ion}=\Gamma_{\rm HI}^{-1}\sim 10^{4}$--$10^{5}\rm\,yr$ once the quasars re-enter the optically/UV bright phase \citep{Davies_2020,Satyavolu_2022}.  Furthermore, for $t_{\rm ep}\lesssim t_{\rm eq}$ the \HI surrounding the quasars never fully equilibriates, and $R_{\rm Ly\alpha}$ remains smaller than predicted for a light bulb light curve with the same \emph{integrated} quasar lifetime.  

However, it is difficult to distinguish between these possibilities using $R_{\rm Ly\alpha}$ alone.  We suggest the proximate 21-cm absorption around sufficiently radio-bright quasars may provide some further insight.   The long adiabatic cooling timescale for neutral gas in the IGM means that, unlike $R_{\rm Ly\alpha}$, $R_{21}$ will be sensitive to the integrated lifetime of the quasars.  To illustrate this point further consider Fig.~\ref{fig:flicker_profiles}, where we use the simplified neutral, homogeneous IGM model discussed in Section~\ref{sec:homog_model} and Fig.~\ref{fig:uniform_profiles} to explore the effect of variable quasar emission on the evolution of $R_{\rm Ly\alpha}$ (left panel) and $R_{21}$ (right panel).  In both panels the orange dotted curves show $R_{\rm Ly\alpha}$ and $R_{21}$ for a light bulb emission model with $M_{1450}=-27$ and the fiducial SED.  For the variable emission model, we instead follow a similar approach to \cite{Davies_2020} and \citet{Satyavolu_2022} and assume the quasar periodically flickers between a bright phase with $M_{1450}=-27$ and faint phase with $M_{1450}=-23$, while keeping the shape of the quasar SED fixed.   We assume an optically/UV bright duty cycle of $f_{\rm duty}=0.5$ and consider episodic lifetimes of $t_{\rm ep}=10^{5}\rm\,yr$ (fuchsia solid curves) and $t_{\rm ep}=10^{6}\rm\,yr$ (blue dashed curves).   Shorter episodic lifetimes, $t_{\rm ep}\ll 10^{5}\rm\,yr$ may also be appropriate for some of the smallest observed near-zones at $z\simeq 6$ with $R_{\rm Ly\alpha,corr}<2\rm\,pMpc$, but the good agreement between the majority of the $R_{\rm Ly\alpha,corr}$ measurements and simple light bulb models with $t_{\rm Q}\sim 10^{6}\rm\,yr$ suggest such short episodic lifetimes are unusual \citep{Morey_2021,Eilers_2021}.    While we find that, as expected, $R_{\rm Ly\alpha}$ varies on timescales $t\simeq t_{\rm eq}$ and can potentially have $R_{\rm Ly\alpha}<1\rm\,pMpc$ for $t_{\rm Q}\sim 10^{7}\rm\,yr$ if the quasar has just re-entered the bright phase, $R_{21}$ instead increases monotonically with $t_{\rm Q}$.  Furthermore, in this example we have assumed the optical/UV and X-ray emission from the quasar become fainter simultaneously.  If instead only the optical/UV emission is reduced -- perhaps due to obscuring material that remains optically thin to X-rays  --  the X-ray heating will continue and $R_{21}$ will evolve similarly to the light bulb model.

Note also that for a homogeneous medium for $t_{\rm Q}\ll t_{\rm rec}$, where $t_{\rm rec}=(\alpha_{\rm A}(T)\langle n_{\rm e}\rangle)^{-1} \equiv t_{\rm eq}/x_{\rm HI}$ is the recombination timescale, the quasar \HII region will have size $R_{\rm HII}=[3\dot{N}f_{\rm duty}t_{\rm Q}/(4\pi \langle n_{\rm H}\rangle)]^{1/3}$, where 
\begin{align} R_{\rm HII} &\simeq 3.5 \mathrm{\,pMpc}\,  \left(\frac{f_{\rm duty}}{x_{\rm HI}}\right)^{1/3}\left(\frac{\dot{N}}{1.64\times 10^{57}\rm\,s^{-1}}\right)^{1/3} \left(\frac{t_{\rm Q}}{10^{7}\rm\,yr}\right)^{1/3} \nonumber \\ &\times \left(\frac{1+z}{8}\right)^{-1}. \label{eq:RHII} \end{align}
Hence, for the example displayed in Fig.~\ref{fig:flicker_profiles}, $R_{\rm Ly\alpha}<R_{\rm HII}$  due to the IGM damping wing, but $R_{21}>R_{\rm HII}$ due to heating by X-rays ahead of the \HII ionization front.  We also expect the ratio $R_{21}/R_{\rm Ly\alpha}$ will typically be larger for flickering quasars with longer integrated lifetimes, $t_{\rm Q}\sim 10^{7}\rm\,yr$, that have just re-entered their bright phase.  As $R_{21}$ is sensitive to the integrated lifetime of the quasar, this suggests a combination of $R_{21}$ and $R_{\rm Ly\alpha}$ -- either for an individual radio-loud quasar or for a population of objects --  could sharpen existing constraints on quasar lifetimes if the uncertainty in the X-ray background efficiency, $f_{\rm X}$, and the optical-to-X-ray spectral index, $\alpha_{\rm OX}$, can be marginalised over.  Evidence for strong 21-cm absorption within a few proper Mpc of a radio-loud quasar would then hint at a short integrated quasar lifetime.

\subsection{Time evolution of Ly\texorpdfstring{$\alpha$}{alpha} and 21-cm near-zones for flickering emission}\label{sec:flicker_tevo}

\begin{figure*}
    \begin{minipage}{2\columnwidth}
 	  \centering
 	  \includegraphics[width=\linewidth]{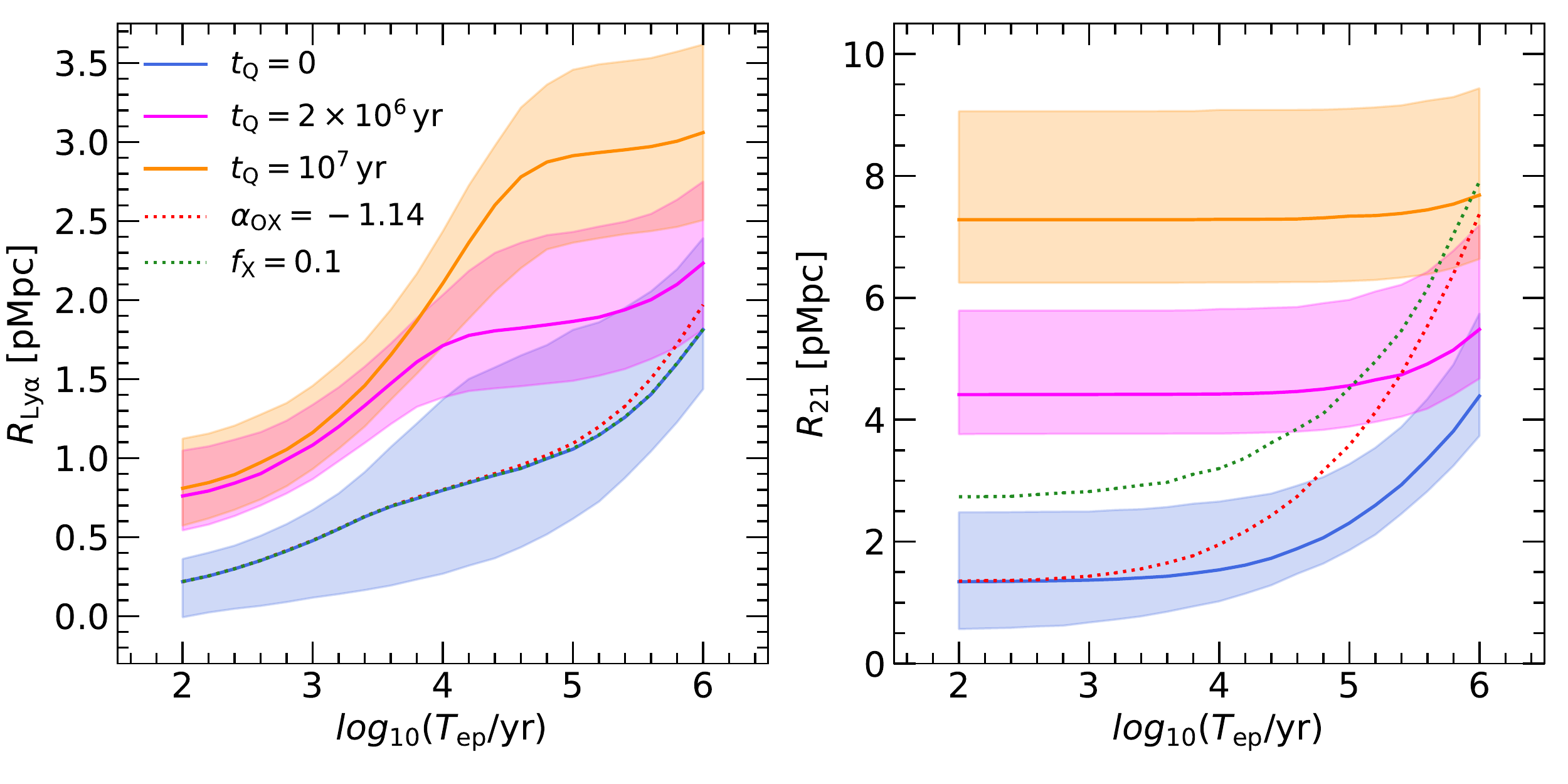}
	\end{minipage}
	\vspace{-0.3cm}
    \caption{The dependence of $R_{\rm Ly\alpha}$ and $R_{21}$ at $z=7$ on the \emph{current} episodic lifetime, $T_{\rm ep}$, in a flickering quasar emission model.  Results are shown for young quasars in their first bright phase ($t_{\rm Q}=0\rm\,yr$, blue curves), for an intermediate case similar to the lifetime inferred by \citet{Morey_2021} ($t_{\rm Q}=10^{6.3}\rm\,yr$, fuchsia curves) and for our fiducial optically/UV bright quasar lifetime  ($t_{\rm Q}=10^{7}\rm\,yr$, orange curves).   The shading corresponds to the $68$ per cent scatter around the median from 2000 simulated sight lines.  Note that while \Lya\ near-zones can be small ($R_{\rm Ly\alpha}<1\rm\,pMpc$) for all $t_{\rm Q}$ when the quasar has recently entered the bright phase, \RCM~increases monotonically and will be considerably larger than $R_{\rm Ly\alpha}$ for $t_{\rm Q}=10^{7}\rm\,yr$.  We also show two additional models in which we boost the X-ray heating in the pre-reionization IGM by setting $f_{\rm X}=0.1$ (dotted green curves) and $\alpha_{\rm OX}=-1.14$ (dotted red curves).  Note these curves are almost indistinguishable in the left panel.}
    \label{fig:tevo_flicker}
\end{figure*}

We further consider the flickering quasar emission model using the RT-late Sherwood-Relics simulation for $f_{\rm X}=0.01$ and our fiducial SED.  In Fig.~\ref{fig:tevo_flicker} we show the dependence of the median $R_{\rm Ly\alpha}$ (left panels) and $R_{21}$ (right panels) at $z=7$ on the current episodic lifetime, $T_{\rm ep}$.  This is just the duration of the most recent optically/UV bright phase with $M_{1450}=-27$ for a quasar that already has an integrated age $t_{\rm Q}$, with $f_{\rm duty}=0.5$ and $t_{\rm ep}=10^{6}\rm\,yr$.  Three different integrated quasar ages are displayed,  where $t_{\rm Q}=0\rm\,yr$ (blue curves), $t_{\rm Q}=2\times 10^{6}\rm\,yr$ (fuchsia curves) and $t_{\rm Q}=10^{7}\rm\,yr$ (orange curves), as measured from the start of the most recent optically/UV bright phase  (i.e. for $0$, $1$ and $5$ earlier episodic cycles with $t_{\rm ep}=10^{6}\rm\,yr$, respectively).   The shaded regions show the 68 per cent scatter around the median.  

First, note the $R_{\rm Ly\alpha}$ and $R_{21}$ values for $t_{\rm Q}=0\rm\,yr$ are almost identical to the light bulb model in Fig.~\ref{fig:tevo_light bulb} (fuchsia curves) for $t_{\rm Q}\leq 10^{6}\rm\,yr$, as should be expected.   However, in the case of older quasars with $t_{\rm Q}>t_{\rm ep}$ that have experienced at least one episodic cycle, we find (within the 68 per cent scatter) that $R_{\rm Ly\alpha}\lesssim 2\rm\,pMpc$ for $T_{\rm ep}\sim t_{\rm ion}< 10^{4.5}\rm\,yr$, and that $R_{\rm Ly\alpha}$ is \emph{insensitive} to the integrated quasar age.  As already discussed, this is a consequence of the re-equilibriation of the neutral hydrogen behind the quasar \HII ionization front during the quasar faint phase.   For an episodic lifetime of $t_{\rm ep}=10^{6}\rm\,yr$, we would therefore expect $R_{\rm Ly\alpha,corr}\lesssim 2\rm\,pMpc$ for $\sim 3$ per cent of $z=7$ quasars, even if the integrated quasar age $t_{\rm Q}>t_{\rm ep}$.  Similar results have been pointed out elsewhere \citep[e.g.][]{Davies_2020}

On the other hand, as a result of the long cooling timescale for neutral gas ahead of the \HII ionization front, \RCM~is $\sim 2$--$5.5$ times larger for $t_{\rm Q}=10^{7}\rm\,yr$ (orange curve) compared to $R_{21}$ for a quasar that has just turned on for the first time (blue curve).  Hence, if invoking flickering quasar emission to reconcile the apparent short optically/UV bright lifetimes of quasars at $z\gtrsim 6$ with the build-up of $\sim 10^{9}\rm\,M_{\odot}$ black holes, we expect $R_{21}>R_{\rm Ly\alpha}$.  Only for the case of a very young quasar do we find proximate 21-cm absorption with $R_{21}\sim 2\rm\,pMpc$.  An important caveat here, however, is the level of X-ray heating in the neutral IGM.  The dotted curves show results for $f_{\rm X}=0.1$ or $\alpha_{\rm OX}=-1.14$ for the case of a $t_{\rm Q}=0\rm\,yr$ (i.e. the blue curves for the fiducial model).  While $R_{\rm Ly\alpha}$ remains unaffected by X-ray heating, \RCM\ increases. Raising the X-ray background efficiency, $f_{\rm X}$, results in a larger initial $R_{21}$, while a harder optical-to-X-ray spectral index, $\alpha_{\rm OX}$, increases $R_{21}$ on timescales  $T_{\rm ep}\gtrsim t_{\rm ion}$.   Nevertheless, for $t_{\rm Q}\lesssim 10^{4}\rm\,yr$ we still expect $R_{21}\lesssim 3\rm\,pMpc$ if the quasar has not undergone earlier episodic cycles for $M_{\rm AB}=-27$, where the magnitude corrected size scales as $R_{\rm 21,corr}\propto 10^{0.4(27+M_{1450})/3}$ (see Appendix~\ref{app:R21_Nion}).  Finally, we point out that a null detection of proximate 21-cm absorption with $F_{21}<0.99$ would be indicative of an X-ray background with $f_{\rm X}\gtrsim 1$ at $z=7$ \citep[see fig. 8 in][]{Soltinsky_2021}.

In summary, we suggest that a measurement of $R_{21}$ along the line of sight to radio-loud quasars could complement existing constraints on the lifetime of quasars obtained from \Lya\ transmission.   Furthermore,  a detection of proximate 21-cm absorption from the diffuse IGM within a few proper Mpc of a bright quasar at $z\simeq 7$ would present yet another challenge for the growth of $\sim 10^{9}\rm\,M_{\odot}$ black holes during the reionization epoch. Our modelling indicates that long range heating by X-ray photons means that for $f_{\rm X}\lesssim 0.1$, $R_{21}\lesssim 2$--$3\rm pMpc$ should only occur for radio-loud quasars that have recently initiated accretion. Larger values of $R_{\rm 21}$ coupled with $R_{\rm Ly\alpha,corr}<2\rm\,pMpc$ would instead hint at black hole growth progressing over timescales much longer than the optically/UV bright lifetimes of $t_{\rm Q}\sim 10^{4}\rm\,yr$ implied by the  smallest \Lya\ near-zone sizes of the quasar population at $z\gtrsim 6$ \citep{Morey_2021}.    


\section{Conclusions}\label{sec:conclusions}

Recent studies have suggested that observed \Lya\ near-zone sizes at $z\gtrsim 6$ \citep{Fan_2006,Carilli_2010,Willott_2010,Venemans_2015,Reed_2015,Eilers_2017,Eilers_2021,Mazzucchelli_2017,Ishimoto_2020} are consistent with an average quasar optically/UV bright lifetime  of $t_{\rm Q}\sim 10^{6}\rm\,yr$, with lifetimes as short as $t_{\rm Q}\lesssim 10^{4}$--$10^{5}\rm\,yr$ preferred by the smallest \Lya\ near-zones at $z\simeq 6$ \citep{Eilers_2017,Eilers_2021,Morey_2021}.  If correct, this presents an apparent challenge for the build-up of $\sim 10^{9}\rm\,M_{\odot}$ supermassive black holes at $z\gtrsim 6$, as the black hole growth e-folding time is at least an order of magnitude larger than $t_{\rm Q}$ if assuming Eddington limited accretion.  These very young quasars would need to have grown from very massive seeds through radiatively inefficient, super Eddington accretion \citep{Madau_2014,Davies_2019}. Note, however, that because the number of black holes implied by the detected optically/UV bright quasars scales inversely with the optically/UV bright lifetime \citep[e.g.][]{Haehnelt_1998}, this would also push the quasars into rather low mass haloes. Alternatively, the quasars could be much older and have only recently entered an optically/UV bright phase.   This is possible if most quasars at $z\gtrsim 6$ grow primarily in an optical/UV obscured phase \citep{Hopkins_2005,Ricci_2017obsc}, or variable accretion causes them to "flicker'' between optically/UV bright and faint states on episodic timescales $t_{\rm ep}\sim 10^{5}$--$10^{6}\rm\,yr$ \citep{Schawinski_2015,Shen_2021}.  Distinguishing between these possibilities with \Lya\ near-zones is difficult, however, due to the relatively short equilibriation timescale, $t_{\rm eq}\sim 10^{5}\rm\,yr$, for the residual neutral hydrogen surrounding the quasar \citep{Davies_2020}.  

In this work, we have therefore used the Sherwood-Relics simulations of inhomogeneous  reionization \citep{Puchwein_2022}, coupled with line of sight radiative transfer calculations, to model the \Lya\ and 21-cm absorption in close proximity to $z\gtrsim 6$ quasars.  The empirically calibrated reionization histories available in the Sherwood-Relics simulation suite and the flexibility of our line of sight radiative transfer algorithm allows us to explore a large parameter space, including variations in the IGM neutral fraction, the X-ray background intensity, and the quasar age and spectral shape.  We suggest that the observation of proximate 21-cm absorption in the spectra of radio-loud quasars at $z\gtrsim 6$ (with e.g. SKA1-low or SKA2) could provide a route for probing the lifetimes of $z\gtrsim 6$ quasars that is complementary to \Lya\ near-zones and proposed analyses of quasar \HII regions using 21-cm tomography \citep[e.g.][]{Wyithe_2004,Kohler_2005,RhookHaehnelt_2006,GeilWyithe_2008,Majumdar_2012,Datta_2012,Kakiichi_2017,Ma_2020,Davies_2021}.  Our main conclusions are as follows:

\begin{itemize}

    \item  If allowing for a distribution of optically/UV bright lifetimes with a median of $t_{\rm Q}\simeq 10^{6}\rm\,yr$ \citep{Morey_2021}, the luminosity corrected sizes of \Lya\ near-zones, $R_{\rm Ly\alpha,corr}$, are reasonably well reproduced within the Sherwood-Relics simulations for a model with late reionization ending at $z=5.3$.    Slightly larger average lifetimes may be allowable within late reionization models \citep[e.g.][]{Satyavolu_2022}, although in the models presented here the effect is modest and differences are within the 68 per cent scatter around the predicted median $R_{\rm Ly\alpha}$ (compare e.g. RT-late and RT-mid in Fig.~\ref{fig:RLya}).  We also confirm that the smallest \Lya\ near-zones at $z\simeq 6$, with quasar luminosity corrected sizes of $R_{\rm Ly\alpha,corr}\lesssim 2\rm\,pMpc$, are consistent with optically/UV bright quasar lifetimes of $t_{\rm Q}\lesssim 10^{4}$--$10^{5}\rm\,yr$ in late reionization models \citep{Eilers_2017,Eilers_2021}.\\
    
    \item We define the ``21-cm near-zone'' size, $R_{21}$, as the distance from a (radio-loud) quasar where the normalised 21-cm forest spectrum first drops below the threshold $F_{\rm 21,th}=0.99$ (i.e $\tau_{21}\gtrsim 10^{-2}$), after smoothing the radio spectrum with a $5\rm\,kHz$ boxcar filter.   Detecting a strong proximate 21-cm absorber with $\tau_{21}\geq 10^{-2}$ requires a minimum source flux density of 17.2 mJy (5.9 mJy) for a 1000 (100) hour integration with SKA1-low (SKA2), assuming a signal-to-noise ratio of $\rm S/N=5$ and bandwidth of $5\rm\,kHz$.  For comparison, the recently discovered radio-loud quasar PSO J172+18 has a $3\sigma$ upper limit on the flux density at $147.5\rm\,MHz$ of $S_{147.5\rm\,MHz}<8.5\rm\,mJy$ \citep{Banados_2021}, and the blazar PSO J0309+27 at $z=6.1$ has $S_{147\rm\, MHz}= 64.2\pm6.2\rm\,mJy$ \citep[][]{Belladitta_2020}. Proximate 21-cm absorption around these or similar radio-loud sources should therefore be within reach of the SKA.\\
    
    \item We show that for modest pre-heating of the IGM by the X-ray background, such that the IGM spin temperature $T_{\rm S}\lesssim 10^{2}\rm\,K$, strong proximate 21-cm absorption from the diffuse IGM should be present in the spectra of radio-loud quasars \citep[see also][]{Soltinsky_2021}.  We demonstrate that $R_{21}$ will depend on the quasar optical-to-X-ray spectral index, $\alpha_{\rm OX}$, and the \emph{integrated} quasar lifetime, $t_{\rm Q}$.  In contrast, the \Lya\ near-zone size remains insensitive to the level of X-ray heating in the IGM.  For very young quasars, $R_{21}$ should trace the extent of the pre-existing \HII regions created by galaxies clustered around the quasar host halo.  \\
    
    \item Unlike the \Lya\ near-zone size --  which can vary over the equilibriation timescale, $t_{\rm eq}\sim 10^{5}\rm\,yr$, for neutral hydrogen in a highly ionized IGM \citep[e.g.][]{Davies_2020} -- $R_{21}$ is sensitive to the integrated lifetime of the quasar and will increase monotonically with quasar age.   This is because the 21-cm optical depth is inversely proportional to the spin temperature of neutral hydrogen, $\tau_{21}\propto T_{\rm S}^{-1}$, and the neutral hydrogen will cool adiabatically on a timescale $t_{\rm H}/2$, where $t_{\rm H}\gg t_{\rm Q}$ is the Hubble time.  A combination of $R_{21}$ and $R_{\rm Ly\alpha}$ may therefore help sharpen constraints on quasar lifetimes if the uncertain heating by X-rays from the quasar and X-ray background can be marginalised over.\\
    
    \item For quasars that exhibit unusually small luminosity corrected \Lya\ near-zone sizes (where evidence for a \Lya\ damping wing from a large neutral column in the IGM may also be limited), proximate 21-cm absorption could help distinguish between very young quasars with $t_{\rm Q}<10^{4}$--$10^{5}\rm\, yr$, or older quasars that have experienced episodic accretion.  We find that proximate 21-cm absorption from the diffuse IGM is only expected within a few proper Mpc of the quasar systemic redshift for very young objects.  Such short lifetimes may point toward massive black hole seeds \citep[e.g.][]{LoebRasio_1994,Dijkstra_2008,Regan_2017} and radiatively inefficient, mildly super-Eddington accretion \citep{Madau_2014,Davies_2019}.
    Larger values of $R_{21}$ coupled with small \Lya\ near-zones with $R_{\rm Ly\alpha,corr}\lesssim 2\rm\,pMpc$ would instead be consistent with time-variable black hole growth occurring over longer periods. 

\end{itemize}

Our results provide further impetus for searching for 21-cm absorption from the diffuse IGM at high redshift.  However, the caveats discussed by our earlier work focusing on 21-cm absorption from the general IGM \citep{Soltinsky_2021} also apply here.  We have not considered any of the practical issues regarding the recovery of 21-cm absorption features from noisy data.   The role of 21-cm absorption from any minihaloes that are unresolved in our simulations (i.e. minihaloes with masses $<2.5\times10^7\rm\,M_{\odot}$) also remains uncertain \citep[][]{Meiksin_2011,Park_2016,Nakatani_2020}.  Soft X-ray heating of the IGM by the transverse quasar proximity effect may also be an important uncertainty, particularly for the large population of faint or obscured quasars that would be implied by short optically/UV bright quasar lifetimes and/or duty cycles. Finally, note that if the neutral IGM is already pre-heated to temperatures $T\gg 10^{2}\rm\,K$ at $z\gtrsim 6$, there will be very little or no detectable 21-cm absorption from the diffuse IGM at all.  Although constraints on the X-ray background and spin temperature in the IGM are still weak \citep{Greig_2020_MWA,HERA_2022_constraints_new}, further progress toward placing limits and/or detecting the 21-cm power spectrum should help narrow parameter space over the next decade.


\section*{Acknowledgements}
We thank Sindhu Satyavolu for comments on a draft version of this work. We also thank an anonymous referee for their constructive comments. The hydrodynamical simulations were performed using the Cambridge Service for Data Driven Discovery (CSD3), part of which is operated by the University of Cambridge Research Computing on behalf of the STFC DiRAC HPC Facility (www.dirac.ac.uk). The DiRAC component of CSD3 was funded by BEIS capital funding via STFC capital grants ST/P002307/1 and ST/R002452/1 and STFC operations grant ST/R00689X/1. This work also used the DiRAC@Durham facility managed by the Institute for Computational Cosmology on behalf of the STFC DiRAC HPC Facility. The equipment was funded by BEIS capital funding via STFC capital grants ST/P002293/1 and ST/R002371/1, Durham University and STFC operations grant ST/R000832/1. DiRAC is part of the National e-Infrastructure.  We also acknowledge the Partnership for Advanced Computing in Europe (PRACE) for awarding us access to the Curie and Irene supercomputers, based in France at the Tr\`es Grand Centre de calcul du CEA, during the 16th Call.  We thank Volker Springel for making \textsc{P-Gadget-3} available. This work has made use of \textsc{matplotlib} \citep{Hunter_2007}, \textsc{astropy} \citep{Robitaille_2013}, \textsc{numpy} \citep{Harris_2020} and \textsc{scipy} \citep{Virtanen_2020}. TŠ is supported by the University of Nottingham Vice Chancellor's Scholarship for Research Excellence (EU). JSB, MM and NH are supported by STFC consolidated grant ST/T000171/1. MGH acknowledges support from UKRI STFC (grant No. ST/N000927/1). Part of this work was supported by FP7 ERC Grant Emergence-320596. LCK was supported by the European Union’s Horizon 2020 research and innovation programme under the Marie Skłodowska-Curie grant agreement No. 885990. GK is partly supported by the Department of Atomic Energy (Government of India) research project with Project Identification Number RTI~4002, and by the Max Planck Society through a Max Planck Partner Group.

\section*{Data Availability}
All data and analysis code used in this work are available from the first author on request. 



\bibliographystyle{mnras}
\bibliography{example} 

\begin{thebibliography}{}
\makeatletter
\relax
\def\mn@urlcharsother{\let\do\@makeother \do\$\do\&\do\#\do\^\do\_\do\%\do\~}
\def\mn@doi{\begingroup\mn@urlcharsother \@ifnextchar [ {\mn@doi@}
  {\mn@doi@[]}}
\def\mn@doi@[#1]#2{\def\@tempa{#1}\ifx\@tempa\@empty \href
  {http://dx.doi.org/#2} {doi:#2}\else \href {http://dx.doi.org/#2} {#1}\fi
  \endgroup}
\def\mn@eprint#1#2{\mn@eprint@#1:#2::\@nil}
\def\mn@eprint@arXiv#1{\href {http://arxiv.org/abs/#1} {{\tt arXiv:#1}}}
\def\mn@eprint@dblp#1{\href {http://dblp.uni-trier.de/rec/bibtex/#1.xml}
  {dblp:#1}}
\def\mn@eprint@#1:#2:#3:#4\@nil{\def\@tempa {#1}\def\@tempb {#2}\def\@tempc
  {#3}\ifx \@tempc \@empty \let \@tempc \@tempb \let \@tempb \@tempa \fi \ifx
  \@tempb \@empty \def\@tempb {arXiv}\fi \@ifundefined
  {mn@eprint@\@tempb}{\@tempb:\@tempc}{\expandafter \expandafter \csname
  mn@eprint@\@tempb\endcsname \expandafter{\@tempc}}}

\bibitem[\protect\citeauthoryear{{Angl{\'e}s-Alc{\'a}zar},
  {Faucher-Gigu{\`e}re}, {Quataert}, {Hopkins}, {Feldmann}, {Torrey}, {Wetzel}
  \& {Kere{\v{s}}}}{{Angl{\'e}s-Alc{\'a}zar} et~al.}{2017}]{AnglesAlcazar_2017}
{Angl{\'e}s-Alc{\'a}zar} D.,  {Faucher-Gigu{\`e}re} C.-A.,  {Quataert} E.,
  {Hopkins} P.~F.,  {Feldmann} R.,  {Torrey} P.,  {Wetzel} A.,   {Kere{\v{s}}}
  D.,  2017, \mn@doi [\mnras] {10.1093/mnrasl/slx161}, \href
  {https://ui.adsabs.harvard.edu/abs/2017MNRAS.472L.109A} {472, L109}

\bibitem[\protect\citeauthoryear{{Astropy Collaboration} et~al.,}{{Astropy
  Collaboration} et~al.}{2013}]{Robitaille_2013}
{Astropy Collaboration} et~al., 2013, \mn@doi [\aap]
  {10.1051/0004-6361/201322068}, \href
  {https://ui.adsabs.harvard.edu/abs/2013A&A...558A..33A} {558, A33}

\bibitem[\protect\citeauthoryear{Aubert \& Teyssier}{Aubert \&
  Teyssier}{2008}]{Aubert_2008}
Aubert D.,  Teyssier R.,  2008, \mn@doi [\mnras]
  {10.1111/j.1365-2966.2008.13223.x}, \href
  {http://dx.doi.org/10.1111/j.1365-2966.2008.13223.x} {387, 295}

\bibitem[\protect\citeauthoryear{{Ba{\~n}ados} et~al.,}{{Ba{\~n}ados}
  et~al.}{2018}]{Banados_2018_z754QSO}
{Ba{\~n}ados} E.,  et~al., 2018, \mn@doi [\nat] {10.1038/nature25180}, \href
  {https://ui.adsabs.harvard.edu/abs/2018Natur.553..473B} {553, 473}

\bibitem[\protect\citeauthoryear{{Ba{\~n}ados} et~al.,}{{Ba{\~n}ados}
  et~al.}{2021}]{Banados_2021}
{Ba{\~n}ados} E.,  et~al., 2021, \mn@doi [\apj] {10.3847/1538-4357/abe239},
  \href {https://ui.adsabs.harvard.edu/abs/2021ApJ...909...80B} {909, 80}

\bibitem[\protect\citeauthoryear{{Bajtlik}, {Duncan}  \& {Ostriker}}{{Bajtlik}
  et~al.}{1988}]{Bajtlik1988}
{Bajtlik} S.,  {Duncan} R.~C.,   {Ostriker} J.~P.,  1988, \mn@doi [\apj]
  {10.1086/166217}, \href
  {https://ui.adsabs.harvard.edu/abs/1988ApJ...327..570B} {327, 570}

\bibitem[\protect\citeauthoryear{{Becker}, {Bolton}  \& {Lidz}}{{Becker}
  et~al.}{2015a}]{Becker_Bolton_Lidz_2015}
{Becker} G.~D.,  {Bolton} J.~S.,   {Lidz} A.,  2015a, \mn@doi [\pasa]
  {10.1017/pasa.2015.45}, \href
  {https://ui.adsabs.harvard.edu/abs/2015PASA...32...45B} {32, e045}

\bibitem[\protect\citeauthoryear{Becker, Bolton, Madau, Pettini, Ryan-Weber  \&
  Venemans}{Becker et~al.}{2015b}]{Becker_2015}
Becker G.~D.,  Bolton J.~S.,  Madau P.,  Pettini M.,  Ryan-Weber E.~V.,
  Venemans B.~P.,  2015b, \mn@doi [\mnras] {10.1093/mnras/stu2646}, \href
  {http://dx.doi.org/10.1093/mnras/stu2646} {447, 3402}

\bibitem[\protect\citeauthoryear{{Belladitta} et~al.,}{{Belladitta}
  et~al.}{2020}]{Belladitta_2020}
{Belladitta} S.,  et~al., 2020, \mn@doi [\aap] {10.1051/0004-6361/201937395},
  \href {https://ui.adsabs.harvard.edu/abs/2020A&A...635L...7B} {635, L7}

\bibitem[\protect\citeauthoryear{{Bolton} \& {Haehnelt}}{{Bolton} \&
  {Haehnelt}}{2007}]{Bolton_Haehnelt_2007}
{Bolton} J.~S.,  {Haehnelt} M.~G.,  2007, \mn@doi [\mnras]
  {10.1111/j.1365-2966.2006.11176.x}, \href
  {http://dx.doi.org/10.1111/j.1365-2966.2006.11176.x} {374, 493}

\bibitem[\protect\citeauthoryear{{Bolton}, {Haehnelt}, {Warren}, {Hewett},
  {Mortlock}, {Venemans}, {McMahon}  \& {Simpson}}{{Bolton}
  et~al.}{2011}]{Bolton2011}
{Bolton} J.~S.,  {Haehnelt} M.~G.,  {Warren} S.~J.,  {Hewett} P.~C.,
  {Mortlock} D.~J.,  {Venemans} B.~P.,  {McMahon} R.~G.,   {Simpson} C.,  2011,
  \mn@doi [\mnras] {10.1111/j.1745-3933.2011.01100.x}, \href
  {https://ui.adsabs.harvard.edu/abs/2011MNRAS.416L..70B} {416, L70}

\bibitem[\protect\citeauthoryear{Bolton, Becker, Raskutti, Wyithe, Haehnelt  \&
  Sargent}{Bolton et~al.}{2012}]{Bolton_2012}
Bolton J.~S.,  Becker G.~D.,  Raskutti S.,  Wyithe J. S.~B.,  Haehnelt M.~G.,
  Sargent W. L.~W.,  2012, \mn@doi [\mnras] {10.1111/j.1365-2966.2011.19929.x},
  \href {http://dx.doi.org/10.1111/j.1365-2966.2011.19929.x} {419, 2880}

\bibitem[\protect\citeauthoryear{Bosman}{Bosman}{2022}]{Bosman_2022}
Bosman S. E.~I.,  2022, All z>5.7 quasars currently known,
  \mn@doi{10.5281/zenodo.6039724}

\bibitem[\protect\citeauthoryear{{Bosman} \& {Becker}}{{Bosman} \&
  {Becker}}{2015}]{BosmanBecker_2015}
{Bosman} S. E.~I.,  {Becker} G.~D.,  2015, \mn@doi [\mnras]
  {10.1093/mnras/stv1336}, \href
  {https://ui.adsabs.harvard.edu/abs/2015MNRAS.452.1105B} {452, 1105}

\bibitem[\protect\citeauthoryear{{Bosman}, {Fan}, {Jiang}, {Reed}, {Matsuoka},
  {Becker}  \& {Haehnelt}}{{Bosman} et~al.}{2018}]{Bosman_2018}
{Bosman} S. E.~I.,  {Fan} X.,  {Jiang} L.,  {Reed} S.,  {Matsuoka} Y.,
  {Becker} G.,   {Haehnelt} M.,  2018, \mn@doi [\mnras]
  {10.1093/mnras/sty1344}, \href
  {https://ui.adsabs.harvard.edu/abs/2018MNRAS.479.1055B} {479, 1055}

\bibitem[\protect\citeauthoryear{{Bosman} et~al.,}{{Bosman}
  et~al.}{2022}]{Bosman2022}
{Bosman} S. E.~I.,  et~al., 2022, \mn@doi [\mnras] {10.1093/mnras/stac1046},
  \href {https://ui.adsabs.harvard.edu/abs/2022MNRAS.514...55B} {514, 55}

\bibitem[\protect\citeauthoryear{{Braun}, {Bonaldi}, {Bourke}, {Keane}  \&
  {Wagg}}{{Braun} et~al.}{2019}]{Braun_2019}
{Braun} R.,  {Bonaldi} A.,  {Bourke} T.,  {Keane} E.,   {Wagg} J.,  2019, arXiv
  e-prints, \href {https://ui.adsabs.harvard.edu/abs/2019arXiv191212699B} {p.
  arXiv:1912.12699}

\bibitem[\protect\citeauthoryear{Calverley, Becker, Haehnelt  \&
  Bolton}{Calverley et~al.}{2011}]{Calverley_2011}
Calverley A.~P.,  Becker G.~D.,  Haehnelt M.~G.,   Bolton J.~S.,  2011, \mn@doi
  [\mnras] {10.1111/j.1365-2966.2010.18072.x}, \href
  {http://dx.doi.org/10.1111/j.1365-2966.2010.18072.x} {412, 2543–2562}

\bibitem[\protect\citeauthoryear{Carilli, Gnedin  \& Owen}{Carilli
  et~al.}{2002}]{Carilli_2002}
Carilli C.~L.,  Gnedin N.~Y.,   Owen F.,  2002, \mn@doi [\apj]
  {10.1086/342179}, \href {http://dx.doi.org/10.1086/342179} {577, 22}

\bibitem[\protect\citeauthoryear{{Carilli} et~al.,}{{Carilli}
  et~al.}{2010}]{Carilli_2010}
{Carilli} C.~L.,  et~al., 2010, \mn@doi [\apj] {10.1088/0004-637X/714/1/834},
  \href {https://ui.adsabs.harvard.edu/abs/2010ApJ...714..834C} {714, 834}

\bibitem[\protect\citeauthoryear{{Cen} \& {Haiman}}{{Cen} \&
  {Haiman}}{2000}]{CenHaiman2000}
{Cen} R.,  {Haiman} Z.,  2000, \mn@doi [\apjl] {10.1086/312937}, \href
  {https://ui.adsabs.harvard.edu/abs/2000ApJ...542L..75C} {542, L75}

\bibitem[\protect\citeauthoryear{{Chen} \& {Gnedin}}{{Chen} \&
  {Gnedin}}{2021}]{Chen_Gnedin_2021_DistEvoPZ}
{Chen} H.,  {Gnedin} N.~Y.,  2021, \mn@doi [\apj] {10.3847/1538-4357/abe7e7},
  \href {https://ui.adsabs.harvard.edu/abs/2021ApJ...911...60C} {911, 60}

\bibitem[\protect\citeauthoryear{{Chen} et~al.,}{{Chen}
  et~al.}{2022}]{Chen2022}
{Chen} H.,  et~al., 2022, \mn@doi [\apj] {10.3847/1538-4357/ac658d}, \href
  {https://ui.adsabs.harvard.edu/abs/2022ApJ...931...29C} {931, 29}

\bibitem[\protect\citeauthoryear{{Choudhury}, {Paranjape}  \&
  {Bosman}}{{Choudhury} et~al.}{2021}]{Choudhury_2021}
{Choudhury} T.~R.,  {Paranjape} A.,   {Bosman} S. E.~I.,  2021, \mn@doi
  [\mnras] {10.1093/mnras/stab045}, \href
  {https://ui.adsabs.harvard.edu/abs/2021MNRAS.501.5782C} {501, 5782}

\bibitem[\protect\citeauthoryear{Ciardi et~al.,}{Ciardi
  et~al.}{2013}]{Ciardi_2013}
Ciardi B.,  et~al., 2013, \mn@doi [\mnras] {10.1093/mnras/sts156}, \href
  {http://dx.doi.org/10.1093/mnras/sts156} {428, 1755}

\bibitem[\protect\citeauthoryear{{Connor} et~al.,}{{Connor}
  et~al.}{2021}]{Connor_2021}
{Connor} T.,  et~al., 2021, \mn@doi [\apj] {10.3847/1538-4357/abe710}, \href
  {https://ui.adsabs.harvard.edu/abs/2021ApJ...911..120C} {911, 120}

\bibitem[\protect\citeauthoryear{{Datta}, {Friedrich}, {Mellema}, {Iliev}  \&
  {Shapiro}}{{Datta} et~al.}{2012}]{Datta_2012}
{Datta} K.~K.,  {Friedrich} M.~M.,  {Mellema} G.,  {Iliev} I.~T.,   {Shapiro}
  P.~R.,  2012, \mn@doi [\mnras] {10.1111/j.1365-2966.2012.21268.x}, \href
  {https://ui.adsabs.harvard.edu/abs/2012MNRAS.424..762D} {424, 762}

\bibitem[\protect\citeauthoryear{Davies et~al.,}{Davies
  et~al.}{2018}]{Davies_2018}
Davies F.~B.,  et~al., 2018, \mn@doi [\apj] {10.3847/1538-4357/aad6dc}, \href
  {http://dx.doi.org/10.3847/1538-4357/aad6dc} {864, 142}

\bibitem[\protect\citeauthoryear{{Davies}, {Hennawi}  \& {Eilers}}{{Davies}
  et~al.}{2019}]{Davies_2019}
{Davies} F.~B.,  {Hennawi} J.~F.,   {Eilers} A.-C.,  2019, \mn@doi [\apjl]
  {10.3847/2041-8213/ab42e3}, \href
  {https://ui.adsabs.harvard.edu/abs/2019ApJ...884L..19D} {884, L19}

\bibitem[\protect\citeauthoryear{{Davies}, {Hennawi}  \& {Eilers}}{{Davies}
  et~al.}{2020}]{Davies_2020}
{Davies} F.~B.,  {Hennawi} J.~F.,   {Eilers} A.-C.,  2020, \mn@doi [\mnras]
  {10.1093/mnras/stz3303}, \href
  {https://ui.adsabs.harvard.edu/abs/2020MNRAS.493.1330D} {493, 1330}

\bibitem[\protect\citeauthoryear{{Davies}, {Croft}, {Di-Matteo}, {Greig},
  {Feng}  \& {Wyithe}}{{Davies} et~al.}{2021}]{Davies_2021}
{Davies} J.~E.,  {Croft} R. A.~C.,  {Di-Matteo} T.,  {Greig} B.,  {Feng} Y.,
  {Wyithe} J. S.~B.,  2021, \mn@doi [\mnras] {10.1093/mnras/staa3531}, \href
  {https://ui.adsabs.harvard.edu/abs/2021MNRAS.501..146D} {501, 146}

\bibitem[\protect\citeauthoryear{{Dijkstra}, {Haiman}, {Mesinger}  \&
  {Wyithe}}{{Dijkstra} et~al.}{2008}]{Dijkstra_2008}
{Dijkstra} M.,  {Haiman} Z.,  {Mesinger} A.,   {Wyithe} J. S.~B.,  2008,
  \mn@doi [\mnras] {10.1111/j.1365-2966.2008.14031.x}, \href
  {https://ui.adsabs.harvard.edu/abs/2008MNRAS.391.1961D} {391, 1961}

\bibitem[\protect\citeauthoryear{D’Aloisio, McQuinn, Maupin, Davies, Trac,
  Fuller  \& Upton~Sanderbeck}{D’Aloisio et~al.}{2019}]{D_Aloisio_2019}
D’Aloisio A.,  McQuinn M.,  Maupin O.,  Davies F.~B.,  Trac H.,  Fuller S.,
  Upton~Sanderbeck P.~R.,  2019, \mn@doi [\apj] {10.3847/1538-4357/ab0d83},
  \href {http://dx.doi.org/10.3847/1538-4357/ab0d83} {874, 154}

\bibitem[\protect\citeauthoryear{{Eilers}, {Davies}, {Hennawi}, {Prochaska},
  {Luki{\'c}}  \& {Mazzucchelli}}{{Eilers} et~al.}{2017}]{Eilers_2017}
{Eilers} A.-C.,  {Davies} F.~B.,  {Hennawi} J.~F.,  {Prochaska} J.~X.,
  {Luki{\'c}} Z.,   {Mazzucchelli} C.,  2017, \mn@doi [\apj]
  {10.3847/1538-4357/aa6c60}, \href
  {https://ui.adsabs.harvard.edu/abs/2017ApJ...840...24E} {840, 24}

\bibitem[\protect\citeauthoryear{{Eilers}, {Davies}  \& {Hennawi}}{{Eilers}
  et~al.}{2018}]{Eilers_2018}
{Eilers} A.-C.,  {Davies} F.~B.,   {Hennawi} J.~F.,  2018, \mn@doi [\apj]
  {10.3847/1538-4357/aad4fd}, \href
  {https://ui.adsabs.harvard.edu/abs/2018ApJ...864...53E} {864, 53}

\bibitem[\protect\citeauthoryear{{Eilers} et~al.,}{{Eilers}
  et~al.}{2020}]{Eilers_2020}
{Eilers} A.-C.,  et~al., 2020, \mn@doi [\apj] {10.3847/1538-4357/aba52e}, \href
  {https://ui.adsabs.harvard.edu/abs/2020ApJ...900...37E} {900, 37}

\bibitem[\protect\citeauthoryear{{Eilers}, {Hennawi}, {Davies}  \&
  {Simcoe}}{{Eilers} et~al.}{2021}]{Eilers_2021}
{Eilers} A.-C.,  {Hennawi} J.~F.,  {Davies} F.~B.,   {Simcoe} R.~A.,  2021,
  \mn@doi [\apj] {10.3847/1538-4357/ac0a76}, \href
  {https://ui.adsabs.harvard.edu/abs/2021ApJ...917...38E} {917, 38}

\bibitem[\protect\citeauthoryear{{Fan} et~al.,}{{Fan} et~al.}{2006}]{Fan_2006}
{Fan} X.,  et~al., 2006, \mn@doi [\aj] {10.1086/504836}, \href
  {https://ui.adsabs.harvard.edu/abs/2006AJ....132..117F} {132, 117}

\bibitem[\protect\citeauthoryear{{Farina} et~al.,}{{Farina}
  et~al.}{2022}]{Farina_2022}
{Farina} E.~P.,  et~al., 2022, arXiv e-prints, \href
  {https://ui.adsabs.harvard.edu/abs/2022arXiv220705113F} {p. arXiv:2207.05113}

\bibitem[\protect\citeauthoryear{{Finlator}, {Keating}, {Oppenheimer},
  {Dav{\'e}}  \& {Zackrisson}}{{Finlator} et~al.}{2018}]{Finlator2018}
{Finlator} K.,  {Keating} L.,  {Oppenheimer} B.~D.,  {Dav{\'e}} R.,
  {Zackrisson} E.,  2018, \mn@doi [\mnras] {10.1093/mnras/sty1949}, \href
  {https://ui.adsabs.harvard.edu/abs/2018MNRAS.480.2628F} {480, 2628}

\bibitem[\protect\citeauthoryear{Furlanetto}{Furlanetto}{2006a}]{Furlanetto_2006a}
Furlanetto S.~R.,  2006a, \mn@doi [\mnras] {10.1111/j.1365-2966.2006.10603.x},
  \href {http://dx.doi.org/10.1111/j.1365-2966.2006.10603.x} {370, 1867}

\bibitem[\protect\citeauthoryear{Furlanetto}{Furlanetto}{2006b}]{Furlanetto_2006b}
Furlanetto S.~R.,  2006b, \mn@doi [\mnras] {10.1111/j.1365-2966.2006.10725.x},
  \href {http://dx.doi.org/10.1111/j.1365-2966.2006.10725.x} {371, 867}

\bibitem[\protect\citeauthoryear{Furlanetto \& Loeb}{Furlanetto \&
  Loeb}{2002}]{Furlanetto_2002}
Furlanetto S.~R.,  Loeb A.,  2002, \mn@doi [\apj] {10.1086/342757}, \href
  {http://dx.doi.org/10.1086/342757} {579, 1}

\bibitem[\protect\citeauthoryear{{Furlanetto} \& {Stoever}}{{Furlanetto} \&
  {Stoever}}{2010}]{FurlanettoStoever2010}
{Furlanetto} S.~R.,  {Stoever} S.~J.,  2010, \mn@doi [\mnras]
  {10.1111/j.1365-2966.2010.16401.x}, \href
  {https://ui.adsabs.harvard.edu/abs/2010MNRAS.404.1869F} {404, 1869}

\bibitem[\protect\citeauthoryear{{Furlanetto}, {Zaldarriaga}  \&
  {Hernquist}}{{Furlanetto} et~al.}{2004}]{Furlanetto_2004}
{Furlanetto} S.~R.,  {Zaldarriaga} M.,   {Hernquist} L.,  2004, \mn@doi [\apj]
  {10.1086/423028}, \href
  {https://ui.adsabs.harvard.edu/abs/2004ApJ...613...16F} {613, 16}

\bibitem[\protect\citeauthoryear{Gaikwad et~al.,}{Gaikwad
  et~al.}{2020}]{Gaikwad_2020}
Gaikwad P.,  et~al., 2020, \mn@doi [\mnras] {10.1093/mnras/staa907}, \href
  {http://dx.doi.org/10.1093/mnras/staa907} {494, 5091}

\bibitem[\protect\citeauthoryear{{Garaldi}, {Kannan}, {Smith}, {Springel},
  {Pakmor}, {Vogelsberger}  \& {Hernquist}}{{Garaldi}
  et~al.}{2022}]{Garaldi2022}
{Garaldi} E.,  {Kannan} R.,  {Smith} A.,  {Springel} V.,  {Pakmor} R.,
  {Vogelsberger} M.,   {Hernquist} L.,  2022, \mn@doi [\mnras]
  {10.1093/mnras/stac257}, \href
  {https://ui.adsabs.harvard.edu/abs/2022MNRAS.512.4909G} {512, 4909}

\bibitem[\protect\citeauthoryear{{Geil} \& {Wyithe}}{{Geil} \&
  {Wyithe}}{2008}]{GeilWyithe_2008}
{Geil} P.~M.,  {Wyithe} J. S.~B.,  2008, \mn@doi [\mnras]
  {10.1111/j.1365-2966.2008.13159.x}, \href
  {https://ui.adsabs.harvard.edu/abs/2008MNRAS.386.1683G} {386, 1683}

\bibitem[\protect\citeauthoryear{{Gloudemans} et~al.,}{{Gloudemans}
  et~al.}{2022}]{Gloudemans2022}
{Gloudemans} A.~J.,  et~al., 2022, \mn@doi [\aap]
  {10.1051/0004-6361/202244763}, \href
  {https://ui.adsabs.harvard.edu/abs/2022A&A...668A..27G} {668, A27}

\bibitem[\protect\citeauthoryear{{Gnedin}}{{Gnedin}}{2014}]{Gnedin2014}
{Gnedin} N.~Y.,  2014, \mn@doi [\apj] {10.1088/0004-637X/793/1/29}, \href
  {https://ui.adsabs.harvard.edu/abs/2014ApJ...793...29G} {793, 29}

\bibitem[\protect\citeauthoryear{{Greig}, {Mesinger}, {Haiman}  \&
  {Simcoe}}{{Greig} et~al.}{2017}]{Greig_2017}
{Greig} B.,  {Mesinger} A.,  {Haiman} Z.,   {Simcoe} R.~A.,  2017, \mn@doi
  [\mnras] {10.1093/mnras/stw3351}, \href
  {https://ui.adsabs.harvard.edu/abs/2017MNRAS.466.4239G} {466, 4239}

\bibitem[\protect\citeauthoryear{{Greig}, {Trott}, {Barry}, {Mutch}, {Pindor},
  {Webster}  \& {Wyithe}}{{Greig} et~al.}{2021}]{Greig_2020_MWA}
{Greig} B.,  {Trott} C.~M.,  {Barry} N.,  {Mutch} S.~J.,  {Pindor} B.,
  {Webster} R.~L.,   {Wyithe} J. S.~B.,  2021, \mn@doi [\mnras]
  {10.1093/mnras/staa3494}, \href
  {https://ui.adsabs.harvard.edu/abs/2021MNRAS.500.5322G} {500, 5322}

\bibitem[\protect\citeauthoryear{{Greig}, {Mesinger}, {Davies}, {Wang}, {Yang}
  \& {Hennawi}}{{Greig} et~al.}{2022}]{Greig_2022}
{Greig} B.,  {Mesinger} A.,  {Davies} F.~B.,  {Wang} F.,  {Yang} J.,
  {Hennawi} J.~F.,  2022, \mn@doi [\mnras] {10.1093/mnras/stac825}, \href
  {https://ui.adsabs.harvard.edu/abs/2022MNRAS.512.5390G} {512, 5390}

\bibitem[\protect\citeauthoryear{{Haehnelt}, {Natarajan}  \& {Rees}}{{Haehnelt}
  et~al.}{1998}]{Haehnelt_1998}
{Haehnelt} M.~G.,  {Natarajan} P.,   {Rees} M.~J.,  1998, \mn@doi [\mnras]
  {10.1046/j.1365-8711.1998.01951.x}, \href
  {https://ui.adsabs.harvard.edu/abs/1998MNRAS.300..817H} {300, 817}

\bibitem[\protect\citeauthoryear{{Harris} et~al.,}{{Harris}
  et~al.}{2020}]{Harris_2020}
{Harris} C.~R.,  et~al., 2020, \mn@doi [Nature] {10.1038/s41586-020-2649-2},
  \href {https://ui.adsabs.harvard.edu/abs/2020arXiv200610256H} {585, 357}

\bibitem[\protect\citeauthoryear{{Hopkins}, {Hernquist}, {Martini}, {Cox},
  {Robertson}, {Di Matteo}  \& {Springel}}{{Hopkins}
  et~al.}{2005}]{Hopkins_2005}
{Hopkins} P.~F.,  {Hernquist} L.,  {Martini} P.,  {Cox} T.~J.,  {Robertson} B.,
   {Di Matteo} T.,   {Springel} V.,  2005, \mn@doi [\apjl] {10.1086/431146},
  \href {https://ui.adsabs.harvard.edu/abs/2005ApJ...625L..71H} {625, L71}

\bibitem[\protect\citeauthoryear{{Hsyu}, {Cooke}, {Prochaska}  \&
  {Bolte}}{{Hsyu} et~al.}{2020}]{Hsyu_2020}
{Hsyu} T.,  {Cooke} R.~J.,  {Prochaska} J.~X.,   {Bolte} M.,  2020, \mn@doi
  [\apj] {10.3847/1538-4357/ab91af}, \href
  {https://ui.adsabs.harvard.edu/abs/2020ApJ...896...77H} {896, 77}

\bibitem[\protect\citeauthoryear{{Hunter}}{{Hunter}}{2007}]{Hunter_2007}
{Hunter} J.~D.,  2007, \mn@doi [Computing in Science and Engineering]
  {10.1109/MCSE.2007.55}, \href
  {https://ui.adsabs.harvard.edu/abs/2007CSE.....9...90H} {9, 90}

\bibitem[\protect\citeauthoryear{{Ighina}, {Belladitta}, {Caccianiga},
  {Broderick}, {Drouart}, {Moretti}  \& {Seymour}}{{Ighina}
  et~al.}{2021}]{Ighina_2021}
{Ighina} L.,  {Belladitta} S.,  {Caccianiga} A.,  {Broderick} J.~W.,  {Drouart}
  G.,  {Moretti} A.,   {Seymour} N.,  2021, \mn@doi [\aap]
  {10.1051/0004-6361/202140362}, \href
  {https://ui.adsabs.harvard.edu/abs/2021A&A...647L..11I} {647, L11}

\bibitem[\protect\citeauthoryear{{Iliev}, {Mellema}, {Ahn}, {Shapiro}, {Mao}
  \& {Pen}}{{Iliev} et~al.}{2014}]{Iliev_2014}
{Iliev} I.~T.,  {Mellema} G.,  {Ahn} K.,  {Shapiro} P.~R.,  {Mao} Y.,   {Pen}
  U.-L.,  2014, \mn@doi [\mnras] {10.1093/mnras/stt2497}, \href
  {https://ui.adsabs.harvard.edu/abs/2014MNRAS.439..725I} {439, 725}

\bibitem[\protect\citeauthoryear{{Inayoshi}, {Visbal}  \& {Haiman}}{{Inayoshi}
  et~al.}{2020}]{Inayoshi_2020}
{Inayoshi} K.,  {Visbal} E.,   {Haiman} Z.,  2020, \mn@doi [\araa]
  {10.1146/annurev-astro-120419-014455}, \href
  {https://ui.adsabs.harvard.edu/abs/2020ARA&A..58...27I} {58, 27}

\bibitem[\protect\citeauthoryear{{Ishimoto} et~al.,}{{Ishimoto}
  et~al.}{2020}]{Ishimoto_2020}
{Ishimoto} R.,  et~al., 2020, \mn@doi [\apj] {10.3847/1538-4357/abb80b}, \href
  {https://ui.adsabs.harvard.edu/abs/2020ApJ...903...60I} {903, 60}

\bibitem[\protect\citeauthoryear{{Kakiichi} et~al.,}{{Kakiichi}
  et~al.}{2017}]{Kakiichi_2017}
{Kakiichi} K.,  et~al., 2017, \mn@doi [\mnras] {10.1093/mnras/stx1568}, \href
  {https://ui.adsabs.harvard.edu/abs/2017MNRAS.471.1936K} {471, 1936}

\bibitem[\protect\citeauthoryear{{Kaur}, {Gillet}  \& {Mesinger}}{{Kaur}
  et~al.}{2020}]{Kaur_2020}
{Kaur} H.~D.,  {Gillet} N.,   {Mesinger} A.,  2020, \mn@doi [\mnras]
  {10.1093/mnras/staa1323}, \href
  {https://ui.adsabs.harvard.edu/abs/2020MNRAS.495.2354K} {495, 2354}

\bibitem[\protect\citeauthoryear{{Keating}, {Haehnelt}, {Cantalupo}  \&
  {Puchwein}}{{Keating} et~al.}{2015}]{Keating_2015}
{Keating} L.~C.,  {Haehnelt} M.~G.,  {Cantalupo} S.,   {Puchwein} E.,  2015,
  \mn@doi [\mnras] {10.1093/mnras/stv2020}, \href
  {https://ui.adsabs.harvard.edu/abs/2015MNRAS.454..681K} {454, 681}

\bibitem[\protect\citeauthoryear{Keating, Weinberger, Kulkarni, Haehnelt,
  Chardin  \& Aubert}{Keating et~al.}{2020}]{Keating_2020}
Keating L.~C.,  Weinberger L.~H.,  Kulkarni G.,  Haehnelt M.~G.,  Chardin J.,
  Aubert D.,  2020, \mn@doi [\mnras] {10.1093/mnras/stz3083}, \href
  {http://dx.doi.org/10.1093/mnras/stz3083} {491, 1736}

\bibitem[\protect\citeauthoryear{{Khrykin}, {Hennawi}  \& {Worseck}}{{Khrykin}
  et~al.}{2019}]{Khrykin_2019}
{Khrykin} I.~S.,  {Hennawi} J.~F.,   {Worseck} G.,  2019, \mn@doi [\mnras]
  {10.1093/mnras/stz135}, \href
  {https://ui.adsabs.harvard.edu/abs/2019MNRAS.484.3897K} {484, 3897}

\bibitem[\protect\citeauthoryear{{Khrykin}, {Hennawi}, {Worseck}  \&
  {Davies}}{{Khrykin} et~al.}{2021}]{Khrykin_2021}
{Khrykin} I.~S.,  {Hennawi} J.~F.,  {Worseck} G.,   {Davies} F.~B.,  2021,
  \mn@doi [\mnras] {10.1093/mnras/stab1288}, \href
  {https://ui.adsabs.harvard.edu/abs/2021MNRAS.505..649K} {505, 649}

\bibitem[\protect\citeauthoryear{{King} \& {Nixon}}{{King} \&
  {Nixon}}{2015}]{King_2015}
{King} A.,  {Nixon} C.,  2015, \mn@doi [\mnras] {10.1093/mnrasl/slv098}, \href
  {https://ui.adsabs.harvard.edu/abs/2015MNRAS.453L..46K} {453, L46}

\bibitem[\protect\citeauthoryear{{Knevitt}, {Wynn}, {Power}  \&
  {Bolton}}{{Knevitt} et~al.}{2014}]{Knevitt2014}
{Knevitt} G.,  {Wynn} G.~A.,  {Power} C.,   {Bolton} J.~S.,  2014, \mn@doi
  [\mnras] {10.1093/mnras/stu1803}, \href
  {https://ui.adsabs.harvard.edu/abs/2014MNRAS.445.2034K} {445, 2034}

\bibitem[\protect\citeauthoryear{{Kohler}, {Gnedin}, {Miralda-Escud{\'e}}  \&
  {Shaver}}{{Kohler} et~al.}{2005}]{Kohler_2005}
{Kohler} K.,  {Gnedin} N.~Y.,  {Miralda-Escud{\'e}} J.,   {Shaver} P.~A.,
  2005, \mn@doi [\apj] {10.1086/444370}, \href
  {https://ui.adsabs.harvard.edu/abs/2005ApJ...633..552K} {633, 552}

\bibitem[\protect\citeauthoryear{{Kroupa}, {Subr}, {Jerabkova}  \&
  {Wang}}{{Kroupa} et~al.}{2020}]{Kroupa_2020}
{Kroupa} P.,  {Subr} L.,  {Jerabkova} T.,   {Wang} L.,  2020, \mn@doi [\mnras]
  {10.1093/mnras/staa2276}, \href
  {https://ui.adsabs.harvard.edu/abs/2020MNRAS.498.5652K} {498, 5652}

\bibitem[\protect\citeauthoryear{Kulkarni, Keating, Haehnelt, Bosman, Puchwein,
  Chardin  \& Aubert}{Kulkarni et~al.}{2019}]{Kulkarni_2019}
Kulkarni G.,  Keating L.~C.,  Haehnelt M.~G.,  Bosman S. E.~I.,  Puchwein E.,
  Chardin J.,   Aubert D.,  2019, \mn@doi [\mnras] {10.1093/mnrasl/slz025},
  \href {http://dx.doi.org/10.1093/mnrasl/slz025} {485, L24}

\bibitem[\protect\citeauthoryear{{Lewis} et~al.,}{{Lewis}
  et~al.}{2022}]{Lewis2022}
{Lewis} J. S.~W.,  et~al., 2022, \mn@doi [\mnras] {10.1093/mnras/stac2383},
  \href {https://ui.adsabs.harvard.edu/abs/2022MNRAS.516.3389L} {516, 3389}

\bibitem[\protect\citeauthoryear{{Lidz}, {McQuinn}, {Zaldarriaga}, {Hernquist}
  \& {Dutta}}{{Lidz} et~al.}{2007}]{Lidz_2007}
{Lidz} A.,  {McQuinn} M.,  {Zaldarriaga} M.,  {Hernquist} L.,   {Dutta} S.,
  2007, \mn@doi [\apj] {10.1086/521974}, \href
  {https://ui.adsabs.harvard.edu/abs/2007ApJ...670...39L} {670, 39}

\bibitem[\protect\citeauthoryear{{Liu} et~al.,}{{Liu} et~al.}{2021}]{Liu_2021}
{Liu} Y.,  et~al., 2021, \mn@doi [\apj] {10.3847/1538-4357/abd3a8}, \href
  {https://ui.adsabs.harvard.edu/abs/2021ApJ...908..124L} {908, 124}

\bibitem[\protect\citeauthoryear{{Loeb} \& {Rasio}}{{Loeb} \&
  {Rasio}}{1994}]{LoebRasio_1994}
{Loeb} A.,  {Rasio} F.~A.,  1994, \mn@doi [\apj] {10.1086/174548}, \href
  {https://ui.adsabs.harvard.edu/abs/1994ApJ...432...52L} {432, 52}

\bibitem[\protect\citeauthoryear{{Lusso} et~al.,}{{Lusso}
  et~al.}{2010}]{Lusso_2010}
{Lusso} E.,  et~al., 2010, \mn@doi [\aap] {10.1051/0004-6361/200913298}, \href
  {https://ui.adsabs.harvard.edu/abs/2010A&A...512A..34L} {512, A34}

\bibitem[\protect\citeauthoryear{{Lusso}, {Worseck}, {Hennawi}, {Prochaska},
  {Vignali}, {Stern}  \& {O'Meara}}{{Lusso} et~al.}{2015}]{Lusso_2015}
{Lusso} E.,  {Worseck} G.,  {Hennawi} J.~F.,  {Prochaska} J.~X.,  {Vignali} C.,
   {Stern} J.,   {O'Meara} J.~M.,  2015, \mn@doi [\mnras]
  {10.1093/mnras/stv516}, \href
  {https://ui.adsabs.harvard.edu/abs/2015MNRAS.449.4204L} {449, 4204}

\bibitem[\protect\citeauthoryear{{Ma}, {Ciardi}, {Kakiichi}, {Zaroubi}, {Zhi}
  \& {Busch}}{{Ma} et~al.}{2020}]{Ma_2020}
{Ma} Q.-B.,  {Ciardi} B.,  {Kakiichi} K.,  {Zaroubi} S.,  {Zhi} Q.-J.,
  {Busch} P.,  2020, \mn@doi [\apj] {10.3847/1538-4357/ab5b95}, \href
  {https://ui.adsabs.harvard.edu/abs/2020ApJ...888..112M} {888, 112}

\bibitem[\protect\citeauthoryear{Mack \& Wyithe}{Mack \&
  Wyithe}{2012}]{Mack_2012}
Mack K.~J.,  Wyithe J. S.~B.,  2012, \mn@doi [\mnras]
  {10.1111/j.1365-2966.2012.21561.x}, \href
  {http://dx.doi.org/10.1111/j.1365-2966.2012.21561.x} {425, 2988}

\bibitem[\protect\citeauthoryear{{Madau} \& {Rees}}{{Madau} \&
  {Rees}}{2000}]{MadauRees2000}
{Madau} P.,  {Rees} M.~J.,  2000, \mn@doi [\apjl] {10.1086/312934}, \href
  {https://ui.adsabs.harvard.edu/abs/2000ApJ...542L..69M} {542, L69}

\bibitem[\protect\citeauthoryear{{Madau}, {Meiksin}  \& {Rees}}{{Madau}
  et~al.}{1997}]{Madau_1997}
{Madau} P.,  {Meiksin} A.,   {Rees} M.~J.,  1997, \mn@doi [\apj]
  {10.1086/303549}, \href
  {https://ui.adsabs.harvard.edu/abs/1997ApJ...475..429M} {475, 429}

\bibitem[\protect\citeauthoryear{{Madau}, {Haardt}  \& {Dotti}}{{Madau}
  et~al.}{2014}]{Madau_2014}
{Madau} P.,  {Haardt} F.,   {Dotti} M.,  2014, \mn@doi [\apjl]
  {10.1088/2041-8205/784/2/L38}, \href
  {https://ui.adsabs.harvard.edu/abs/2014ApJ...784L..38M} {784, L38}

\bibitem[\protect\citeauthoryear{{Majumdar}, {Bharadwaj}  \&
  {Choudhury}}{{Majumdar} et~al.}{2012}]{Majumdar_2012}
{Majumdar} S.,  {Bharadwaj} S.,   {Choudhury} T.~R.,  2012, \mn@doi [\mnras]
  {10.1111/j.1365-2966.2012.21914.x}, \href
  {https://ui.adsabs.harvard.edu/abs/2012MNRAS.426.3178M} {426, 3178}

\bibitem[\protect\citeauthoryear{{Martini}}{{Martini}}{2004}]{Martini_2004}
{Martini} P.,  2004, in {Ho} L.~C.,  ed., Coevolution of Black Holes and
  Galaxies. p.~169 (\mn@eprint {arXiv} {astro-ph/0304009})

\bibitem[\protect\citeauthoryear{{Maselli}, {Gallerani}, {Ferrara}  \&
  {Choudhury}}{{Maselli} et~al.}{2007}]{Maselli2007}
{Maselli} A.,  {Gallerani} S.,  {Ferrara} A.,   {Choudhury} T.~R.,  2007,
  \mn@doi [\mnras] {10.1111/j.1745-3933.2007.00283.x}, \href
  {https://ui.adsabs.harvard.edu/abs/2007MNRAS.376L..34M} {376, L34}

\bibitem[\protect\citeauthoryear{{Mazzucchelli} et~al.,}{{Mazzucchelli}
  et~al.}{2017}]{Mazzucchelli_2017}
{Mazzucchelli} C.,  et~al., 2017, \mn@doi [\apj] {10.3847/1538-4357/aa9185},
  \href {https://ui.adsabs.harvard.edu/abs/2017ApJ...849...91M} {849, 91}

\bibitem[\protect\citeauthoryear{{Meiksin}}{{Meiksin}}{2011}]{Meiksin_2011}
{Meiksin} A.,  2011, \mn@doi [\mnras] {10.1111/j.1365-2966.2011.19362.x}, \href
  {https://ui.adsabs.harvard.edu/abs/2011MNRAS.417.1480M} {417, 1480}

\bibitem[\protect\citeauthoryear{{Mesinger} \& {Furlanetto}}{{Mesinger} \&
  {Furlanetto}}{2008}]{MesingerFurlanetto2008}
{Mesinger} A.,  {Furlanetto} S.~R.,  2008, \mn@doi [\mnras]
  {10.1111/j.1365-2966.2007.12836.x}, \href
  {https://ui.adsabs.harvard.edu/abs/2008MNRAS.385.1348M} {385, 1348}

\bibitem[\protect\citeauthoryear{{Miralda-Escud{\'e}} \&
  {Rees}}{{Miralda-Escud{\'e}} \& {Rees}}{1998}]{MiraldaEscudeRees_1998}
{Miralda-Escud{\'e}} J.,  {Rees} M.~J.,  1998, \mn@doi [\apj] {10.1086/305458},
  \href {https://ui.adsabs.harvard.edu/abs/1998ApJ...497...21M} {497, 21}

\bibitem[\protect\citeauthoryear{{Molaro} et~al.,}{{Molaro}
  et~al.}{2022}]{Molaro_2022}
{Molaro} M.,  et~al., 2022, \mn@doi [\mnras] {10.1093/mnras/stab3416}, \href
  {https://ui.adsabs.harvard.edu/abs/2022MNRAS.509.6119M} {509, 6119}

\bibitem[\protect\citeauthoryear{{Morey}, {Eilers}, {Davies}, {Hennawi}  \&
  {Simcoe}}{{Morey} et~al.}{2021}]{Morey_2021}
{Morey} K.~A.,  {Eilers} A.-C.,  {Davies} F.~B.,  {Hennawi} J.~F.,   {Simcoe}
  R.~A.,  2021, \mn@doi [\apj] {10.3847/1538-4357/ac1c70}, \href
  {https://ui.adsabs.harvard.edu/abs/2021ApJ...921...88M} {921, 88}

\bibitem[\protect\citeauthoryear{{Mortlock} et~al.,}{{Mortlock}
  et~al.}{2011}]{Mortlock_2011}
{Mortlock} D.~J.,  et~al., 2011, \mn@doi [\nat] {10.1038/nature10159}, \href
  {https://ui.adsabs.harvard.edu/abs/2011Natur.474..616M} {474, 616}

\bibitem[\protect\citeauthoryear{{Murdoch}, {Hunstead}, {Pettini}  \&
  {Blades}}{{Murdoch} et~al.}{1986}]{Murdoch1986}
{Murdoch} H.~S.,  {Hunstead} R.~W.,  {Pettini} M.,   {Blades} J.~C.,  1986,
  \mn@doi [\apj] {10.1086/164573}, \href
  {https://ui.adsabs.harvard.edu/abs/1986ApJ...309...19M} {309, 19}

\bibitem[\protect\citeauthoryear{{Nakatani}, {Fialkov}  \&
  {Yoshida}}{{Nakatani} et~al.}{2020}]{Nakatani_2020}
{Nakatani} R.,  {Fialkov} A.,   {Yoshida} N.,  2020, \mn@doi [\apj]
  {10.3847/1538-4357/abc5b4}, \href
  {https://ui.adsabs.harvard.edu/abs/2020ApJ...905..151N} {905, 151}

\bibitem[\protect\citeauthoryear{{Nasir} \& {D'Aloisio}}{{Nasir} \&
  {D'Aloisio}}{2020}]{Nasir_2020}
{Nasir} F.,  {D'Aloisio} A.,  2020, \mn@doi [\mnras] {10.1093/mnras/staa894},
  \href {https://ui.adsabs.harvard.edu/abs/2020MNRAS.494.3080N} {494, 3080}

\bibitem[\protect\citeauthoryear{{O{\~n}orbe}, {Davies}, {Luki{\'c}}, {},
  {Hennawi}  \& {Sorini}}{{O{\~n}orbe} et~al.}{2019}]{Onorbe_2019}
{O{\~n}orbe} J.,  {Davies} F.~B.,  {Luki{\'c}} {} Z.,  {Hennawi} J.~F.,
  {Sorini} D.,  2019, \mn@doi [\mnras] {10.1093/mnras/stz984}, \href
  {https://ui.adsabs.harvard.edu/abs/2019MNRAS.486.4075O} {486, 4075}

\bibitem[\protect\citeauthoryear{{Ocvirk}, {Lewis}, {Gillet}, {Chardin},
  {Aubert}, {Deparis}  \& {Th{\'e}lie}}{{Ocvirk} et~al.}{2021}]{Ocvirk2021}
{Ocvirk} P.,  {Lewis} J. S.~W.,  {Gillet} N.,  {Chardin} J.,  {Aubert} D.,
  {Deparis} N.,   {Th{\'e}lie} {\'E}.,  2021, \mn@doi [\mnras]
  {10.1093/mnras/stab2502}, \href
  {https://ui.adsabs.harvard.edu/abs/2021MNRAS.507.6108O} {507, 6108}

\bibitem[\protect\citeauthoryear{{Park}, {Shapiro}, {Choi}, {Yoshida}, {Hirano}
   \& {Ahn}}{{Park} et~al.}{2016}]{Park_2016}
{Park} H.,  {Shapiro} P.~R.,  {Choi} J.-h.,  {Yoshida} N.,  {Hirano} S.,
  {Ahn} K.,  2016, \mn@doi [\apj] {10.3847/0004-637X/831/1/86}, \href
  {https://ui.adsabs.harvard.edu/abs/2016ApJ...831...86P} {831, 86}

\bibitem[\protect\citeauthoryear{{Planck Collaboration}}{{Planck
  Collaboration}}{2014}]{planck2014}
{Planck Collaboration} 2014, \mn@doi [A&A] {10.1051/0004-6361/201321591}, \href
  {http://dx.doi.org/10.1051/0004-6361/201321591} {571, A16}

\bibitem[\protect\citeauthoryear{{Puchwein}, {Haardt}, {Haehnelt}  \&
  {Madau}}{{Puchwein} et~al.}{2019}]{Puchwein_2019}
{Puchwein} E.,  {Haardt} F.,  {Haehnelt} M.~G.,   {Madau} P.,  2019, \mn@doi
  [\mnras] {10.1093/mnras/stz222}, \href
  {https://ui.adsabs.harvard.edu/abs/2019MNRAS.485...47P} {485, 47}

\bibitem[\protect\citeauthoryear{{Puchwein} et~al.,}{{Puchwein}
  et~al.}{2022}]{Puchwein_2022}
{Puchwein} E.,  et~al., 2022, arXiv e-prints, \href
  {https://ui.adsabs.harvard.edu/abs/2022arXiv220713098P} {p. arXiv:2207.13098}

\bibitem[\protect\citeauthoryear{{Qin}, {Mesinger}, {Bosman}  \& {Viel}}{{Qin}
  et~al.}{2021}]{Qin_2021}
{Qin} Y.,  {Mesinger} A.,  {Bosman} S. E.~I.,   {Viel} M.,  2021, \mn@doi
  [\mnras] {10.1093/mnras/stab1833}, \href
  {https://ui.adsabs.harvard.edu/abs/2021MNRAS.506.2390Q} {506, 2390}

\bibitem[\protect\citeauthoryear{{Reed} et~al.,}{{Reed}
  et~al.}{2015}]{Reed_2015}
{Reed} S.~L.,  et~al., 2015, \mn@doi [\mnras] {10.1093/mnras/stv2031}, \href
  {https://ui.adsabs.harvard.edu/abs/2015MNRAS.454.3952R} {454, 3952}

\bibitem[\protect\citeauthoryear{{Regan}, {Visbal}, {Wise}, {Haiman},
  {Johansson}  \& {Bryan}}{{Regan} et~al.}{2017}]{Regan_2017}
{Regan} J.~A.,  {Visbal} E.,  {Wise} J.~H.,  {Haiman} Z.,  {Johansson} P.~H.,
  {Bryan} G.~L.,  2017, \mn@doi [Nature Astronomy] {10.1038/s41550-017-0075},
  \href {https://ui.adsabs.harvard.edu/abs/2017NatAs...1E..75R} {1, 0075}

\bibitem[\protect\citeauthoryear{{Rhook} \& {Haehnelt}}{{Rhook} \&
  {Haehnelt}}{2006}]{RhookHaehnelt_2006}
{Rhook} K.~J.,  {Haehnelt} M.~G.,  2006, \mn@doi [\mnras]
  {10.1111/j.1365-2966.2006.11003.x}, \href
  {https://ui.adsabs.harvard.edu/abs/2006MNRAS.373..623R} {373, 623}

\bibitem[\protect\citeauthoryear{{Ricci} et~al.,}{{Ricci}
  et~al.}{2017}]{Ricci_2017obsc}
{Ricci} C.,  et~al., 2017, \mn@doi [\mnras] {10.1093/mnras/stx173}, \href
  {https://ui.adsabs.harvard.edu/abs/2017MNRAS.468.1273R} {468, 1273}

\bibitem[\protect\citeauthoryear{{Salpeter}}{{Salpeter}}{1964}]{Salpeter_1964}
{Salpeter} E.~E.,  1964, \mn@doi [\apj] {10.1086/147973}, \href
  {https://ui.adsabs.harvard.edu/abs/1964ApJ...140..796S} {140, 796}

\bibitem[\protect\citeauthoryear{{Satyavolu}, {Kulkarni}, {Keating}  \&
  {Haehnelt}}{{Satyavolu} et~al.}{2022}]{Satyavolu_2022}
{Satyavolu} S.,  {Kulkarni} G.,  {Keating} L.~C.,   {Haehnelt} M.~G.,  2022,
  arXiv e-prints, \href {https://ui.adsabs.harvard.edu/abs/2022arXiv220908103S}
  {p. arXiv:2209.08103}

\bibitem[\protect\citeauthoryear{{Schawinski}, {Koss}, {Berney}  \&
  {Sartori}}{{Schawinski} et~al.}{2015}]{Schawinski_2015}
{Schawinski} K.,  {Koss} M.,  {Berney} S.,   {Sartori} L.~F.,  2015, \mn@doi
  [\mnras] {10.1093/mnras/stv1136}, \href
  {https://ui.adsabs.harvard.edu/abs/2015MNRAS.451.2517S} {451, 2517}

\bibitem[\protect\citeauthoryear{Semelin}{Semelin}{2016}]{Semelin_2016}
Semelin B.,  2016, \mn@doi [\mnras] {10.1093/mnras/stv2312}, \href
  {http://dx.doi.org/10.1093/mnras/stv2312} {455, 962}

\bibitem[\protect\citeauthoryear{{Shakura} \& {Sunyaev}}{{Shakura} \&
  {Sunyaev}}{1973}]{ShakuraSunyaev_1973}
{Shakura} N.~I.,  {Sunyaev} R.~A.,  1973, \aap, \href
  {https://ui.adsabs.harvard.edu/abs/1973A&A....24..337S} {24, 337}

\bibitem[\protect\citeauthoryear{{Shapiro} \& {Giroux}}{{Shapiro} \&
  {Giroux}}{1987}]{ShapiroGiroux1987}
{Shapiro} P.~R.,  {Giroux} M.~L.,  1987, \mn@doi [\apjl] {10.1086/185015},
  \href {https://ui.adsabs.harvard.edu/abs/1987ApJ...321L.107S} {321, L107}

\bibitem[\protect\citeauthoryear{{Shen}}{{Shen}}{2021}]{Shen_2021}
{Shen} Y.,  2021, \mn@doi [\apj] {10.3847/1538-4357/ac1ce4}, \href
  {https://ui.adsabs.harvard.edu/abs/2021ApJ...921...70S} {921, 70}

\bibitem[\protect\citeauthoryear{{Shen} et~al.,}{{Shen}
  et~al.}{2007}]{Shen2007}
{Shen} Y.,  et~al., 2007, \mn@doi [\aj] {10.1086/513517}, \href
  {https://ui.adsabs.harvard.edu/abs/2007AJ....133.2222S} {133, 2222}

\bibitem[\protect\citeauthoryear{{Shen}, {Hopkins}, {Faucher-Gigu{\`e}re},
  {Alexander}, {Richards}, {Ross}  \& {Hickox}}{{Shen}
  et~al.}{2020}]{Shen_2020}
{Shen} X.,  {Hopkins} P.~F.,  {Faucher-Gigu{\`e}re} C.-A.,  {Alexander} D.~M.,
  {Richards} G.~T.,  {Ross} N.~P.,   {Hickox} R.~C.,  2020, \mn@doi [\mnras]
  {10.1093/mnras/staa1381}, \href
  {https://ui.adsabs.harvard.edu/abs/2020MNRAS.495.3252S} {495, 3252}

\bibitem[\protect\citeauthoryear{Springel}{Springel}{2005}]{Springel_2005}
Springel V.,  2005, \mn@doi [\mnras] {10.1111/j.1365-2966.2005.09655.x}, \href
  {http://dx.doi.org/10.1111/j.1365-2966.2005.09655.x} {364, 1105}

\bibitem[\protect\citeauthoryear{{Steffen}, {Strateva}, {Brandt}, {Alexander},
  {Koekemoer}, {Lehmer}, {Schneider}  \& {Vignali}}{{Steffen}
  et~al.}{2006}]{Steffen_2006}
{Steffen} A.~T.,  {Strateva} I.,  {Brandt} W.~N.,  {Alexander} D.~M.,
  {Koekemoer} A.~M.,  {Lehmer} B.~D.,  {Schneider} D.~P.,   {Vignali} C.,
  2006, \mn@doi [\aj] {10.1086/503627}, \href
  {https://ui.adsabs.harvard.edu/abs/2006AJ....131.2826S} {131, 2826}

\bibitem[\protect\citeauthoryear{{\VAN{Szoltinsk{\'y}}{\v{{S}}oltinsk{\'y}}{\v{{S}}oltinsk{\'y}}}
  et~al.,}{{\VAN{Szoltinsk{\'y}}{\v{{S}}oltinsk{\'y}}{\v{{S}}oltinsk{\'y}}}
  et~al.}{2021}]{Soltinsky_2021}
{\VAN{Szoltinsk{\'y}}{\v{{S}}oltinsk{\'y}}{\v{{S}}oltinsk{\'y}}} T.,  et~al.,
  2021, \mn@doi [\mnras] {10.1093/mnras/stab1830}, \href
  {https://ui.adsabs.harvard.edu/abs/2021MNRAS.506.5818S} {506, 5818}

\bibitem[\protect\citeauthoryear{{Tepper-Garc{\'\i}a}}{{Tepper-Garc{\'\i}a}}{2006}]{TepperGarcia_2006}
{Tepper-Garc{\'\i}a} T.,  2006, \mn@doi [\mnras]
  {10.1111/j.1365-2966.2006.10450.x}, \href
  {https://ui.adsabs.harvard.edu/abs/2006MNRAS.369.2025T} {369, 2025}

\bibitem[\protect\citeauthoryear{{The HERA Collaboration}}{{The HERA
  Collaboration}}{2022}]{HERA_2022_constraints_new}
{The HERA Collaboration} 2022, arXiv e-prints, \href
  {https://ui.adsabs.harvard.edu/abs/2022arXiv221004912T} {p. arXiv:2210.04912}

\bibitem[\protect\citeauthoryear{{Venemans} et~al.,}{{Venemans}
  et~al.}{2015}]{Venemans_2015}
{Venemans} B.~P.,  et~al., 2015, \mn@doi [\apjl] {10.1088/2041-8205/801/1/L11},
  \href {https://ui.adsabs.harvard.edu/abs/2015ApJ...801L..11V} {801, L11}

\bibitem[\protect\citeauthoryear{{Viel}, {Haehnelt}  \& {Springel}}{{Viel}
  et~al.}{2004}]{Viel_2004}
{Viel} M.,  {Haehnelt} M.~G.,   {Springel} V.,  2004, \mn@doi [\mnras]
  {10.1111/j.1365-2966.2004.08224.x}, \href
  {https://ui.adsabs.harvard.edu/abs/2004MNRAS.354..684V} {354, 684}

\bibitem[\protect\citeauthoryear{{Villanueva-Domingo} \&
  {Ichiki}}{{Villanueva-Domingo} \& {Ichiki}}{2022}]{VillanuevaDomingo2021}
{Villanueva-Domingo} P.,  {Ichiki} K.,  2022, \mn@doi [\pasj]
  {10.1093/pasj/psab119}

\bibitem[\protect\citeauthoryear{{Virtanen} et~al.,}{{Virtanen}
  et~al.}{2020}]{Virtanen_2020}
{Virtanen} P.,  et~al., 2020, \mn@doi [Nature Methods]
  {10.1038/s41592-019-0686-2}, \href
  {https://ui.adsabs.harvard.edu/abs/2020NatMe..17..261V} {17, 261}

\bibitem[\protect\citeauthoryear{{Vito} et~al.,}{{Vito}
  et~al.}{2019}]{Vito_2019}
{Vito} F.,  et~al., 2019, \mn@doi [\aap] {10.1051/0004-6361/201936217}, \href
  {https://ui.adsabs.harvard.edu/abs/2019A&A...630A.118V} {630, A118}

\bibitem[\protect\citeauthoryear{{Volonteri}, {Silk}  \& {Dubus}}{{Volonteri}
  et~al.}{2015}]{Volonteri_2015}
{Volonteri} M.,  {Silk} J.,   {Dubus} G.,  2015, \mn@doi [\apj]
  {10.1088/0004-637X/804/2/148}, \href
  {https://ui.adsabs.harvard.edu/abs/2015ApJ...804..148V} {804, 148}

\bibitem[\protect\citeauthoryear{{Wang} et~al.,}{{Wang}
  et~al.}{2020}]{Wang_2020}
{Wang} F.,  et~al., 2020, \mn@doi [\apj] {10.3847/1538-4357/ab8c45}, \href
  {https://ui.adsabs.harvard.edu/abs/2020ApJ...896...23W} {896, 23}

\bibitem[\protect\citeauthoryear{{Wang} et~al.,}{{Wang}
  et~al.}{2021}]{Wang_2021}
{Wang} F.,  et~al., 2021, \mn@doi [\apj] {10.3847/1538-4357/abcc5e}, \href
  {https://ui.adsabs.harvard.edu/abs/2021ApJ...908...53W} {908, 53}

\bibitem[\protect\citeauthoryear{{Willott} et~al.,}{{Willott}
  et~al.}{2010}]{Willott_2010}
{Willott} C.~J.,  et~al., 2010, \mn@doi [\aj] {10.1088/0004-6256/140/2/546},
  \href {https://ui.adsabs.harvard.edu/abs/2010AJ....140..546W} {140, 546}

\bibitem[\protect\citeauthoryear{{Worseck}, {Khrykin}, {Hennawi}, {Prochaska}
  \& {Farina}}{{Worseck} et~al.}{2021}]{Worseck_2021}
{Worseck} G.,  {Khrykin} I.~S.,  {Hennawi} J.~F.,  {Prochaska} J.~X.,
  {Farina} E.~P.,  2021, \mn@doi [\mnras] {10.1093/mnras/stab1685}, \href
  {https://ui.adsabs.harvard.edu/abs/2021MNRAS.505.5084W} {505, 5084}

\bibitem[\protect\citeauthoryear{{Wyithe}}{{Wyithe}}{2008}]{Wyithe_2008}
{Wyithe} J. S.~B.,  2008, \mn@doi [\mnras] {10.1111/j.1365-2966.2008.13267.x},
  \href {https://ui.adsabs.harvard.edu/abs/2008MNRAS.387..469W} {387, 469}

\bibitem[\protect\citeauthoryear{{Wyithe} \& {Loeb}}{{Wyithe} \&
  {Loeb}}{2004a}]{WyitheLoeb2004}
{Wyithe} J. S.~B.,  {Loeb} A.,  2004a, \mn@doi [\nat] {10.1038/nature02336},
  \href {https://ui.adsabs.harvard.edu/abs/2004Natur.427..815W} {427, 815}

\bibitem[\protect\citeauthoryear{{Wyithe} \& {Loeb}}{{Wyithe} \&
  {Loeb}}{2004b}]{Wyithe_2004}
{Wyithe} J. S.~B.,  {Loeb} A.,  2004b, \mn@doi [\nat] {10.1038/nature03033},
  \href {https://ui.adsabs.harvard.edu/abs/2004Natur.432..194W} {432, 194}

\bibitem[\protect\citeauthoryear{{Wyithe}, {Bolton}  \& {Haehnelt}}{{Wyithe}
  et~al.}{2008}]{Wyithe2008_NZ}
{Wyithe} J. S.~B.,  {Bolton} J.~S.,   {Haehnelt} M.~G.,  2008, \mn@doi [\mnras]
  {10.1111/j.1365-2966.2007.12578.x}, \href
  {https://ui.adsabs.harvard.edu/abs/2008MNRAS.383..691W} {383, 691}

\bibitem[\protect\citeauthoryear{{Xu}, {Ferrara}  \& {Chen}}{{Xu}
  et~al.}{2011}]{Xu_2011}
{Xu} Y.,  {Ferrara} A.,   {Chen} X.,  2011, \mn@doi [\mnras]
  {10.1111/j.1365-2966.2010.17579.x}, \href
  {https://ui.adsabs.harvard.edu/abs/2011MNRAS.410.2025X} {410, 2025}

\bibitem[\protect\citeauthoryear{{Yang} et~al.,}{{Yang}
  et~al.}{2020a}]{Yang_2020_z75}
{Yang} J.,  et~al., 2020a, \mn@doi [\apjl] {10.3847/2041-8213/ab9c26}, \href
  {https://ui.adsabs.harvard.edu/abs/2020ApJ...897L..14Y} {897, L14}

\bibitem[\protect\citeauthoryear{{Yang} et~al.,}{{Yang}
  et~al.}{2020b}]{Yang_2020_z63}
{Yang} J.,  et~al., 2020b, \mn@doi [\apj] {10.3847/1538-4357/abbc1b}, \href
  {https://ui.adsabs.harvard.edu/abs/2020ApJ...904...26Y} {904, 26}

\bibitem[\protect\citeauthoryear{{Yu} \& {Tremaine}}{{Yu} \&
  {Tremaine}}{2002}]{YuTremaine_2002}
{Yu} Q.,  {Tremaine} S.,  2002, \mn@doi [\mnras]
  {10.1046/j.1365-8711.2002.05532.x}, \href
  {https://ui.adsabs.harvard.edu/abs/2002MNRAS.335..965Y} {335, 965}

\bibitem[\protect\citeauthoryear{{Zhu} et~al.,}{{Zhu} et~al.}{2022}]{Zhu2022}
{Zhu} Y.,  et~al., 2022, \mn@doi [\apj] {10.3847/1538-4357/ac6e60}, \href
  {https://ui.adsabs.harvard.edu/abs/2022ApJ...932...76Z} {932, 76}

\makeatother
\end{thebibliography}



\appendix

\section{The dependence of \texorpdfstring{$R_{21}$}{R21} on transmission threshold} \label{app:R21}

\begin{figure}
    \begin{minipage}{\columnwidth}
 	  \centering
 	  \includegraphics[width=\linewidth]{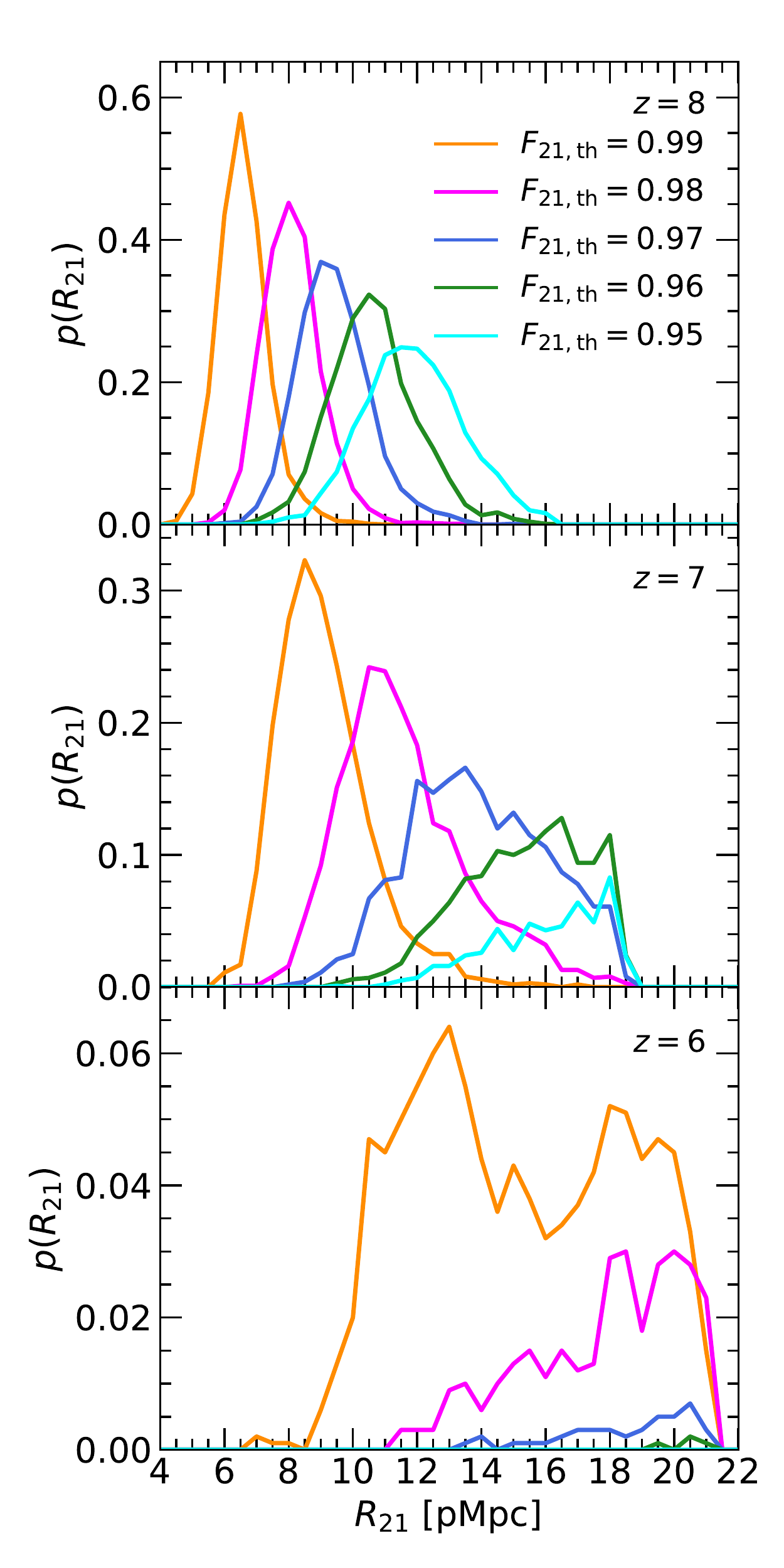}
	\end{minipage}
	\vspace{-0.5cm}
    \caption{The probability distribution of \RCM~assuming different values for distance from the quasar at which the 21-cm transmission first drops below $F_{\rm 21,th}$, after smoothing the 21-cm spectrum with a boxcar filter of width $5\rm\,kHz$.  The results are shown for our fiducial model for 2000 sight lines at $z=8$, 7 and 6.  The orange curves for $F_{\rm 21,th}=0.99$ are the same as the solid curves shown in the lower left panel of Fig.~\ref{fig:R21_RLya_hist_param}.  Note the different scale on the vertical axes of each panel; many sight lines at $z=6$ show no 21-cm absorption with $F<F_{\rm 21,th}$.  Additionally, the length of the simulated sight-lines is $100 h^{-1}\rm\,cMpc$, so there is an artificial cut-off in the distributions at $R_{21}=[16.4,\,18.4,\,21.1]\rm\,pMpc$ at $z=[8,\,7,\,6]$.}
    \label{fig:R21_hist_F21th}
\end{figure}

\begin{table}
    \centering
    \caption{The minimum flux density required to detect a 21-cm forest absorption feature with $F_{\rm 21,th}$ with $\rm S/N=5$ using SKA1-low (middle column) or SKA2 (right column). This has been calculated from Eq.~(\ref{eq:Smin}) assuming a bandwidth of $\Delta\nu=5\rm\,kHz$, sensitivity $A_{\rm eff}/T_{\rm sys}=600\rm\, m^{2}\,K^{-1}$ $\left(5500\rm\, m^{2}\,K^{-1}\right)$ \citep[][]{Braun_2019} and an integration time of $t_{\rm int}=1000\rm\,hr$ $\left(100\rm\,hr\right)$ for SKA1-low (SKA2).}
    \label{tab:Smin}
    \begin{tabular}{c c c}
    \hline
    $F_{\rm 21,th}$ & $S_{\rm min}/\rm\,mJy,\, SKA1-low$ & $S_{\rm min}/\rm\,mJy,\, SKA2$ \\     
    \hline
    0.99 & 17.2 & 5.9 \\
    0.98 & 8.6 & 3.0 \\
    0.97 & 5.7 & 2.0 \\
    0.96 & 4.3 & 1.5 \\
    0.95 & 3.4 & 1.2 \\
    \hline
    \end{tabular}
\end{table}

In analogy to the widely used definition for \RLya~\citep[e.g.][]{Fan_2006}, our definition of \RCM~is practical rather than physically motivated. The choice of $F_{\rm 21,th}=0.99$ as the transmission threshold where we define \RCM~is somewhat arbitrary.  Here we show how a different choice of $F_{\rm 21,th}$ affects our results. Fig. \ref{fig:R21_hist_F21th} shows the distribution of \RCM~in our fiducial RT-late reionization model at redshift $z=8$, 7 and 6, assuming a range of $F_{\rm 21,th}$ values.  We have assumed $M_{1450}=-27$, $f_{\rm X}=0.01$, $t_{\rm Q}=10^{7}\rm\,yr$ and our fiducial quasar SED in the models.  Decreasing $F_{\rm 21,th}$ shifts the $R_{21}$ distribution to larger values, consistent with the expectation that stronger 21-cm absorption features should appear further from the quasar due to the lower spin temperatures (see e.g. Fig.~\ref{fig:uniform_profiles}).

In addition, note that while we find absorption features with $F_{\rm 21,th}\geq0.98$ in almost all sight lines at $z=7$, only $62$ per cent contain features with $F_{\rm 21,th}=0.96$, and this further decreases to $26$ per cent for $F_{\rm 21,th}=0.95$.  In Table~\ref{tab:Smin}, we list the minimum intrinsic flux density that a radio source must have for SKA1-low or SKA2 to detect a 21-cm forest absorber with $F_{\rm 21,th}$ at a signal-to-noise ratio of $\rm S/N=5$.  Here we use Eq.~(\ref{eq:Smin}), and assume $A_{\rm eff}/T_{\rm sys}=600\rm\, m^{2}\,K^{-1}$ and $t_{\rm int}=1000\rm\,hr$ for SKA1-low and $A_{\rm eff}/T_{\rm sys}=5500\rm\, m^{2}\,K^{-1}$ and $t_{\rm int}=100\rm\,hr$ for SKA2, and a bandwidth of $\Delta\nu=5\rm\,kHz$.

\section{The dependence of \texorpdfstring{$R_{21}$}{R21} on quasar magnitude} \label{app:R21_Nion}

\begin{figure}
    \begin{minipage}{\columnwidth}
 	  \centering
 	  \includegraphics[width=\linewidth]{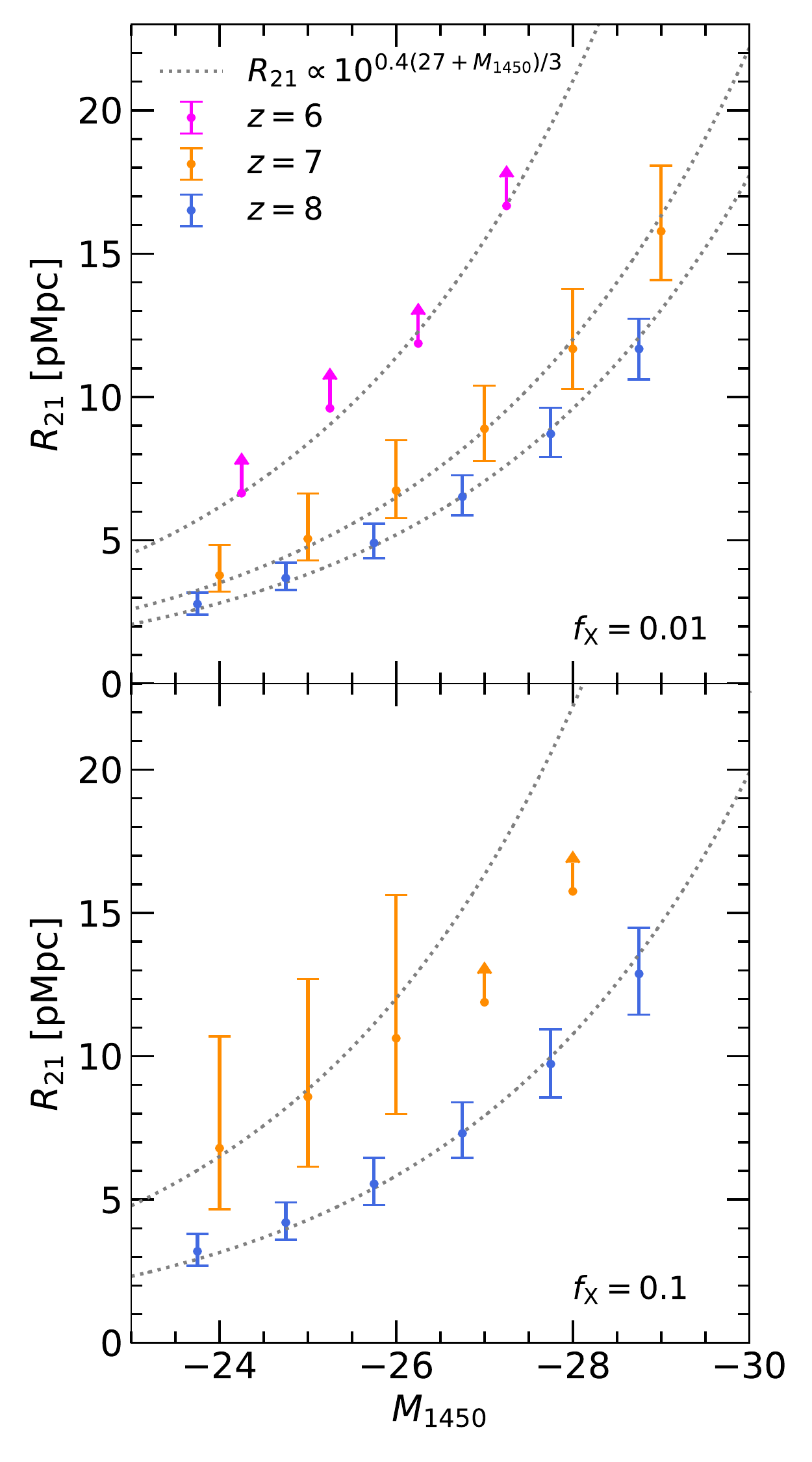}
	\end{minipage}
	\vspace{-0.5cm}
    \caption{The 21-cm near-zone size, $R_{21}$, as a function of the quasar magnitude, $M_{1450}$, at $\rm z=6$ (fuchsia points), $z=7$ (orange points) and $z=8$ (blue points) in the RT-late model.  The fiducial SED and optically/UV bright lifetime of $t_{\rm Q}=10^{7}\rm\,yr$ are assumed, for an X-ray background efficiency $f_{\rm X}=0.01$ (upper panel) and $f_{\rm X}=0.1$ (lower panel).  The data points correspond to the median and $68$ per cent range for $2000$ simulated quasar sight-lines.  Arrows indicate the $68$ per cent lower limit for $\rm R_{21}$ when multiple sight-lines have no pixels with $F_{\rm 21,th}<0.99$.  The points are slightly offset on the horizontal axes for presentation purposes.  The grey dotted curves show $R_{21}\propto 10^{0.4(27+M_{1450})/3}$, which is the expected scaling for an \HII region (i.e. $R_{21}\propto \dot{N}^{1/3}$).  Note also there are no sight-lines with $F_{\rm 21,th}<0.99$ for $f_{\rm X}=0.1$ at $z=6$.}
    \label{fig:R21_vs_M1450}
\end{figure}

The dependence of $R_{\rm Ly\alpha}$ on the quasar magnitude, $M_{1450}$ (or equivalently the ionizing photon emission rate, $\dot{N}$) has been discussed extensively elsewhere \citep[e.g.][]{Bolton_Haehnelt_2007,Davies_2020,Ishimoto_2020,Satyavolu_2022}.   In particular, \citet{Eilers_2017} derived the scaling relation in Eq.~(\ref{eq:RNZ_cor}) using their radiative transfer simulations.  Analogously, we present the dependence of \RCM~on $M_{1450}$ in Fig.~\ref{fig:R21_vs_M1450} for $f_{\rm X}=0.01$ (top panel) and $f_{\rm X}=0.1$ (bottom panel) at $z=6$ (fuchsia points), $z=7$ (orange points) and $z=8$ (blue points) for a quasar with an optically/UV bright lifetime of $t_{\rm Q}=10^7\rm\,yr$. The error bars show the 68 per cent scatter around the median obtained from $2000$ simulated sight lines, and the arrows show $68$ per cent lower limits. 

We find $R_{21}\propto 10^{0.4(27+M_{1450})/3}\propto \dot{N}^{1/3}$ (dashed grey curves) is consistent with the simulations, in agreement with the expected scaling for the expansion of a quasar \HII region given by Eq.~(\ref{eq:RHII}) (although note, as discussed earlier, $R_{21}$ does not necessarily correspond to $R_{\rm HII}$ -- it instead roughly corresponds to the size of the region heated to $T_{\rm S}\gtrsim 100\rm\,K$ by the quasar). The only exception is for $f_{\rm X}=0.1$ at $z=6$, where proximate 21-cm absorption is very rare due to the heating of the remaining neutral gas in the IGM to spin temperatures $T_{\rm S}\gtrsim 10^{2}\rm\,K$. In this case only $\sim 0.2$ per cent of our $2000$ synthetic spectra have $R_{21}<21\rm\,pMpc$ for $M_{1450}>-27$, and even fewer for more luminous quasars.  For comparison, \citet[][]{Soltinsky_2021} infer a lower limit of $f_{\rm X}>0.109$ assuming a null detection of 21-cm absorption with $F_{21}\leq0.99$ over a path length of $ 5.8h^{-1}\rm\,cGpc$ ($\Delta z = 20$) at $z=6$ (see their table 2).  However, these numbers are for the general IGM, and exclude the effect of localised ionization and heating in close proximity to bright sources.  Here, over our simulated path length of  $200h^{-1}\rm\,cGpc$ ($\Delta z = 687.9$) at $z=6$, from \citet{Soltinsky_2021} we would naively expect $\sim 34$ 21-cm absorbers with $F_{\rm 21}<0.99$.  Instead, we find only 3 absorbers.  This difference is largely due to the soft X-ray heating by the quasars reducing the incidence of the proximate 21-cm absorbers, and the rapid redshift evolution of the average IGM neutral fraction along our $100h^{-1}\rm\,cMpc$ sight lines.

\section{The quasar lifetime distribution obtained from \texorpdfstring{L\lowercase{y}$\alpha$}{alpha} near-zone sizes} \label{app:tq_obs}

\begin{figure}
    \begin{minipage}{\columnwidth}
 	  \centering
 	  \includegraphics[width=\linewidth]{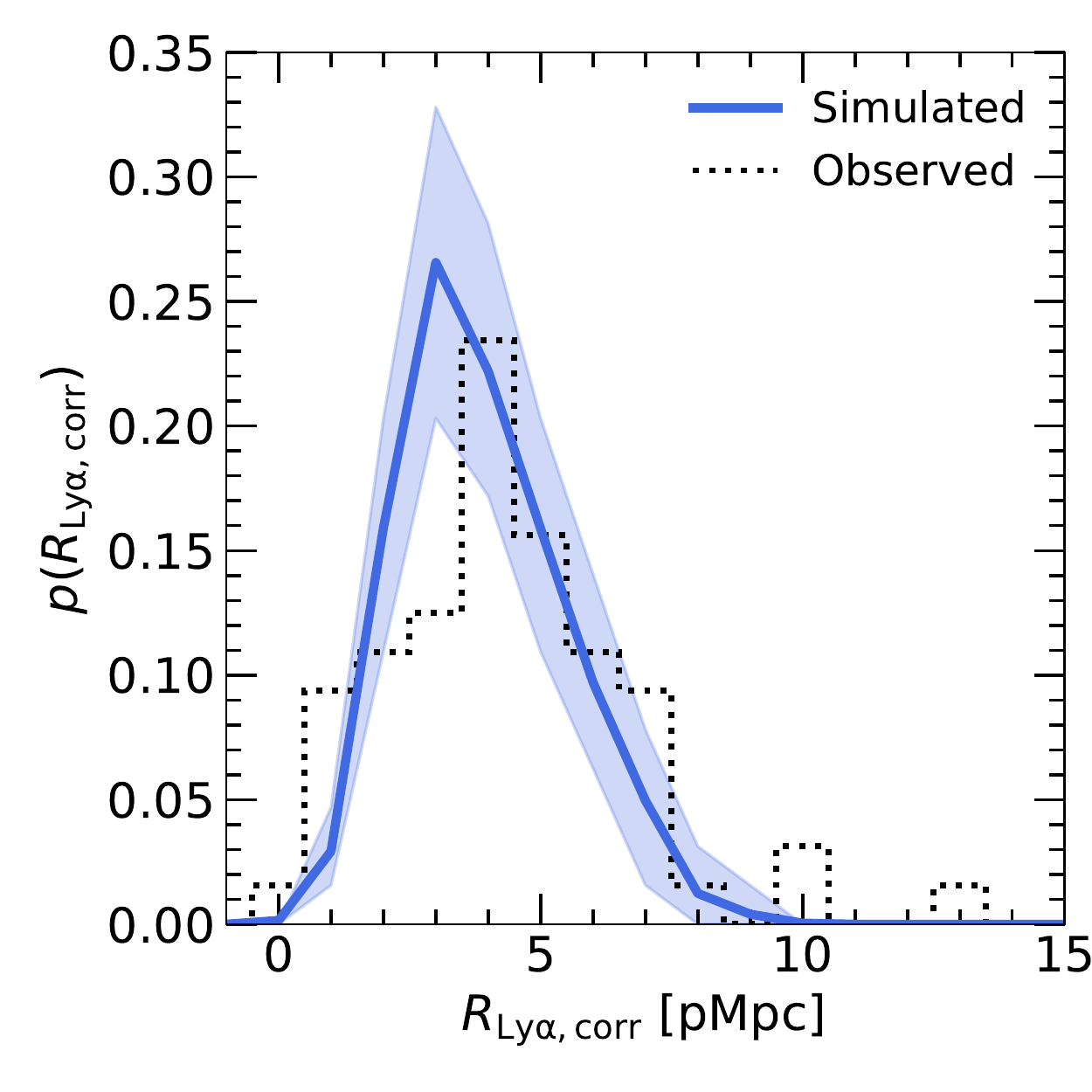}
	\end{minipage}
	\vspace{-0.4cm}
    \caption{The probability distribution for (luminosity corrected) \Lya\ near-zone sizes (blue solid curve) at $z=6$ from radiative transfer simulations using our fiducial model the quasar lifetime distribution from \citet[][]{Morey_2021}.  The shaded region shows the $1\sigma$ uncertainty obtained by bootstrapping.  For comparison, the $R_{\rm Ly\alpha,corr}$ distribution from observed quasars in the redshift range $5.8\leq z\leq6.6$ is shown by the dotted histogram.}
    \label{fig:tq_dist}
\end{figure}

\cite{Morey_2021} have recently demonstrated that the majority of $R_{\rm Ly\alpha,\rm corr}$ measurements at $z\simeq 6$ are reproduced assuming a median optically/UV bright lifetime of $t_{\rm Q}=10^{5.7}\rm\,yr$ with a $95$ per cent confidence interval $t_{\rm Q}=10^{5.3}$--$10^{6.5}\rm\,yr$ (see their fig. 6).  We test this in Fig.~\ref{fig:tq_dist}, where instead of using a single value for $t_{\rm Q}$ in our simulations, we adopt values using the posterior probability distribution for the quasar lifetimes inferred by \citet[][]{Morey_2021}.  We select 2000 quasar lifetime values from their distribution using a Monte Carlo rejection method. Each simulated sight line was then randomly assigned a different $t_{\rm Q}$ from this sample.  We then performed 2000 radiative transfer simulations of our fiducial model at $z=6$, and bootstrapped $10^{4}$ sets of sight lines from these simulations to obtain a $1\sigma$ uncertainty. Each bootstrapped set contains 64 synthetic sight lines, corresponding to the number of quasars in the compiled observational sample we use for quasars at $5.8\leq z\leq 6.6$.

The dotted black curve in Fig. \ref{fig:tq_dist} shows the observed distribution of luminosity corrected  \Lya\ near-zone sizes at $5.8\leq z\leq 6.6$.  The solid blue curve corresponds to the median and $1\sigma$ uncertainty obtained by bootstrapping our simulations.  A two-sided Kolmogorov-Smirnov test yields a p-value of $0.055$, which remains consistent ($p>0.05$) with the null-hypothesis that the samples are drawn from the same distribution.  There is a hint that the simulated near-zone sizes are slightly smaller than the observational data, which may be a result of applying the \citet{Morey_2021} $t_{\rm Q}$ distribution to our late reionization model \citep[see also][]{Satyavolu_2022}.  Our RT-late simulation has a larger average IGM neutral fraction at $z=6$ compared to the models used by \citet{Morey_2021}, which assumes a fully ionized IGM.  However, this difference is not highly significant.

\bsp	
\label{lastpage}
\end{document}